\documentclass[useAMS,usenatbib]{mn2e}
\usepackage[pdfpagemode={UseOutlines},bookmarks,bookmarksopen,colorlinks,citecolor={blue},urlcolor={red}]{hyperref}
\usepackage[flushleft]{threeparttable}
\usepackage{booktabs,caption}
\usepackage{graphicx}
\usepackage{amsmath}
\usepackage{subfigure}
\usepackage{amssymb}
\usepackage{tabularx}
\usepackage{natbib}
\usepackage{float}
\usepackage{etoolbox}
\usepackage{natbib}
\usepackage[usenames,dvipsnames]{xcolor}
\usepackage[normalem]{ulem}
\usepackage{multirow}
\newcommand{\Ms}{M$_\odot$ }
\newcommand{\msun}{\ensuremath{\,\textrm{M}_{\odot}}}
\newcommand{\pc}{\ensuremath{\,\textrm{pc}}}
\newcommand{\pcc}{\ensuremath{\,\textrm{pc}^3}}
\newcommand{\kpc}{\ensuremath{\,\textrm{kpc}}}
\newcommand\kms{\, \rm km\,s^{-1}}
\newcommand{\gpcc}{\ensuremath{\,\textrm{Gpc}^{-3}}}

\newcommand{\bh}{{\rm BH}}
\newcommand{\imbh}{{\rm IMBH}}
\newcommand{\emri}{{\rm EMRI}}
\newcommand{\gc}{{\rm GC}}

\newcommand{\bea}{\begin{eqnarray}}
\newcommand{\eea}{\end{eqnarray}}

\newcommand{\gae}{\lower 2pt \hbox{$\, \buildrel {\scriptstyle >}\over {\scriptstyle \sim}\,$}} % % \gtrsim 
\newcommand{\lae}{\lower 2pt \hbox{$\, \buildrel {\scriptstyle <}\over {\scriptstyle \sim}\,$}} % % \lesssim 

\makeatletter

\patchcmd{\NAT@citex}
  {\@citea\NAT@hyper@{%
     \NAT@nmfmt{\NAT@nm}%
     \hyper@natlinkbreak{\NAT@aysep\NAT@spacechar}{\@citeb\@extra@b@citeb}%
     \NAT@date}}
  {\@citea\NAT@nmfmt{\NAT@nm}%
   \NAT@aysep\NAT@spacechar\NAT@hyper@{\NAT@date}}{}{}

\patchcmd{\NAT@citex}
  {\@citea\NAT@hyper@{%
     \NAT@nmfmt{\NAT@nm}%
     \hyper@natlinkbreak{\NAT@spacechar\NAT@@open\if*#1*\else#1\NAT@spacechar\fi}%
       {\@citeb\@extra@b@citeb}%
     \NAT@date}}
  {\@citea\NAT@nmfmt{\NAT@nm}%
   \NAT@spacechar\NAT@@open\if*#1*\else#1\NAT@spacechar\fi\NAT@hyper@{\NAT@date}}
  {}{}

\makeatother

\title[Gravitational wave sources from inspiralling GCs]{Gravitational wave sources from inspiralling globular clusters in the Galactic Centre and similar environments}

\author[M. Arca-Sedda and A. Gualandris]{Manuel
  Arca-Sedda$^{1}$\thanks{E-mail: m.arcasedda@gmail.com} and Alessia
  Gualandris$^{2}$\\ $^{1}$Zentrum f\"{u}r Astronomie der Universit\"{a}t Heidelberg, Astronomisches Rechen-Institut, M\"{o}nchhofstr. 12-14, D-69120, Heidelberg (Germany)\\
$^{2}$Department of Physics, Faculty of Engineering and Physical Sciences, University of Surrey, Guildford GU2 7XH, UK \\ }

\begin{document}
%\date{Revised to }
\date{}

\pagerange{\pageref{firstpage}--\pageref{lastpage}} \pubyear{2018}

\maketitle

%\label{firstpage}

\begin{abstract}
We model the inspiral of globular clusters (GCs) towards a galactic
nucleus harboring a supermassive black hole (SMBH), a leading scenario
for the formation of nuclear star clusters. We consider the case of
GCs containing either an intermediate-mass black hole (IMBH) or a
population of stellar mass black holes (BHs), and study the formation
of gravitational wave (GW) sources. We perform direct summation
$N$-body simulations of the infall of GCs with different orbital
eccentricities in the live background of a galaxy with either a
shallow or steep density profile. We find that the GC acts as an
efficient carrier for the IMBH, facilitating the formation of a bound
pair. The hardening and evolution of the binary depends sensitively on
the galaxy's density profile.  If the host galaxy has a shallow
profile the hardening is too slow to allow for coalescence within a
Hubble time, unless the initial cluster orbit is highly eccentric. If
the galaxy hosts a nuclear star cluster, the hardening leads to
coalescence by emission of GWs within $3-4$ Gyr.  In this case, we
find a IMBH-SMBH merger rate of $\Gamma_{\rm IMBH-SMBH} = 2.8\times
10^{-3}$ yr$^{-1}\gpcc$.  If the GC hosts a population of
stellar BHs, these are deposited close enough to the SMBH to form
extreme-mass-ratio-inspirals with a merger rate of $\Gamma_{\rm EMRI}
= 0.25$ yr$^{-1}\gpcc$.  Finally, the SMBH tidal field can boost
the coalescence of stellar black hole binaries delivered from the
infalling GCs. The merger rate for this merging channel is
$\Gamma_{\rm BHB} = 0.4-4$ yr$^{-1}\gpcc$.
\end{abstract}


\section{Introduction}
\label{sec:intro}
It is now well established that the majority of galaxies -
possibly with the exception of very low mass galaxies and dwarf
irregulars - harbour supermassive black holes (SMBHs) at their
centres, and their presence is expected to be ubiquitous in massive
galaxies $(M \gae 10^9\msun)$ \citep[e.g.][]{ff2005}.
Similarly, there is strong evidence for the existence of dense stellar
 nuclei, usually referred to as nuclear star clusters (NSCs)
\citep{cote06,Neum12,Turetal12,georgiev14}. With effective radii of a
few parsecs they are similar in size to galactic globular clusters
(GCs) but significantly more massive $(M \sim 10^6 - 10^8\msun)$.
They are among the densest known stellar systems in the Universe,
which makes them ideal locations for stellar dynamical encounters and
nurseries of gravitational waves (GWs) sources.  Like SMBHs, NSCs obey
scaling relations with the properties of their hosts, including galaxy
mass and stellar velocity dispersion \citep[e.g.][]{graham2007}.

While SMBHs and NSCs are found in galaxies of different type, there is
evidence for SMBHs to be found predominantly above a threshold stellar
mass of about $10^{10}\msun$, with NSCs found preferentially below such
limit \citep{scot}. However, the two are not mutually exclusive and
have been observed to coexist in at least a dozen galaxies with
stellar masses ranging from $10^8\msun$ to $10^{10}\msun$
\citep{Graham2009,georgiev16,melo17}. The Milky Way is one such case: it hosts a central
SMBH with mass $\sim 4 \times 10^6\msun$ \citep{schodel07,ghez08,gillessen09} and a massive
NSC, with a mass $\sim 2.5 \times 10^7\msun$ \citep{schodel14,gallegocano2017,schodel2017}.
SMBHs and NSCs are often referred to as compact massive objects (CMOs).

The transition from NSC to SMBH dominated galaxies is likely a
consequence of their formation histories. Evidence for different
formation histories is provided by the shallower slope observed in the
$M-\sigma$ relation between CMO mass and stellar velocity dispersion
of NSCs with respect to SMBHs \citep{LGH,scot,ASCD14b}.

A leading scenario for the formation of NSCs is the {\it dry-merger}
scenario, according to which NSCs form by migration and merger of a
population of GCs that inspiral from large galactocentric distances
due to dynamical friction \citep{Trem75,Dolc93}. Theoretical arguments
and numerical simulations show that the scenario well reproduces the
observed scaling relations of NSCs
\citep{antonini13,gnedin14,ASCD14b}.  

The main observational properties of the Milky Way NSC can be
  successfully reproduced through this formation mechanism, as widely
  shown by numerical simulations \citep{AMB,Tsatsi17}, although a
  contribution from in-situ star formation cannot be completely ruled
  out \citep{baumgardt2018}.  Moreover, the dry-merger scenario provides
  a suitable explanation for the intense flux of $\gamma$ rays coming
  from the Galactic Centre, which would be due to millisecond pulsars
  delivered by the infalling clusters
  \citep{brandt15,abbate18,Fragioneantonini,arcakocsis18}, and the
  Galactic central X-ray excess, which would be due to cataclysmic
  variables \citep{arcakocsis18}.  

The dry-merger scenario has interesting consequences for the evolution of dwarf galaxies, 
possibly connecting with the evolution of their dark matter content
and the missing formation of SMBHs in these low-mass systems
\citep{ASCD16a, ASCD17b}.

The buildup of a NSC is possible even in nuclei with a pre-existing
SMBH \citep{AMB,ASCD15He}, though in this case tidal disruption of the
clusters leads to lower density NSCs and possibly explains the lack of
NSCs in galaxies with SMBHs more massive than $\sim 10^8\msun$
\citep{antonini13,ASCDS15}. In addition, in giant elliptical galaxies
the inspiral time-scale for the clusters may be too long to allow for
the formation of a NSC \citep{antonini13,ASCDS15,ASCD17a}.

Internal evolution of the GC, driven by the interplay of dynamics and
stellar evolution, may lead to mass segregation as the GC migrates
towards the galactic centre. As a result, the most massive stars
segregate to the centre and form a dense massive stellar system (MSS),
likely dominated by the presence of stellar mass black holes (BHs) and
other heavy remnants \citep{freitag06a,as16}.   As recently
  discussed by \cite{ArcaSeddaetal18}, MSSs can survive up to 12 Gyr,
  leaving an observable fingerprint on their parent cluster. Using the
  correlations connecting the MSS and the corresponding GC properties,
  \cite{askar18} have shown that at least 29 Galactic GCs can contain
  at their centre a noticeable MSS, comprised of 10-200 BHs with
  average masses of $\sim 10-20\msun$.  

In dense and compact clusters mass segregation may initiate a phase of
runaway collisions among stars and binaries that eventually lead to
the formation of an intermediate-mass black hole (IMBH)
\citep{zwart02,freitag06c,gaburov08,Giersz15} or a very massive star
(VMS) \citep{freitag06c, mapelli16}.

 As suggested by \cite{Giersz15}, the IMBH assembly can occur
  through either a fast or slow process. According to the ``FAST''
  scenario, the BH population forms a very dense subsystem in which
  collisions between single and binary BHs are greatly enhanced,
  leading rapidly to the formation of an IMBH seed. Anisotropic GW
  emission can result in significant recoil velocities for the IMBH
  which may be ejected from the cluster as a result, at least for IMBH
  masses $\lesssim 10^3\msun$ \citep{bockelmann08,fragione17a}.

  In the ``SLOW'' formation process, the cluster density is much lower
  ($\sim 10^5-10^6\msun$) and BHs are mostly ejected in binary-binary
  and binary-single interactions, while BH-BH mergers are strongly
  suppressed. In this case, the few-body interactions result in the
  ejection of all BHs but one, which starts growing slowly through
  repeated mergers with surrounding stars.  In this case, the IMBH
  that forms is likely to be retained in the cluster. The SLOW process seems to have a higher probability than the FAST process \citep{Giersz15}, however we caution that the formation channel strongly depends on the initial cluster properties.

However, \cite{petts17} show that the collapse of a VMS into an IMBH
can be prevented by strong stellar winds.  If the time-scale for the
formation of an IMBH or VMS is shorter than the inspiral time-scale of
the cluster, the massive object will be deposited close to the SMBH.

Such interactions are particularly interesting in the context of GWs
emission \citep[e.g.][]{baumgardt06,mapelli12}, especially in light
of the recent detections of black hole
\citep{abbott16a,abbott16b,abbott16c,abbott17a,abbott17b} and neutron
star \citep{abbott17c,abbott17d,abbott17e} mergers by the LIGO
experiment, and the recent success of the LISA Pathfinder testing
mission whose results have validated beyond expectations the
feasibility of LISA's detection principle \citep{Amaro13}.

In this study, we model the orbital evolution of a GC inspiralling
towards the centre of a host galaxy with a central SMBH.  Earlier
studies suggest that if all the GCs that contributed to the formation
of the Milky Way nuclear cluster delivered an IMBH to the Galactic
Centre, they should have given rise to well detectable kinematical
signatures \citep{mastrobuono14}. On the another hand, the dynamical
formation of IMBHs in GCs seems to have low efficiency, with a
``success'' probability of $\sim 20\%$ \citep{Giersz15}.

 In our Milky Way, NSC formation likely resulted from the rapid merger
of a few massive clusters, either formed in the inner galactic bulge
($\simeq 300\pc$) or segregated from larger distances due to dynamical friction.
As shown by \cite{ASCD15He}, star clusters with masses below $5\times
10^5\msun$ are strongly affected by the tidal forces of a SMBH with mass $\gae10^6\msun$
even though they formed in the galactic nucleus, and contribute little to the NSC formation.  As a
consequence, the typical number of clusters expected to contribute to
the NSC formation in a Milky Way type galaxy is $\sim 10$
\citep{AMB,ASCD14b,gnedin14}.

Given the IMBH formation probability, we expect that during the
assembly of the NSC $\sim 2$ IMBHs were dragged to the centre in the
dry merger scenario.

On the hand, as discussed above, the NSC could have formed through
in-situ star formation \citep{king03,king05,Mil04,nayakshin}. In this
case, it is still possible for a GC to deliver its IMBH close to the
SMBH, either during the NSC assembly or much later, depending on the
GC birth location.

We consider GCs harboring either an IMBH or a BHs sub-system in the
centre, and investigate the formation of GW sources, either a
SMBH-IMBH binary or an extreme mass ratio inspiral (EMRI) resulting
from the interaction between the SMBH and stellar BHs.  We employ
high-accuracy $N$-body simulations of the GC inspiral in a live galaxy
background, varying the CMO mass, GC orbit and galaxy properties.  We
find that the late evolution of the IMBH-SMBH binary depends strongly
on the environment in which it lives. Indeed, the interaction with
galactic stars and the GC stellar debris can drive the binary towards
coalescence on time-scales of $1-10$ Gyr. The IMBH-SMBH shrinkage is
maximized for GCs moving on highly eccentric orbits and galactic
nuclei with a high central density.  On the another hand, in the case
in which the GC delivers stellar BHs to the galactic centre (both
single and in binaries), we find a non-negligible merger rate in terms
of EMRIs ($\Gamma = 0.25 {\,\rm yr}^{-1} \gpcc$) and black
hole binaries (BHBs) ($\Gamma = 0.4-4 {\,\rm yr}^{-1}\gpcc$)
coalescence.

The paper is organized as follows: in Sect. \ref{sec:method} we
discuss the adopted models for the GC and the galaxy as well as the
numerical methods; in Sect. \ref{sec:result} we introduce our results,
discussing the impact of a GC CMO on the overall evolution of the GC
and the galactic nucleus; Sect. \ref{sec:gws} is devoted to discuss
the subsequent evolution of the binary system composed by the IMBH and
the SMBH, and the evolution of a population of stellar BHs around the
SMBH. Finally, in Sect. \ref{end} we present our conclusions.

\section{Method}
\label{sec:method}
In order to study the possible formation of GW sources during the
inspiral of a GC towards the centre of its host galaxy, we perform
high-accuracy direct summation $N$-body simulations of the orbital
evolution of a GC by means of HiGPUs, a highly parallel direct
$N$-body integrator implementing a 4-th order Hermite scheme with
block time-steps running on Graphic Processing Units \citep{Spera}.

We consider three main sets of simulations, distinguished by the
presence of either an IMBH or a population of stellar mass BHs in the
infalling GC, and by the slope of the density profile of the host
galaxy.

The different models reflect the fact that the time-scale of NSC
formation in the dry-merger scenario is not well constrained.
Indeed, the infalling GCs can either: i) form far from the galactic
nucleus and slowly inspiral toward the centre, or ii) form already
within the galactic nucleus and segregate rapidly.
None of the above possibilities is at odds with our current knowledge
of nuclear cluster formation and evolution. In fact, the MW NSC
contains a large fraction of old, metal poor stars typical of the
Galactic GCs. While this suggests a direct connection between
infalling GCs and NSCs, it does not constrain the time of NSC
formation in the lifetime of the host galaxy.

Recent observations of star-burst galaxies support the possibility that
massive star clusters form within the inner $100-500\pc$ of a galactic
nucleus \citep{nguyen14}.  Since the assembly of an NSC can occur
within $100$ Myr of the GCs formation, i.e. while the star clusters
are still dynamically evolving \citep{ASCD15He}, it is possible that
the population of stellar mass BHs in the clusters is still largely
unaffected by dynamics while the NSC is being built. The time-scale
over which stellar BHs form and segregate to the cluster centre is
dictated by the stellar evolution and dynamical friction
time-scales. A star with an initial mass of $40\msun$ evolves into a
BH in $\sim 10$ Myr, and segregates toward the cluster centre on a
dynamical friction time-scale \citep{as16,antonini16}. This is $\sim
10-100$ Myr for a $10^6\msun$ GC having a population of stellar BHs
distributed around the cluster core \citep{ASCD17b}.  Therefore, most
of the BHs will be inside their parent clusters during NSC formation,
unless natal kicks are large enough to expel them at birth. The
resulting NSC will be characterised by a significant population of
BHs.  In this case, NSC formation occurs on a time-scale much shorter
than the time needed for IMBH formation \citep{Giersz15}. This, in
turn, implies that an IMBH can be brought to the galactic centre only
if it is transported from a GC initially orbiting outside the galactic
nucleus, characterised by a longer dynamical friction time.

On the other hand, if the clusters form outside the galactic nucleus, their
orbital decay will be much slower. Thus, the time for NSC build up 
can become sufficiently long for an IMBH to form and bind to the SMBH
while the NSC is still growing, and for the stellar BHs population to
have been significantly reduced by internal dynamics.

Our models capture both the possibility that an NSC forms rapidly, and
subsequently an infalling GC drags its IMBH to the centre, and that
the NSC forms slowly, after an inspiral sufficiently slow to allow for
IMBH formation in the cluster core. 
The main properties of the models
are given in the first four columns of Table \ref{tab:sims}.
\begin{table*}
  \begin{center}
    \caption{Main simulations properties:
  simulation name, galaxy density profile, number of IMBHs, number of
  stellar BHs, number of GC stars, GC stars mass, GC mass, dimensionless potential,
  core radius, initial distance from SMBH.}
    \label{tab:sims}.
\begin{tabular}{cccccccccc}
\hline
Name & Galaxy profile & $N_{\rm IMBH}$ & $N_{\rm BH}$ &   $N_s$ & $m_*$ & $M_{\rm GC}$ &  $W_0$ & $r_c$ & $r_a$ \\
  & & & & & ($\msun$) & ($\msun$) & & (pc) & (pc) \\
\hline
S & shallow  & 1 & 0 & 29832 & 33 & $10^6$ & 6 & 0.24 & 50\\
M & steep & 1 & 0 & 22262 &  45 & $10^6$ & 6 & 0.24 & 50\\
B & shallow  & 0 & 114 & 98713 & 10 & $10^6$ & 6 & 0.24 & 50\\
\hline
\end{tabular}
  \end{center}
\end{table*}

We model both the globular cluster and the background galaxy as an
$N$-body system with direct summation of all gravitational
forces. This approach is extremely demanding from a computational
point of view and, despite the high degree of parallelism allowed by
the code, we are limited to a total particle number $N \sim 2^{20}\sim
10^6$.

Following previous numerical studies \citep{ASCD15He,ASCDS15}, we
distribute particles between the galaxy and the GC, so that $N = N_g
+ N_{\rm GC}$, where $N_g$ represents the number of particles in the
galaxy and $N_{\rm GC}$ represents the total number of particles in
the GC. In general, for the galaxy we have $N_g = N_{\rm SMBH} +
N_{\rm gs}$, with $N_{\rm gs}$ the number of stars and $N_{\rm SMBH}
=1$ the number of SMBHs.  For the GC we have $N_{\rm GC} = N_s +
N_{\rm IMBH} + N_{\rm BH}$, where $N_s$ is the number of stars,
$N_{\rm IMBH}$ is the number of IMBHs and $N_{\rm BH}$ is the number
of stellar mass BHs. The particle numbers adopted in the simulations
are given in Table\,\ref{tab:sims} and Table \ref{tab:GALpar},
\ref{tab:GALparM}. With a stellar mass resolution of $10-45\msun$,
these direct summation simulations can be considered state-of-the-art.
In addition, the simultaneous interaction of heavy objects (IMBH,
stellar BHs) with GC stars, galactic particles and the central SMBH
makes these simulations extremely time-consuming. The runs have been
performed over a time span of about 2 years, taking advantage of
ASTROC9, a high-performance workstation hosting 1 RADEON HD7990 and 2
RADEON HD7970 GPUs, ASTROC15, which hosts 4 RADEON HD7970 GPUs, and
ASTROC16b, hosting 4 NVIDIA GTX Titan X GPUs. Morever, the few-body models described in Section \ref{sec:bhs} have been carried out on the Milky Way cluster, hosted at the Heidelberg University in the framework of the SFB881 collaborative research project\footnote{http://sfb881.zah.uni-heidelberg.de/}.

\subsection{The galaxy model}
\label{sec:galaxy}
 As discussed above, the NSC formation process is quite rapid,
  lasting $0.1-1$ Gyr. Over this time interval, GCs segregate and
  accumulate into the growing nucleus, giving rise to two extreme
  possibilities: i) the i-th GC, containing an IMBH, reaches the
  galactic centre when the NSC main body is still not assembled; or
  ii) the GC forms farther away and arrives at late times, merging with
  an already fully grown NSC.

In order to capture these extremes, we provide two different galaxy
models, namely model S and B, in which the galaxy is characterised by
a shallow density profile and the NSC is not yet formed, and model M,
in which the galaxy hosts a central NSC characterised by a fully
relaxed cusp.
For the sake of clarity, letter B denotes those models in which the GC
hosts an MSS, while letters S and M label the cases in which an IMBH
is taken into account.

The simulations with a shallow galaxy profile (S and B) assume a
Dehnen mass density law \citep{Deh93} truncated via a hyperbolic cosine:
\begin{equation}
\rho(r) = \frac{(3-\gamma)M_g}{4\pi r_g^3}\left(\frac{r}{r_g}
\right)^{-\gamma} \left(1+ \frac{r}{r_g} \right)^{-4+\gamma}
\cosh(r/r_{\rm tr}).
\label{rhoD}
\end{equation}
Here, $M_g$ represents the total galaxy mass, $r_g$ the model's scale
length and $\gamma$ the inner density slope of the profile.  In both
models S and B, we set $M_g = 10^{10}\msun$, $r_g = 995\pc$ and
$\gamma=0.3$. However, while in model S we assume a truncation radius
of $r_{\rm tr} = 70\pc$, in model B we choose a smaller value of
$r_{\rm tr} = 40\pc$ in order to correctly reproduce the stellar BH
population against other GC stars. We verified that this choice is
appropriate and does not impact the GC overall orbital evolution.  A
shallow density profile well describes massive elliptical galaxies
\citep[e.g.][]{cote06}.

\begin{figure}
\centering
\includegraphics[width=8cm]{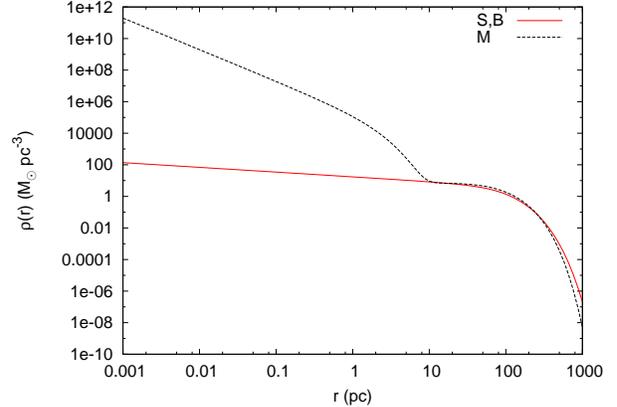}
\caption{Density profile of models S and B (red solid line) and M
  (dashed black line). The profiles are quite similar in the outermost
  regions but differ significantly in the inner $\sim10\pc$ due to the
  presence of a NSC in model M.}
\label{fig:SBMdensity}
\end{figure}
Model M, on the other hand, adopts a steeper galaxy density profile,
typical for example of the Milky Way (MW) nucleus. We assume a
two-components model for the MW comprised of: (i) a nuclear star
cluster (NSC), extending out to about $2\pc$ from the centre and with
a density profile steeply rising towards the SMBH, and (ii) a nuclear
bulge (NB), extending out to about $300\pc$.  For the nuclear bulge,
we consider a Plummer model with total mass $M_{\rm NB} =
3\times10^{10}\msun$ \citep{valenti16}, a scale radius $r_{\rm
  NB}=1\kpc$ and a cut-off radius $r_{\rm o}=50\pc$, thus producing a
circular velocity profile in agreement with that provided by
\citet[][see their Fig.9]{portail15}.  For the NSC we choose instead a
Dehnen profile with $\gamma=2$ \citep{schodel14}.  The mass of the
central SMBH is set to $M_{\rm SMBH}= 5\times10^6\msun$.  This choice
is consistent with both the mass of the MW central SMBH and the
typical MBH expected in a galaxy of $10^{10}\msun$, according to
observational scaling laws \citep{scot}.

 The formation of a steep cusp around the SMBH, with slope $\gamma
  = 1.75$, is expected as a result of two-body relaxation, as
  predicted by \cite{bahcall76} (see \cite{Merri13} and
  \cite{alexander17} for a review).  However, observations of the
  Galactic Centre fail to reveal such a cusp in the older population
  of observable giants \citep{buchholz2009,do2009}. Recent observations
  of the Galactic NSC show that the population of old bright giants is
  characterised by a distribution with slope $\gamma = 1.43$
  \citep{gallegocano2017}, thus shallower than the Bahcall-Wolf
  solution, and the population of old main sequence stars has an even
  shallower distribution $\gamma = 1.23$ \citep{schodel2017}. However,
  the presence of compact remnants can lead to an increase in the
  density slope, bringing it to $\gamma \simeq 1.55$
  \citep{baumgardt2018}. These recent studies seem to suggest that
  some mechanisms are at work in the Galactic Centre preventing, or
  delaying, the formation of a Bahcall-Wolf cusp.  In these regards,
  our galaxy models represent two extreme cases: an almost flat
  distribution (models S) and a steeply cuspy distribution (models
  M).

The density profiles of the galaxy models
are shown in Fig.\,\ref{fig:SBMdensity}, while the parameters of the
galaxy models are given in Table \ref{tab:GALpar} and Table
\ref{tab:GALparM}.
\begin{table}
  \begin{center}
    \caption{Host galaxy parameters adopted in the simulations with a shallow profile:
      simulation name, galaxy mass, scale radius, inner slope, truncation radius,
      number of stars, star mass.}
    \label{tab:GALpar}
\begin{tabular}{ccccccc}
\hline
Name &$M_g$ & $r_g$ & $\gamma$ & $r_t$ & $N_{\rm gs}$ &$m_{g*}$\\
     &($\msun$) & (pc) & & (pc) & &($\msun$)\\
\hline
S&$ 10^{10}$ & $ 995$ & $0.3$ & $ 70$ &$ 1018473$  &$33$\\
B&$ 10^{10}$ & $ 995$ & $0.3$ & $ 40$   &$ 949857$ &$10$ \\
\hline
\end{tabular}
\end{center}
\end{table}

\begin{table}
  \begin{center}
    \caption{Host galaxy parameters adopted in the simulations with a
      steep profile: component, component mass, scale radius,
      truncation radius, number of stars, star mass.}
    \label{tab:GALparM}
\begin{tabular}{cccccc}
\hline
Component &$M$ & $r_g$ & $r_t$ & $N_{\rm gs}$ &$m_*$\\
     & ($\msun$) & (pc) & (pc) & &($\msun$)\\
\hline
NB  & $3\times 10^{10}$ & $10^3$ & 50  & \multirow{2}{*}{1026313} & 45\\
NSC & $1\times 10^{7}$  & 4 &  50  &  & 45\\
\hline
\end{tabular}
\begin{tablenotes}
\item {In Column 5 we provide the total number of particles used to
    model both the NB and the NSC. Since we sampled these two
    components using their collective distribution function, we cannot
    distinguish between NB and NSC stars.}
\end{tablenotes}
\end{center}
\end{table}

In order to generate a self-consistent model, we calculated
numerically the distribution function associated with the global
density profile given by the sum of the NB and NSC profiles. We then
randomly sampled particles from the distribution function to produce
initial conditions. This ensures that the system is stable over a
time-scale comparable to its own relaxation time.

The gravitational interactions among the stars in the galaxy are
smoothed via a softening length $\epsilon = 0.1\pc$, whereas we
assumed $\epsilon = 0.03\pc$ for SMBH-IMBH and SMBH-GC stars
interactions, and $\epsilon = 0.01\pc$ for IMBH-GC stars interactions.

We selected three possible orbits for the GC: circular ($e=0$),
eccentric ($e=0.7$), and radial ($e=1$), assuming for the eccentricity
the usual relation $e=(r_a-r_p)/(r_a+r_p)$, where $r_p$ is the
pericentre and $r_a$ the apocentre.  All the orbits have initial
apocentre $r_a = 50\pc$.

Similarly to the case of model S, we append the letter ``a'' to
simulations with $e=0$, the letter ``b'' to simulations with $e=0.7$
and the letter ``c'' for simulations with $e=1$.

\subsection{The globular cluster model}
\label{sec:cluster}
We adopt a King model \citep{King} for all simulated GCs, with
central dimensionless potential $W_0=6$, core radius $r_c=0.24\pc$
and total mass $M_{\rm GC} = 10^6\msun$. The choice of a relatively
large mass is due to the fact that only massive clusters can reach the
galaxy centre without being disrupted by the tidal forces exerted by
the SMBH and the galactic background (see for instance \cite{ASCD15He}).

 The choice of $r_c$ ensures that the GC King tidal radius
  equals the actual tidal radius, determined by the gravitational
  field of the SMBH and the galaxy, at the GC pericentre in eccentric
  models (``b models'').  We decided to use only one GC model in
  different configurations, in order to focus on the role of the GC
  infall on the IMBH-SMBH binary formation.  A larger core radius
  would imply that the GC model over-fills its Roche lobe at
  pericentre. This could potentially boost the GC dissolution leading
  to an earlier deposition of the IMBH. We note that this can have an
  impact on the time-scale of the IMBH-SMBH binary formation only in
  the ``a models'', since in all the other cases the pairing occurs
  after the first pericentre passage. In the case of an immediate
  disruption of the GC, the the dynamical friction time-scale would
  increase by a factor $\sim (M_{\rm GC}/M_{\rm IMBH})^{0.67}$
  \citep{ASCD14a}.  A smaller core, on the other hand, would imply an
  underfilling model, leading to a later tidal disruption of the GC.
  This would slightly reduce the IMBH-SMBH ``pairing time'', while
  could potentially enhance the mass ejected in the eccentric orbits
  \citep{ASCDS15}.

It is worthy nothing the complexity of the system studied, which
depends on the GC internal properties (mass, potential well, core
radius), the GC orbital parameters (apocentre, eccentricity), the IMBH
mass or the properties of the stellar BHs population, and the galaxy
structural properties (SMBH mass, density slope, total mass and
effective radius). Such a large parameter space is beyond current
state-of-the-art computational capabilities, and we have therefore
decided to restrict our analysis to a single GC model, leaving further
exploration of the parameters to future works.

The main parameters of the GC models in the three sets of simulations
are listed in Table\,\ref{tab:sims}.

We investigated the orbital evolution of this GC model in the live
potential of the host galaxy assuming either a steep (model M) or a
shallow density profile (models S and B). The GC hosts either an IMBH
or a cluster of stellar mass BHs in the centre. The number of BHs in
model B is chosen following \citet{as16}. Mass segregation drives the
formation of a massive stellar system (MSS) composed of white dwarfs,
neutron stars and black holes. The number of these objects depends on
the GC mass, size and metallicity. For a relatively old, metal poor
($Z =10^{-4}$) GC characterised by a Kroupa mass function, \cite{as16}
find that the MSS mass is linked to the total GC mass through the
relation
\begin{equation}
{\rm Log} M_{\rm MSS} = 0.999{\rm Log} M_{\rm GC}-2.238.
\end{equation}
For $M_{\rm GC} = 10^6\msun$, this implies $M_{\rm MSS} \simeq
5700\msun$, of which $\sim 60\%$ is due to the stellar mass BHs,
$M_{\rm BH} \sim 3800\msun$.  Assuming a mean mass for the BHs $m_{\rm
  BH} = 30\msun$, we obtain $N_{\rm BH} = 114$. In the following, we
make the simplifying assumption that all BHs are retained in the
cluster and no ejections occur due to natal kicks. While this is not
fully realistic, it is qualitatively supported by several recent studies 
\citep{Morscher15,peuten16,Weatherford17,ArcaSeddaetal18}.

\begin{table}
    \begin{center}
      \caption{Properties of the simulations performed with an IMBH
        and a shallow galaxy profile. We consider two values for the
        mass of the IMBH hosted by the GC and three values of the
        initial orbital eccentricity.}
      \label{tab:imbh}
      \begin{tabular}{ccc}
        \hline
        Name & $M_{\rm IMBH} (\msun)$ & $e$\\
        \hline
        S1a & $10^4$ & 0.0\\ 
        S1b & $10^4$ & 0.7\\
        S1c & $10^4$ & 1.0\\
        \hline
        S2a & $10^3$ & 0.0\\
        S2b & $10^3$ & 0.7\\
        S2c & $10^3$ & 1.0\\ 
        \hline
      \end{tabular}
    \end{center}
\end{table}
We repeat simulation S with an IMBH mass of either $M_{\rm
  IMBH}=10^4\msun$ (S1) or $M_{\rm IMBH}=10^3\msun$ (S2). In all
cases, the GC starts out at an apocentre distance of $50\pc$ from the
SMBH but we consider three values of the initial eccentricity, as
illustrated in Table \ref{tab:imbh}.

 The choice of a BH mass of $30\msun$, larger than typical values
  found in earlier studies \citep[see for instance][]{antonini14},
  stems from the following assumptions: i) the GC has low metallicity,
  and ii) the maximum initial stellar mass is $100-150\msun$.  Upon
  these choices and according to a standard \cite{kroupa01} mass
  function, the average mass of stars that can turn into BHs,
  i.e. having $m_*>20\msun$, is $35-40\msun$, depending on the maximum
  stellar mass allowed.  The corresponding BH mass is then
  $15-35\msun$, depending on the stellar evolution recipes adopted
  \citep{hurley00,spera15}.  Our assumption for the BH population mass
  therefore complements earlier studies and offers a new perspective
  on the evolution of massive stellar BHs, also in light of the recent
  discovery of GWs by massive BHBs.

\section{Results}
\label{sec:result}
We followed the evolution of all GC models in the live background of
their host galaxy for a time  which is at least 2.5 times longer than the time over which the GC disruption occurred.
As we will discuss in detail in the next section, for circular orbits
this time-scale can be very long, $\gtrsim 100$ Myr, thus requiring
huge computational resources. For eccentric and radial orbits the
disruption times are much shorter, and we followed these models for 5-10 times the GC disruption time-scale.

The GCs inspiral towards the galaxy centre due to dynamical friction exerted
by the background stars and deposit stars and BHs around the SMBH,
leading to the formation of potential GW sources detectable by ground
based interferometers such as Advanced LIGO \citep{abbott16a}. Models
containing an IMBH lead to the formation of a bound SMBH-IMBH pair,
which then hardens due to encounters with stars.  Such systems are
potential sources of low frequency GWs detectable by upcoming space
based missions such as LISA \citep{seoane07,Barausse2015} or the Chinese experiment TianQin \citep{tianqin16}.

\subsection{Simulations of GCs with a central IMBH}
\label{sec:imbh}
We first consider simulations of GCs hosting a central IMBH in a
shallow (model S) and steep (model M) galaxy model. 

%SNAPSHOTS
\begin{figure}
\centering
\includegraphics[width=8.5cm]{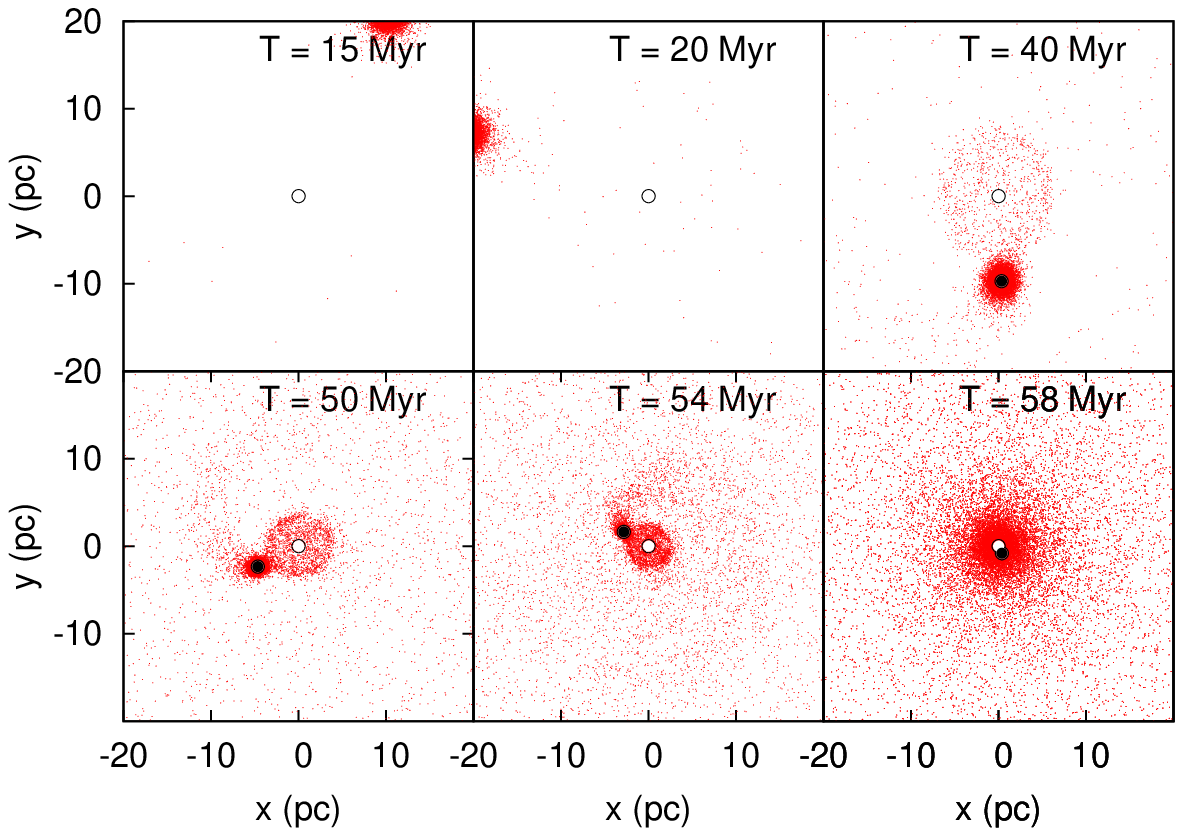}\\
\includegraphics[width=8.5cm]{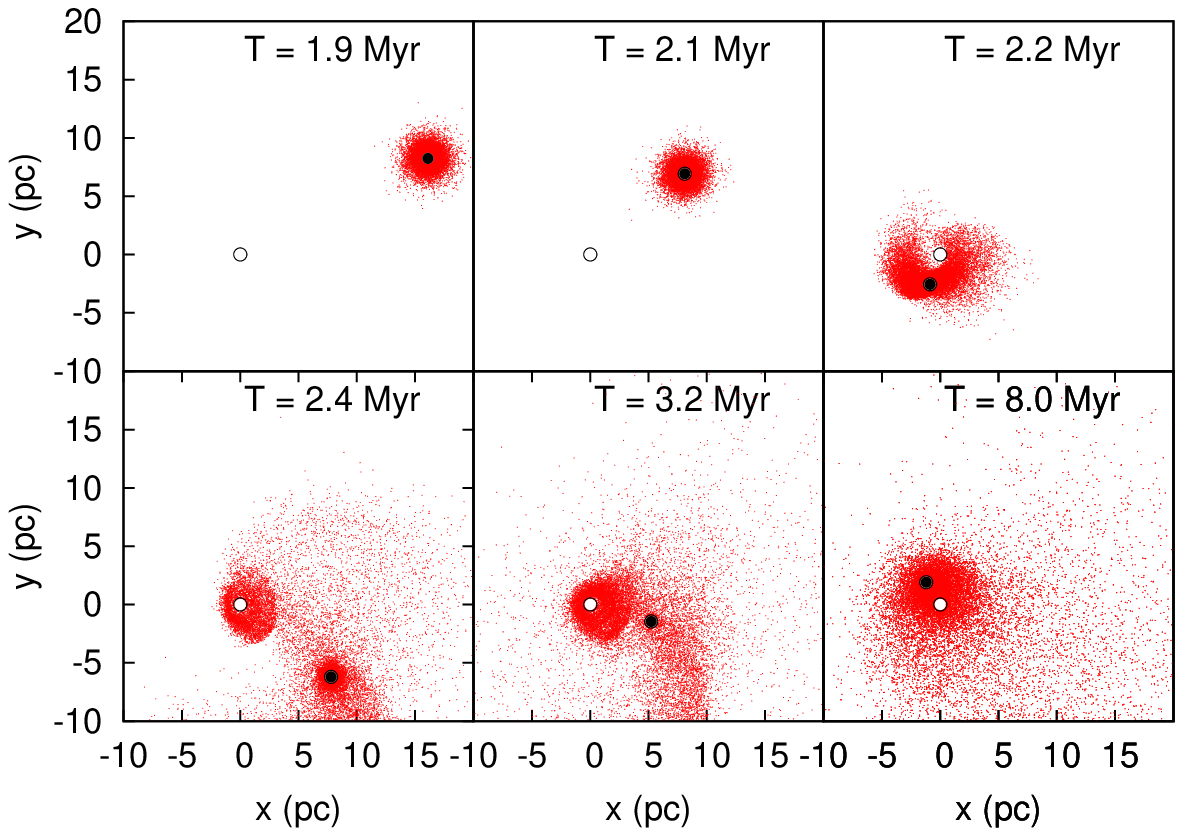}
\caption{Snapshots of simulations S1a (top panels) and S1b (bottom
  panels) in the inspiral plane. The empty white circle represents the
  central SMBH, the filled black circle represents the IMBH whereas
  the smaller red dots indicate the GC particles.}
\label{fig:traj}
\end{figure}
Figure \ref{fig:traj} shows snapshots from simulations S1a and S1b,
which differ only for the initial eccentricity of the orbit.  We see
that the evolution consists of three distinct phases. At early times,
the GC inspirals towards the galaxy centre due to dynamical friction
from the background stars. When the cluster starts losing stars from
the Lagrangian point L1, an inner structure forms around the SMBH made
of deposited GC stars. Tidal torques become efficient and lead to the
formation of tidal tails and streams from the GC. Eventually, the GC
can be considered disrupted and stars have been deposited around the
SMBH. At the same time, stars are lost from the system through the
Lagrangian point L2. The same qualitative behaviour is seen in models
S1b and S1c, however the inspiral is much faster in the case of the
eccentric orbit, as expected. From the snapshots, the time of GC
disruption is about 50 Myr in model S1a and about 3 Myr in model
S1b.

\begin{figure}
\centering \includegraphics[width=8cm]{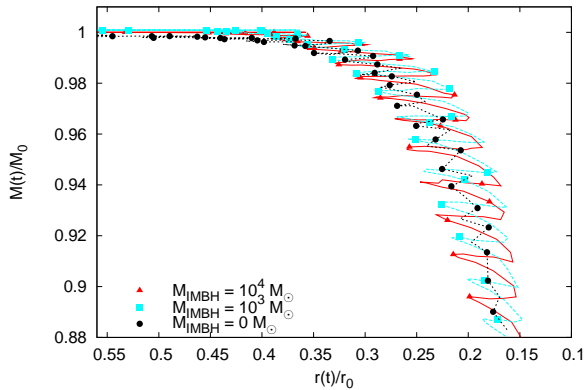}
\caption{GC mass enclosed within its tidal radius as a function of its
  radial distance to the SMBH, for three different values of the IMBH
  mass: $10^4\msun$ (red triangles), $10^3\msun$ (cyan squares) and no
  IMBH (black filled circles). The GC mass and distance are normalised
  to their initial values.}
\label{fig:mloss}
\end{figure}

% GC MASS LOSS
As the cluster inspirals, it loses mass due to tidal stripping.
This can be seen in Fig.\,\ref{fig:mloss}, where we show the GC mass
enclosed within the cluster's tidal radius as a function of
galactocentric distance during the inspiral.
We compare simulations S1a, S2a and an additional simulation performed
with no IMBH but with the same configuration. We find that the mass
loss experienced by the cluster is modest and completely unaffected by
the presence of an IMBH, regardless of its mass (for the mass values
considered here). This is due to the fact that at early times the orbital
decay is due to dynamical friction acting on the cluster as a whole,
and the presence of a central IMBH does not affect this phase.  As we
will show, it is only after the GC is disrupted that the presence and
mass of an IMBH affects the evolution of the cluster remnant.

\begin{figure}
\centering
\includegraphics[width=8cm]{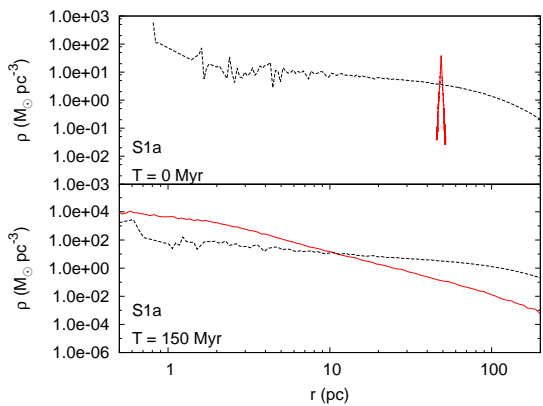}
\caption{Density profile of the galaxy (dashed black line), and of the
  GC (solid red line), for model S1a at $T=0$ (top panel) and
  $T=150$ Myr (bottom panel).}
\label{fig:density}
\end{figure}

% DENSITY PROFILE
The mass deposited by the cluster around the SMBH in the centre can be
quite significant, and is well visible in the density profile of the
system.  Fig.\,\ref{fig:density} shows the radial density profile for
model S1a at the start of the simulations and at a late time,
comparing the mass density of the GC and galaxy.
We find that the GC deposits stars along the inspiral and dominates
the total mass density in the inner few parsecs at late times.

% MASS PROFILE - MASS DEPOSITED
\begin{figure*}
\centering
\includegraphics[width=5.5cm]{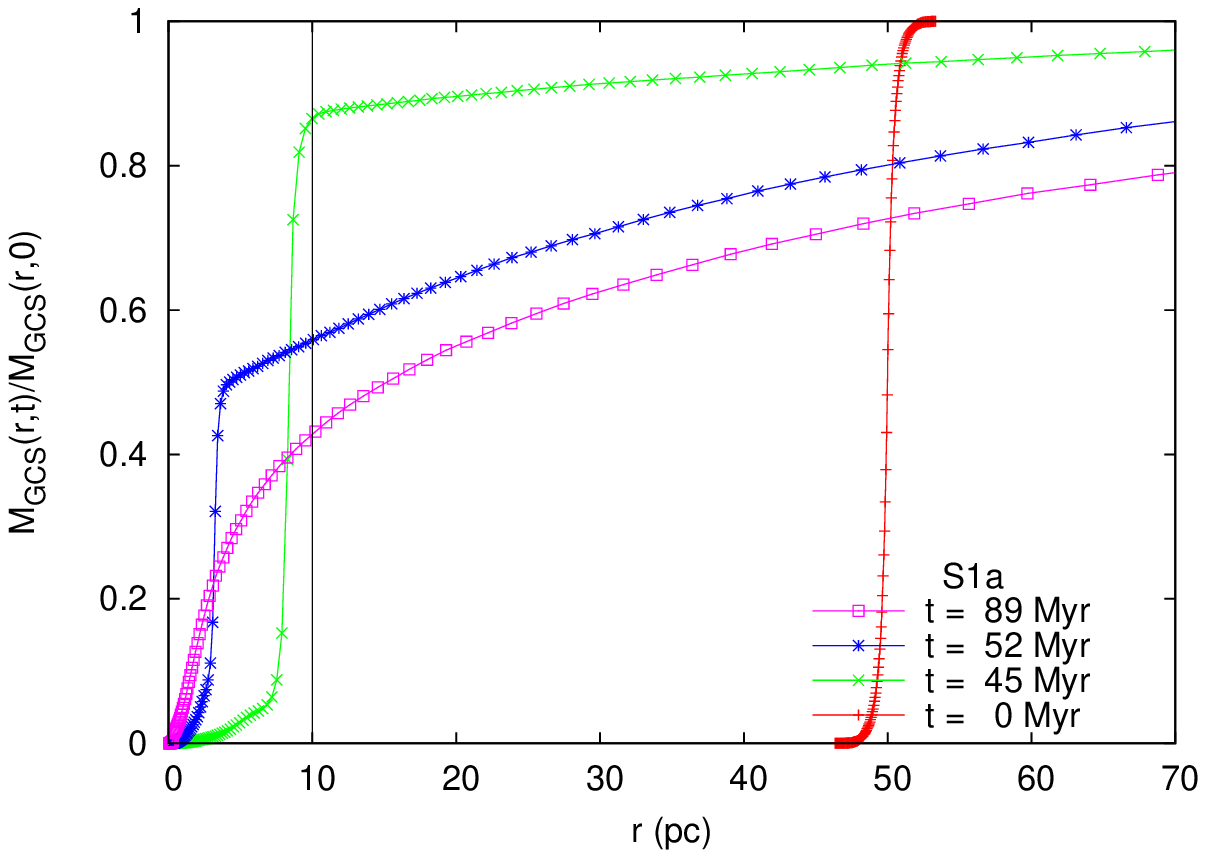}
\includegraphics[width=5.5cm]{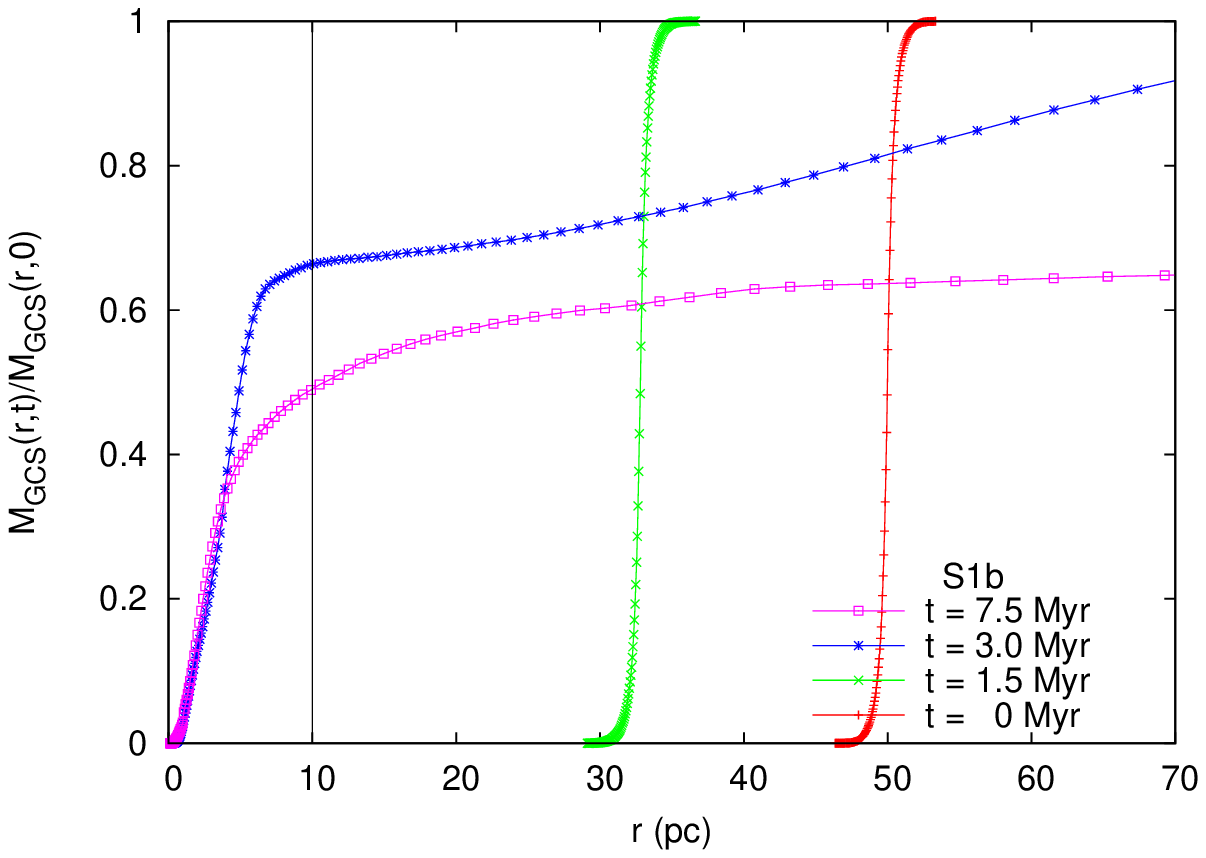}
\includegraphics[width=5.5cm]{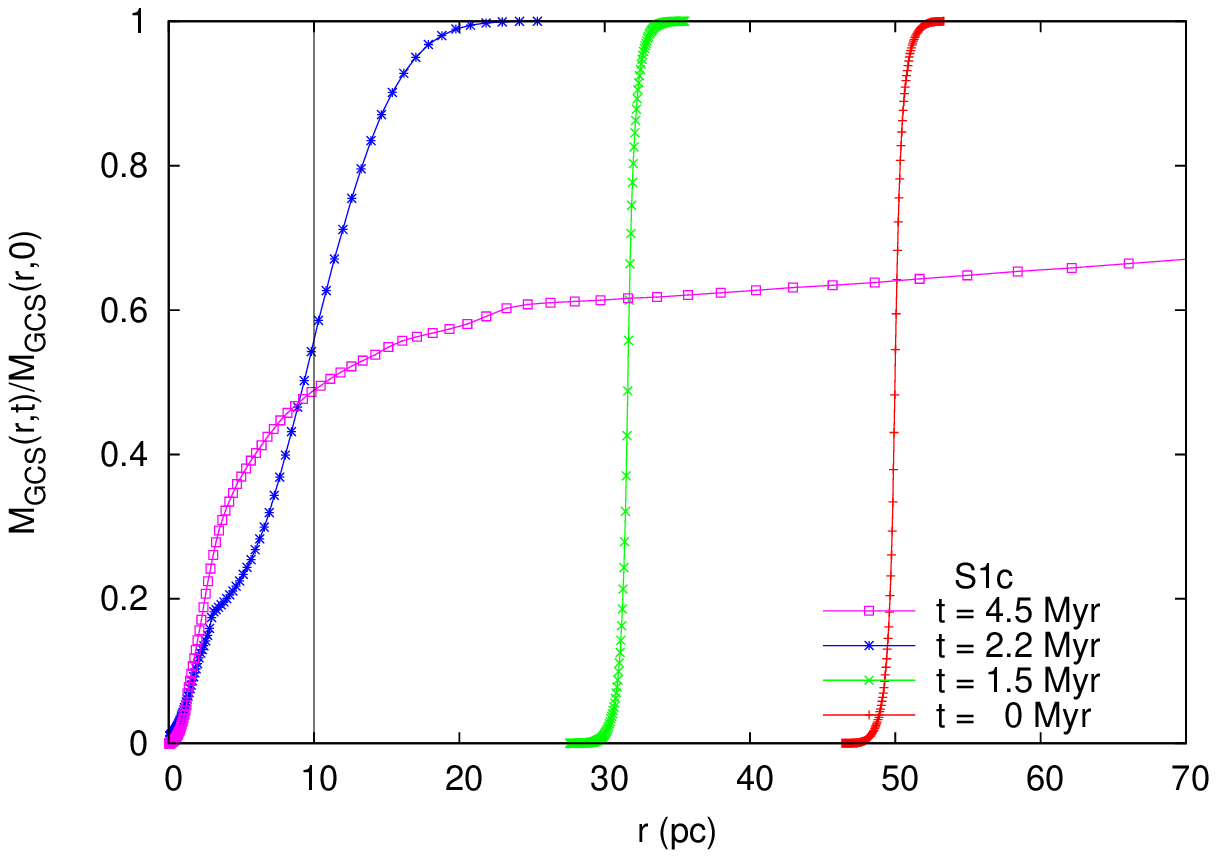}
\caption{Mass profile of GC stars as a function of distance from the
  SMBH in simulations S1a (left), S1b (middle) and S1c (right), at
  different times during the inspiral. The simulations considered
  here differ only for the initial eccentricity of the GC orbit.}
\label{fig:masspS1}
\end{figure*}

\begin{figure*}
\centering
\includegraphics[width=5.5cm]{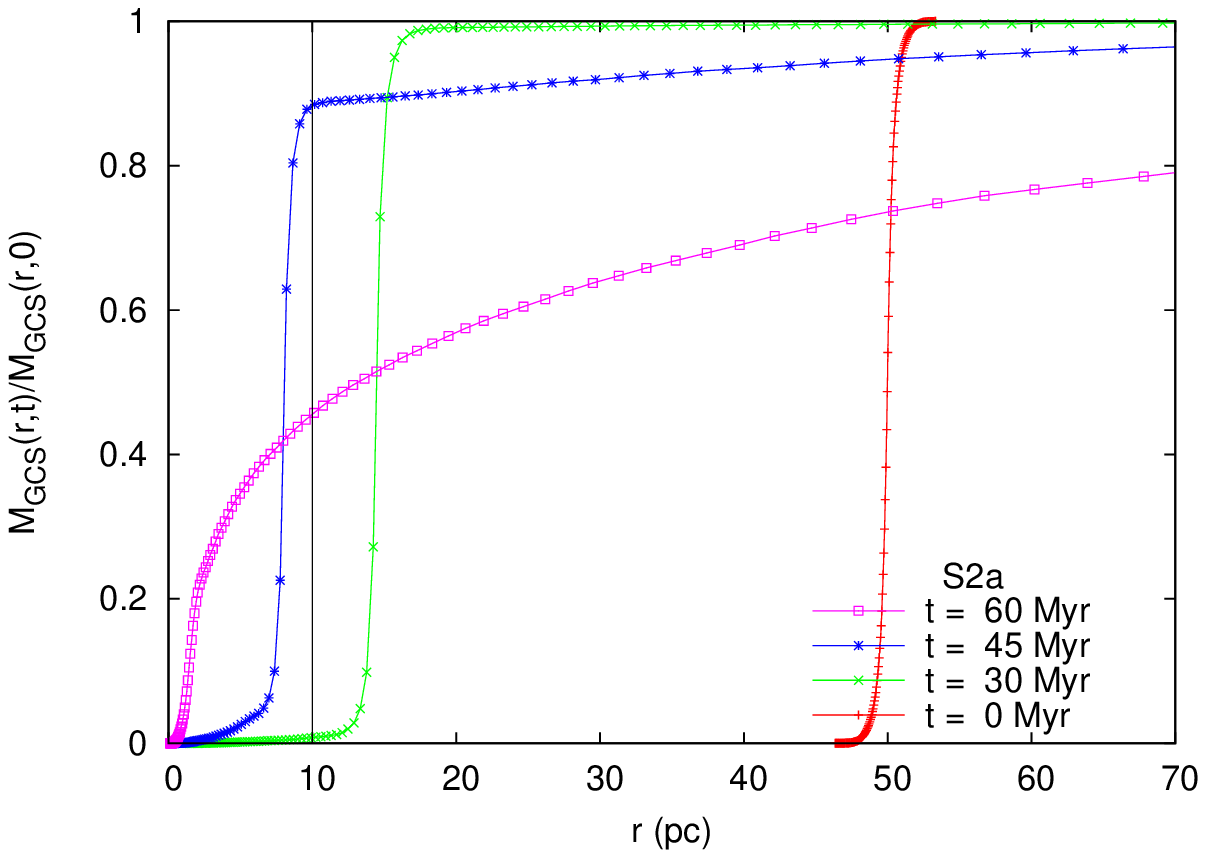}
\includegraphics[width=5.5cm]{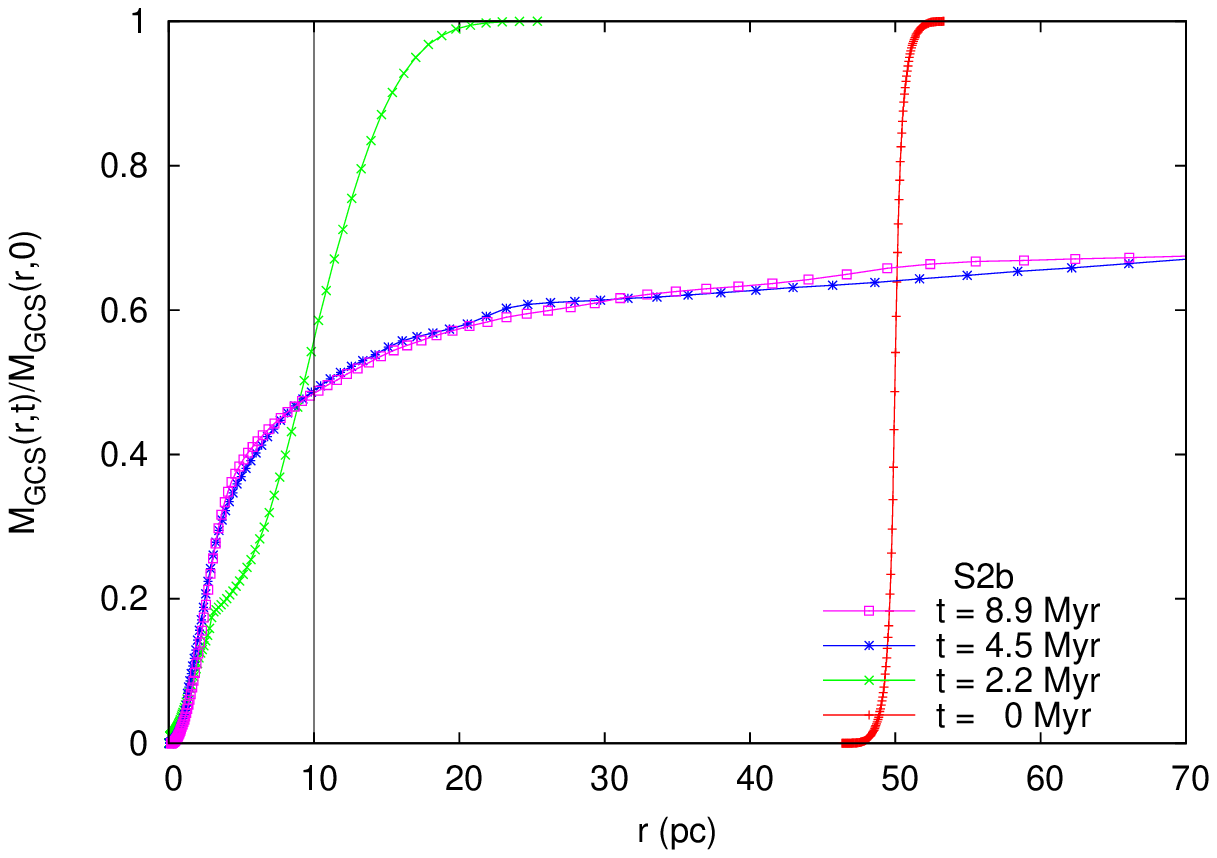}
\includegraphics[width=5.5cm]{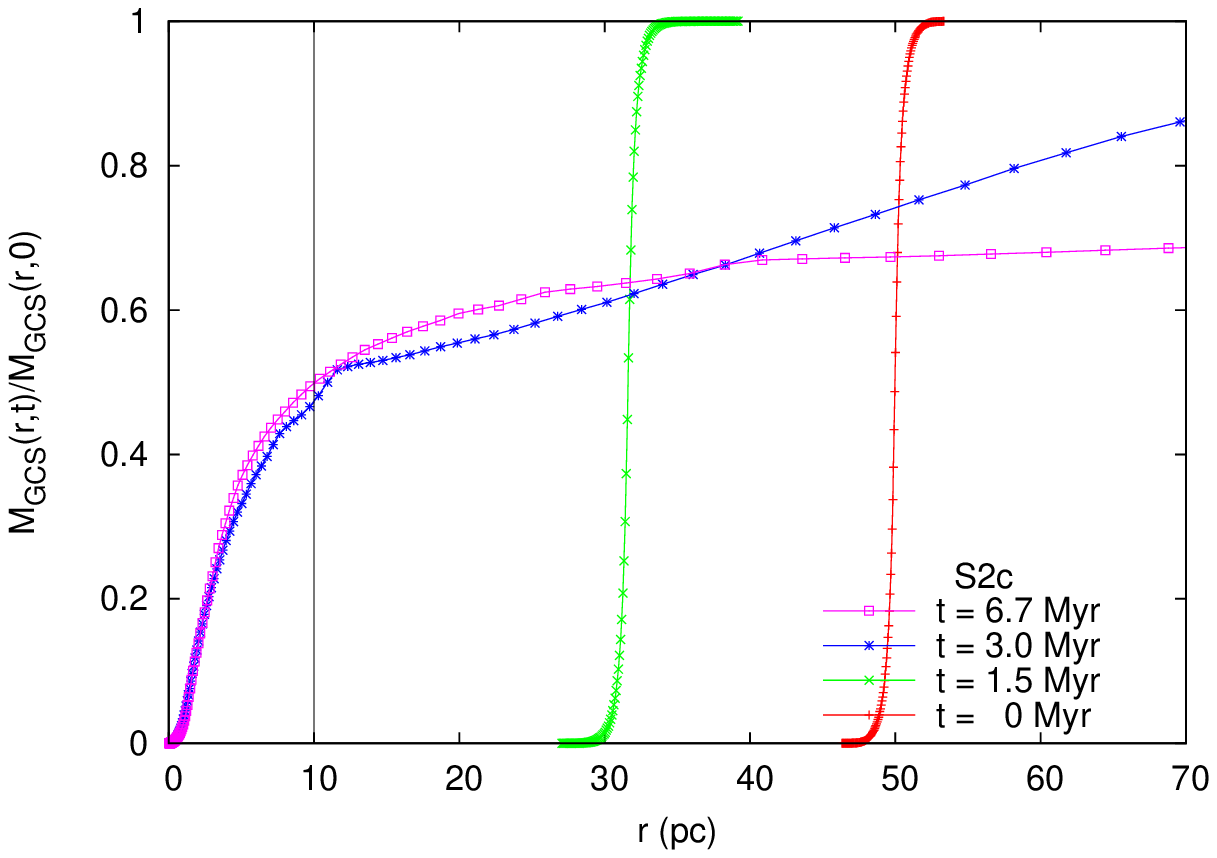}
\caption{Same as Fig.\,\ref{fig:masspS1}, but for simulations S2a
  (left), S2b (middle) and S2c (right).}
\label{fig:masspS2}
\end{figure*}
The radial mass distribution of stars in the GC is shown in
Fig.\,\ref{fig:masspS1} and \ref{fig:masspS2} for models S1 and S2,
where each panel refers to a different orbital eccentricity.  As time
progresses and the GC inspirals, more and more mass is deposited
around the SMBH, but mass is also lost from the system. At late times,
the mass deposited in the central $10\pc$ of the galaxy is about
$50\%$ of the initial GC mass, with little or no dependence on the
mass of the IMBH.  The initial orbital eccentricity of the cluster has
some effect, with the circular orbits resulting in slightly lower
($\sim 40\%$) fractions of deposited mass at late times in the
inspiral.

% MASS DEPOSITED WITHIN 10 pc
\begin{figure}
\centering
\includegraphics[width=8cm]{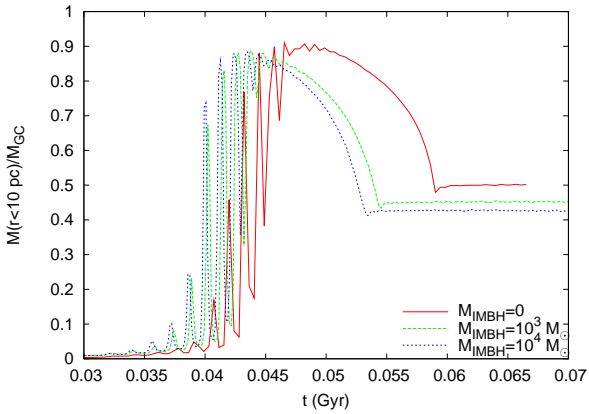}\\
\caption{Mass in GC stars enclosed within $10\pc$ of the central SMBH
  as a function of time during cluster inspiral, for the simulations
  with an initial circular orbit: S1a (blue dotted line), S2a (green
  dashed line) and the simulation without an IMBH (red solid line).}
\label{fig:menc}
\end{figure}
Similarly, fig.\,\ref{fig:menc} shows the mass in GC stars within
$10\pc$ of the SMBH as a function of time during the inspiral for the
simulations with an initial circular orbit.  Initially, the enclosed
mass oscillates significantly due to the motion of the GC and reaches
a peak value as high as $90\%$ of the total GC mass. It then decreases
steadily as the cluster settles in the centre, reaching values of
about $50\%$ after $\sim 40$ Myr.  This is due to the fact that, while
the cluster inspirals and deposits stars around the SMBH through the
Lagrangian point L1, stars are lost through the Lagrangian point L2,
due to the effects of tidal forces and the presence of an IMBH (see
also Fig.\,\ref{fig:traj}). At later times, slingshot encounters with
stars become important.  These are strong encounters with low angular
momentum stars that typically result in the ejection of the stars to
large distances, while the binary shrinks its separation. The mass in
stars that can be ejected with slingshot encounters is of the order of
the total mass of the binary, and is therefore higher in the case of
the most massive IMBH. This likely explains the observed small
dependence of the final deposited mass on IMBH mass.

%GC SHAPE at late times
\begin{figure}
\centering 
\includegraphics[width=8.5cm]{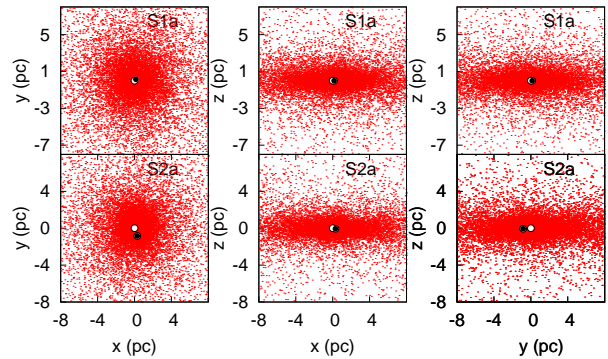}
\caption{GC orbits after $140$ Myr in models S1a (left panels) and S2a
  (right panels). The empty white circle represents the central SMBH, the
  filled black circle represents the IMBH whereas the smaller red dots
  indicate the GC particles.}
\label{fig:tilt}
\end{figure}
Figure \ref{fig:tilt} shows snapshots of the GC in models S1a and S2a
(in all three orbital projections) at the late time of $140$ Myr. The
models differ only for the mass of the IMBH.  Both models appear
flattened in the direction perpendicular to the orbital plane as a
result of the inspiral, with a similar scale height.  Model S2 appears
more centrally concentrated, likely a consequence of the slingshot
ejections phase. Since the amount of ejected mass is proportional to
(and of the order of) the total binary black hole mass, we expect a
larger mass scouring in the case of the most massive IMBH.

Mass deficits due to slingshot ejections from a SMBH-IMBH binary might
have interesting implications in the context of the formation and
evolution of a galactic nucleus.  For instance, it has been suggested
that the SMBH residing in the centre of the Milky Way is the primary
component of a massive black hole binary 
\citep{hansen03,merritt09,gualandris09}. Constraints due to
theoretical \citep{gualandris09,gualandris2010} as well as
observational arguments \citep{RD2004} exclude the presence of an IMBH
more massive than $10^4\msun$ unless the binary separation is very
small ($\lae 0.1\pc$).  Model S1 suggests that the formation of a
SMBH-IMBH binary following the inspiral of a GC has an efficient,
disruptive action on the surrounding environment due to slingshot
ejections, limiting the mass that a growing nucleus could achieve in
this scenario of NSC formation. On the other hand, if we assume that
NSCs originate primarily in this way, our results imply that the SMBH
at the centre of the MW can't have a companion more massive than $\sim
10^4\msun$ with a small separation, unless the IMBH binds to the SMBH
the NSC formation.

% BHB distance
\begin{figure}
\centering
\includegraphics[width=8cm]{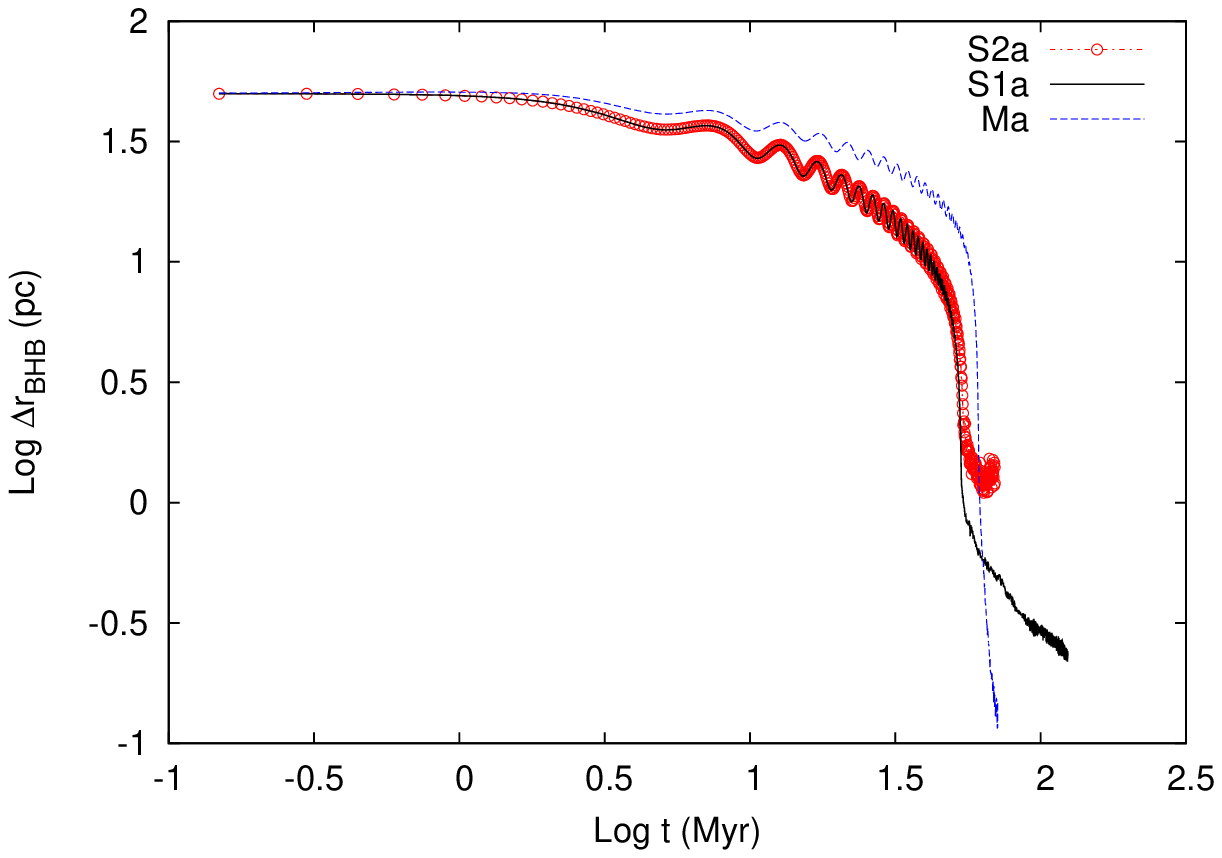}\\
\includegraphics[width=8cm]{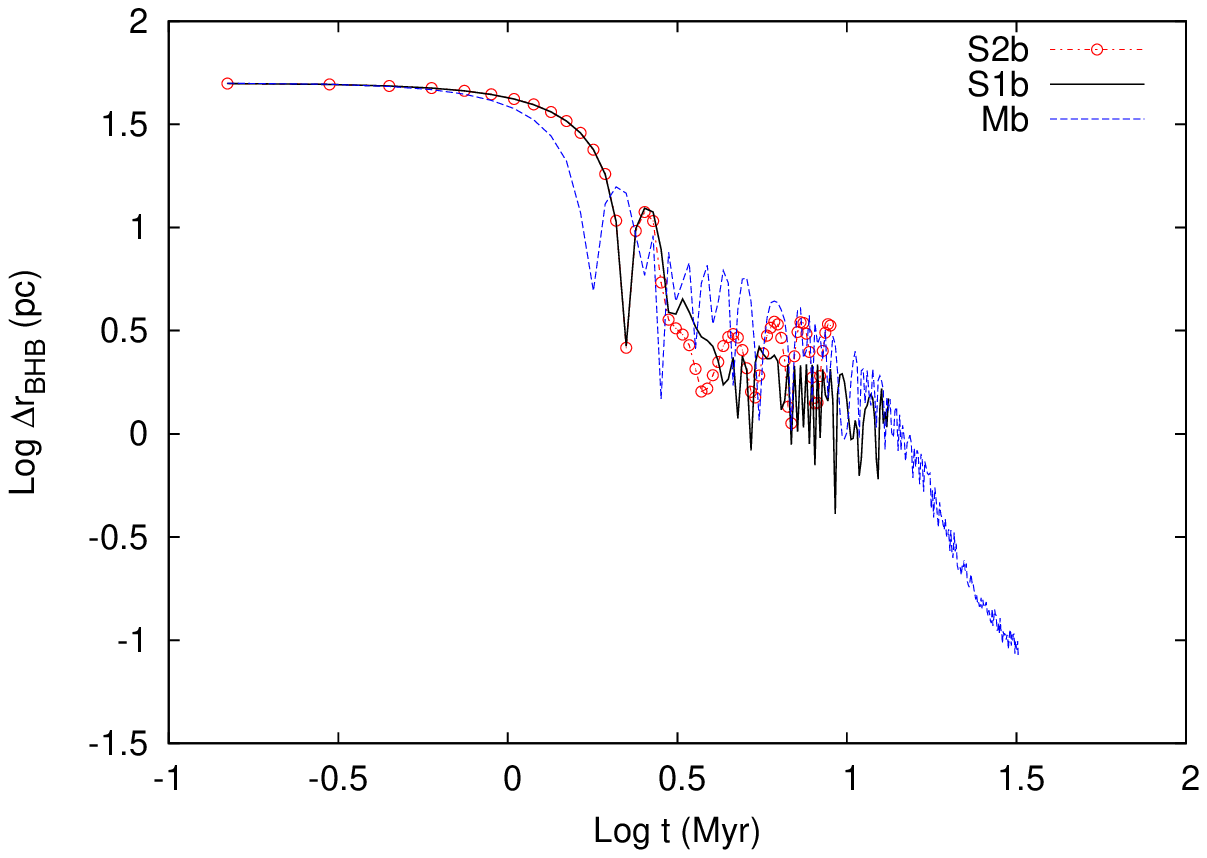}\\
\includegraphics[width=8cm]{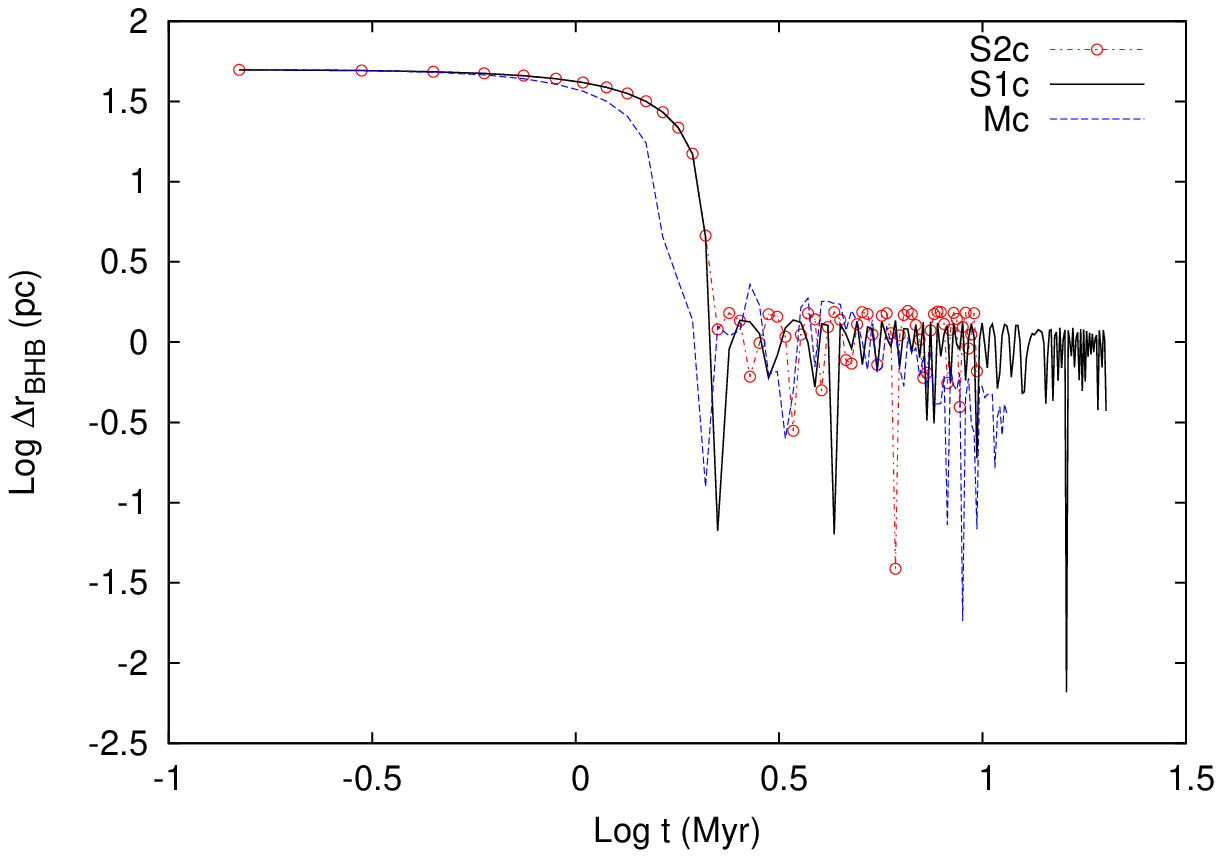}\\
\caption{Distance between the IMBH and the SMBH in models with a
  circular orbit (top panel), an eccentric orbit (middle panel) and a
  radial orbit (bottom panel). In each panel, we consider models with
  both a shallow and a steep galaxy profile.}
  \label{fig:BHBsep}
\end{figure}
After the cluster is disrupted, dynamical friction becomes ineffective
and further orbital decay of the SMBH-IMBH pair is due mainly to
encounters with stars.  The time of GC disruption can be seen in
Fig.\,\ref{fig:BHBsep}, which shows the distance between the SMBH and
the IMBH as a function of time for all simulations with an IMBH.  For
the circular orbit simulations, the time of transition is about 50 Myr,
and this markedly decreases to about 2-3 Myr for the eccentric and
radial orbits, in accordance with our early estimates based on a
visual inspection of the snapshots (Fig.\,\ref{fig:traj}).

The figure also shows the effect of the adopted galaxy
model. Simulations M represent a galaxy with a steeper density profile
and with a nuclear star cluster, similar to the case of the Milky
Way. The early orbital decay is similar in all models. This is due to
the fact that the galaxy profile in the outer region ($50-100\pc$) is
quite similar in both cases (see Fig.\,\ref{fig:SBMdensity}). The main
differences arise in the innermost $10\pc$, where the NSC dominates
the mass distribution in the M models.  In the case of a circular
orbit, the GC orbital decay is less efficient in the M model than in
the S model. This may appear counter-intuitive, but is due to the fact
that the GC and the NSC form a sort of binary system and start
orbiting the common centre of mass. This effect is less pronounced in
the case of the eccentric and radial orbits, due to the much faster
inspiral.

%HIGH VELOCITY STARS
\begin{figure*}
\centering
\includegraphics[width=5.5cm]{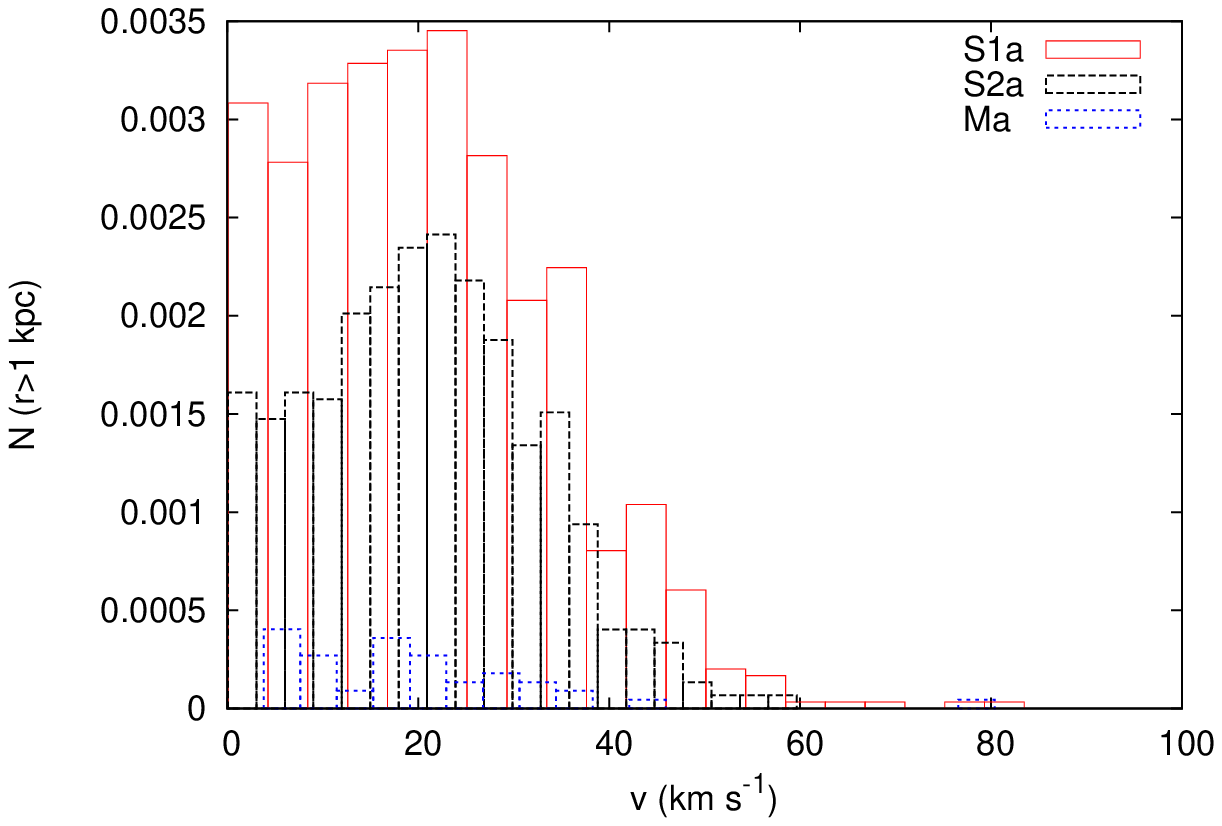}
\includegraphics[width=5.5cm]{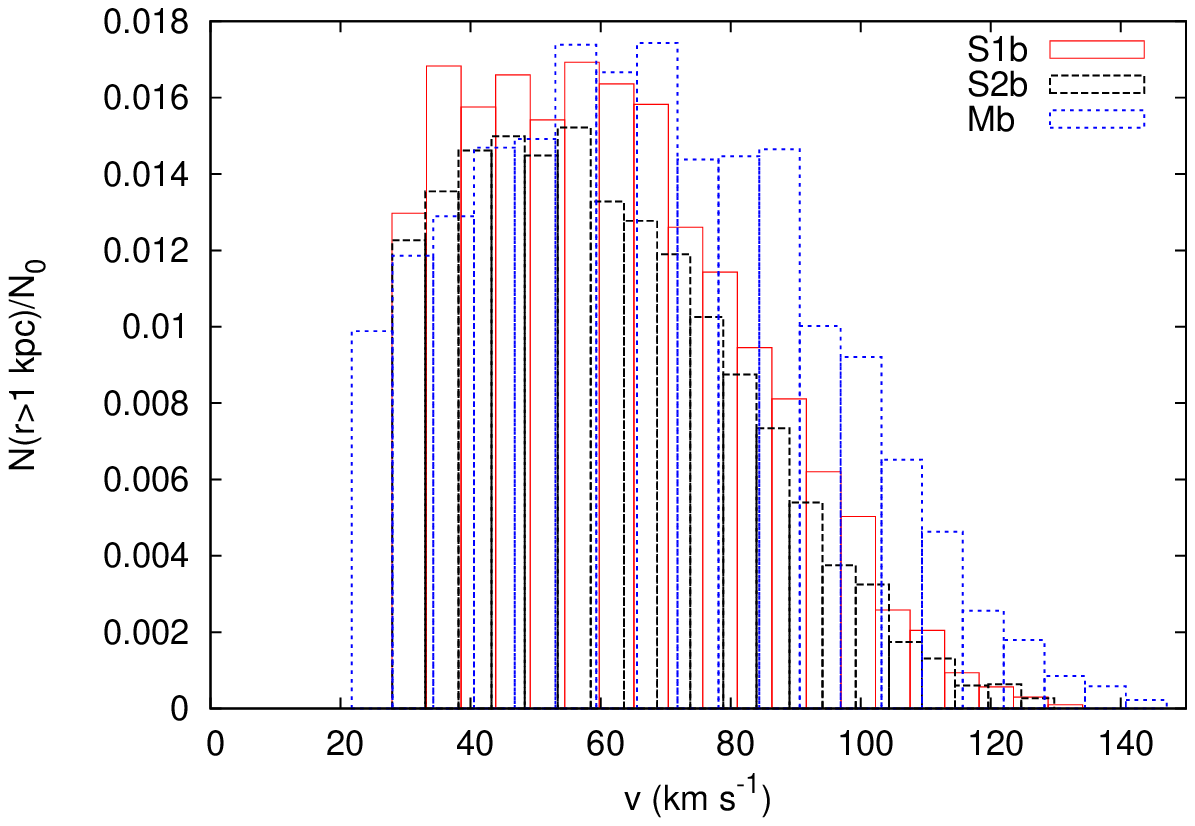}
\includegraphics[width=5.5cm]{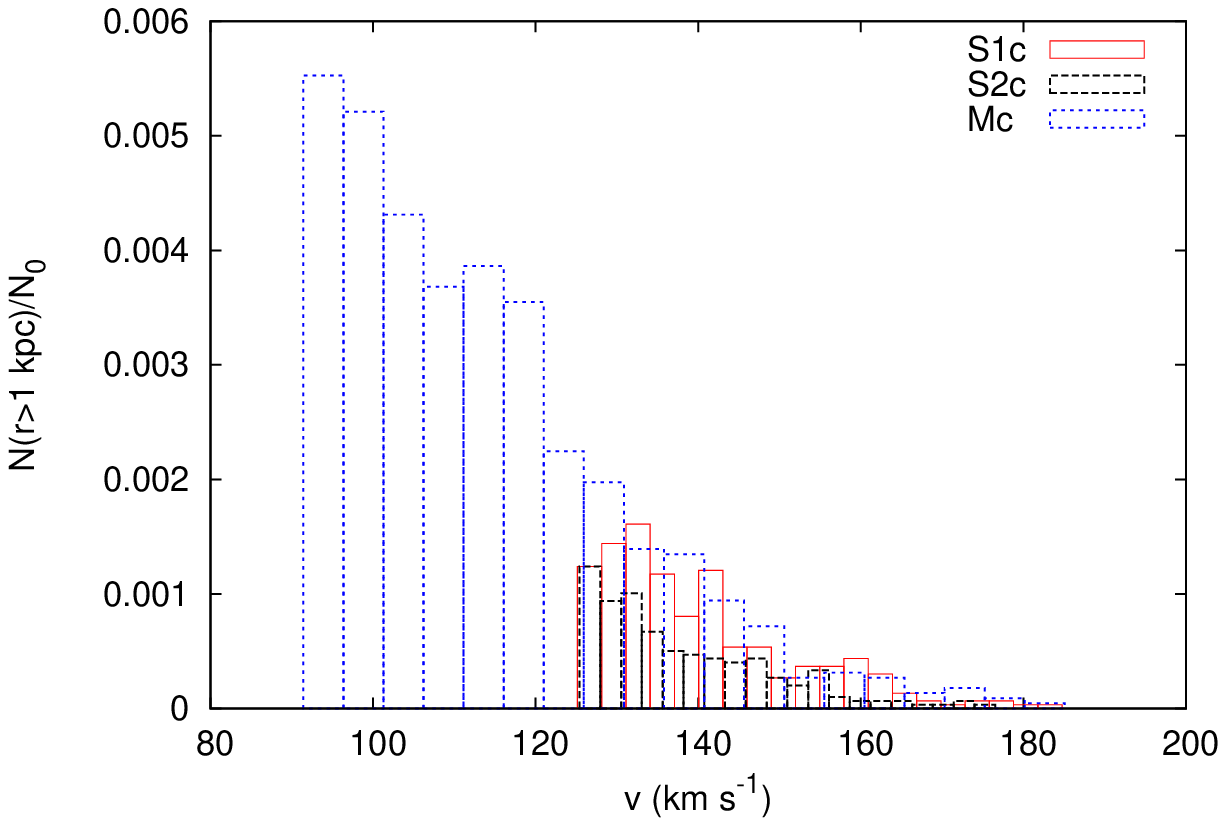}
\caption{Velocity distribution of escapers reaching a distance of at
  least $1\kpc$ from the centre by the end of the simulations, for
  circular (left), eccentric (middle) and radial (right) models S1, S2 and M.}
\label{fig:hvs}
\end{figure*}

The inspiral of a GC also results in the ejection of high velocity
cluster stars.  This is due to a three-body interaction involving the
GC, the SMBH and a star, as described in \citet{ASCDS15} and tested in
\citet{CF2015} by means of scattering experiments  (but see also
  \citet{fragione17c}). For SMBHs more massive than $\sim10^8\msun$,
this mechanism represents an efficient source of high velocity stars.
Following \cite{ASCDS15}, we computed the number of stars ejected to a
distance of at least $1\kpc$ by the end of the simulation, a distance
much larger than the scale length of our simulated nuclei. The
velocity distribution of the GC escapers is shown in
Fig.\,\ref{fig:hvs}. We find a dependence of ejection velocity on the
initial orbital eccentricity of the cluster, with more escapers
produced in the eccentric models than in the circular ones. On the
other hand, the radial case produces very few escapers, likely due to
the very short inspiralling time. The ejection velocities do not
depend on the mass of the IMBH present in the GC, as the ejection
mechanism involves an interaction between a star, the SMBH and the
whole GC. The number of escapers, however, is higher in the case of
the most massive IMBH.  After an interaction with the SMBH, a star can
either be ejected promptly, be captured by the SMBH in a bound orbit
or simply remain bound to the GC. In the latter case, the star can
undergo further interactions with the SMBH and be ejected at a later
time. A larger fraction of ejected stars is expected the more
concentrated the cluster is. In our simulations, this corresponds to
the cases with the most massive IMBH.

The presence of an NSC already surrounding the SMBH (M models) seems
to cause a smaller number of escapers with a lower minimum velocity
compared to other models. Model Mc, differently from the ``a'' and
``b'' cases, is characterised by a larger fraction of ejected stars,
having velocities significantly lower than S1 and S2 escapers.

A potential additional source of escaping stars is represented by
encounters with the IMBH-SMBH binary. Escapers produced by slingshot
ejection can be identified based on their later ejection time.
\begin{figure}
\includegraphics[width=8cm]{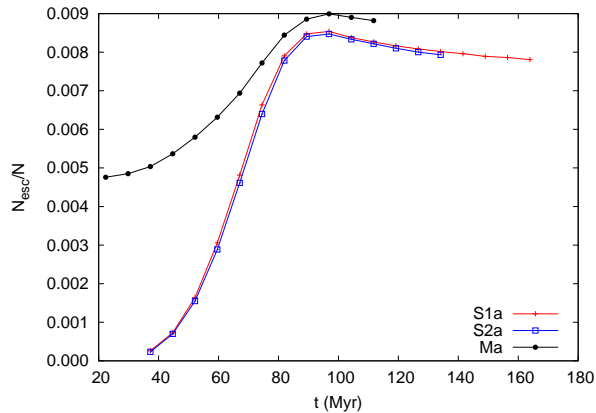}
\caption{Number of escaper stars, normalized to the total particle
  number, as a function of the time in models S1a, S2a and
  Ma.}
\label{S1esc}
\end{figure}
Figure \ref{S1esc} shows the fraction of escapers, i.e. the number of
stars with a positive total energy reaching distances above 1 kpc,
normalized to the total number of stars in models S1a, S2a and
Ma. There is no significant difference between model S1a and S2a,
suggesting that escapers are mostly produced in the 3-body interaction
discussed above. The larger fraction of ejected stars in model M
reflects the higher density that characterises the inner region of the
galaxy in this case.  As shown in Fig. \ref{S1esc}, the combined
  GC+SMBH interactions lead to star ejections with velocities in the
  range $20-200\kms$, depending on the IMBH mass and the galaxy
  environment.

 The leading scenario for the production of hypervelocity stars is the tidal breakup of stellar binaries by a SMBH, as first suggested by \cite{hills}. In this scenario, a tight
binary undergoes a close flyby with an SMBH, with one component of the
binary being captured and the other ejected at a velocity \citep{hills,bromley06}
\begin{align}
v_{\rm ej} \sim & 1800\kms \left(\frac{a_{\rm bin}}{0.01~{\rm AU}}\right)^{-1/2} 
            \left(\frac{M_{\rm bin}}{2\msun}\right)^{1/3} \times \nonumber \\
            & \times \left(\frac{M_{\rm SMBH}}{4\times 10^6\msun}\right)^{1/6} \left(\frac{M_c}{M_e+M_c}\right)^{1/2}, 
\end{align}
being $M_c$ the mass of the captured component and $M_e$ the mass of
the ejected star. In order to be disrupted, the binary must have a
pericentre smaller than the SMBH tidal radius $r_t = (M_{\rm
  SMBH}/M_{\rm bin})^{1/3}a_{\rm bin}$.

A similar process can be thought to occur when infalling GCs impact on
a SMBH.  In this case, the binary mass can be replaced with the GC
mass $M_{\rm bin}\sim M_\gc$ and the binary semi-major axis with the
GC core $a_{\rm bin} \simeq r_c$.  When the GC passes at the pericentre, it
fills its Roche robe, satisfying the condition $r_p<r_t$ for a star to
be ejected.

\begin{figure}
\centering
\includegraphics[width=8cm]{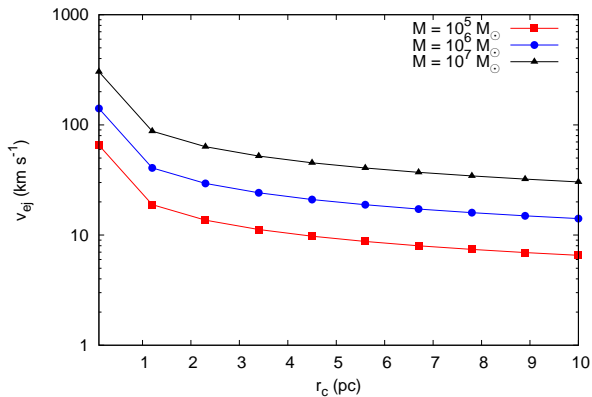}
\caption{Ejection velocity for stars at varying GC core radii and for
  different GC masses, assuming $M_{\rm SMBH} = 5\times 10^6\msun$.}
\label{esct}
\end{figure}

Figure \ref{esct} shows the ejection velocity for stars at different
cluster core radii$r_c$ and for different GC masses, $M_\gc =
10^5-10^6-10^7\msun$, assuming that the escaping star is orbiting at
$r\sim r_c$ at the time of ejection.  We note here that this
requirement implies the maximum ejection velocity, as suggested by
\cite{ASCDS15}.  For a $10^6\msun$ SMBH and a GC core radius $\sim
0.24\pc$, the expected ejection velocity is $\simeq 80-100\kms$, quite
similar to the values observed in our models.

\subsection{Simulations of GCs with a cluster of stellar mass black holes}
\label{sec:bhs}
We now consider the simulations in which the GC harbours a cluster of
stellar mass black holes instead of an IMBH at its centre.  For models
B, we assume BH masses in the range $20-40\msun$ distributed within
the GC according to the overall density profile. Therefore, our BH
population is not initially mass-segregated. Similarly to models S and
M, we consider three initial orbital eccentricities for the GC,
labelled with the letter ``a'' (circular), the letter ``b''
(eccentric) and the letter ``c'' (radial).

\begin{figure}
\centering
\includegraphics[width=8cm]{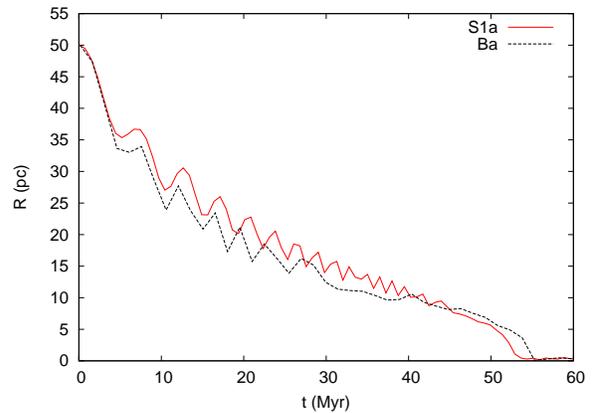}
\caption{Distance of the GC centre of density as a function of time in models S1a and Ba.}
\label{Rt}
\end{figure} 
The inspiral of the GC in simulations with a cluster of BHs is similar
to the case of a central IMBH, as illustrated in Figure
\ref{Rt}. 
\begin{figure}
\centering
\includegraphics[width=8cm]{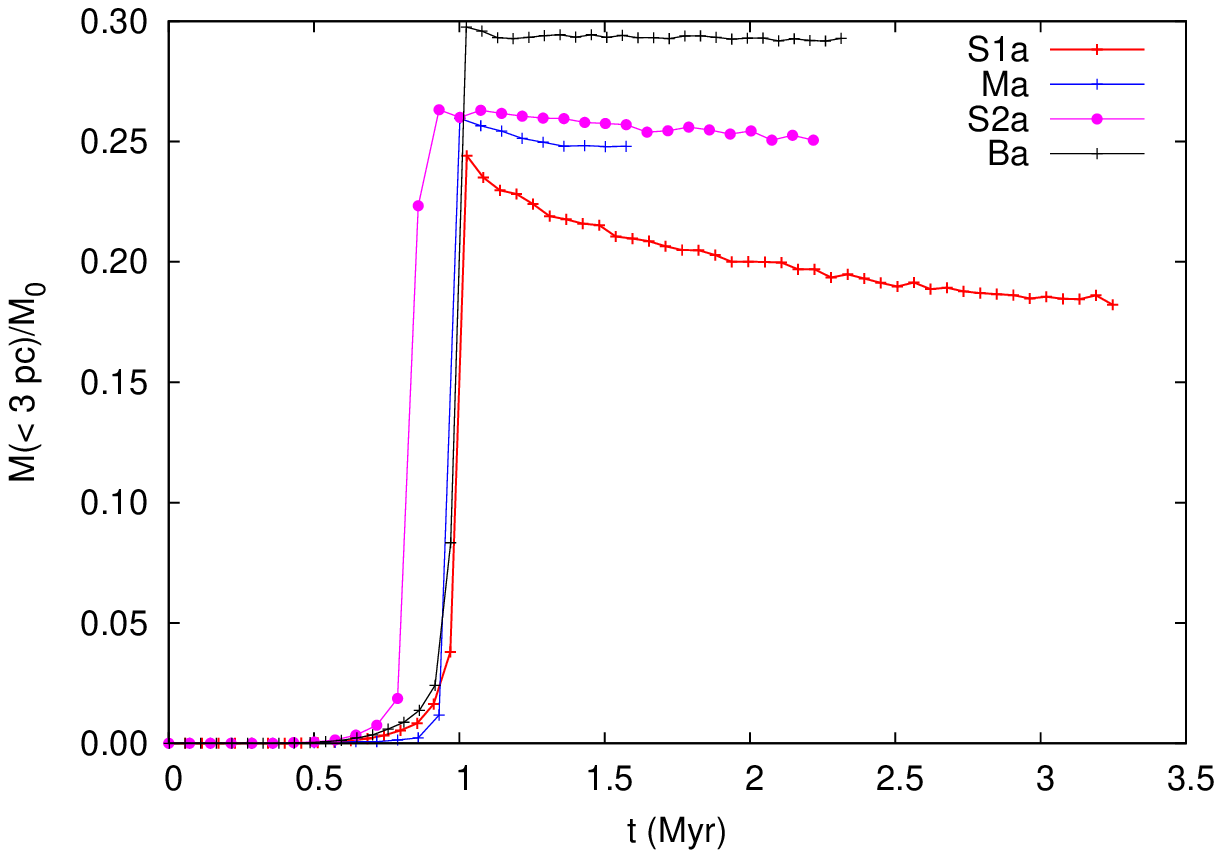}\\
\includegraphics[width=8cm]{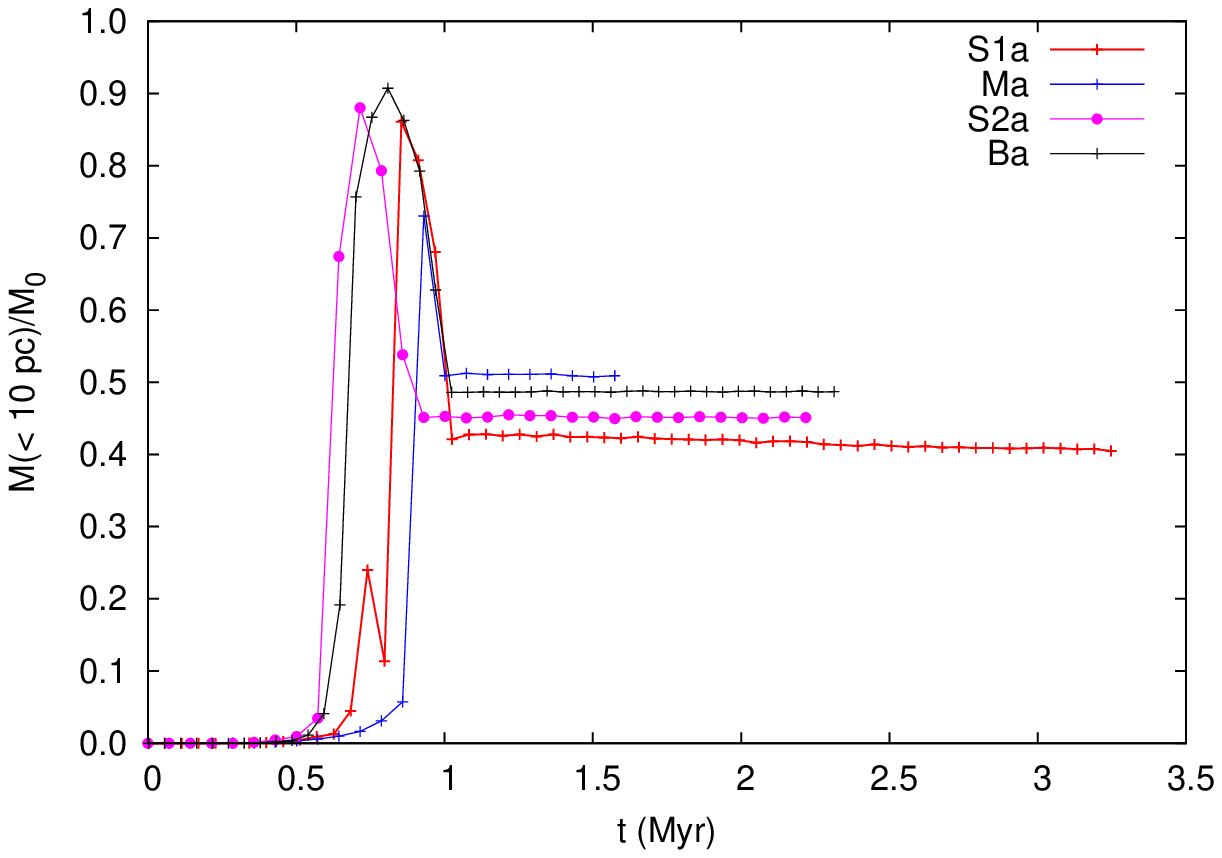}
\caption{Mass enclosed within $3\pc$ (top panel) and $10\pc$ (bottom
  panel) from the SMBH in models S1a, Ma and Ba. It is evident that
  the absence of the IMBH allows a more efficient mass deposit around
  the SMBH.}
\label{Mt}
\end{figure} 
However, a larger amount of mass is deposited around the
SMBH in the B models (see Figure \ref{Mt}). This is due to the mass
scouring of the SMBH-IMBH binary and subsequent ejection of stars, a
feature that is absent in the B models. 

\begin{figure}\centering 
\includegraphics[width=8.5cm]{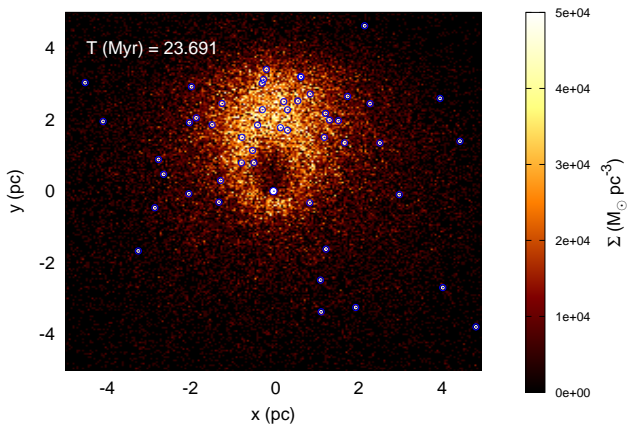}
\caption{Surface density map of the GC in model Bb after inspiral. The central larger white dot represents the SMBH while
  the smaller white dots represent the stellar BHs.}
\label{BHmap}
\end{figure}
Figure \ref{BHmap} shows a surface density map of the GC in model Bb
after a few pericentre passages, with the larger dot marking the SMBH
and the smaller dots representing the stellar mass BHs.  The GC debris
distributes in a disc configuration that undergoes precession around
the SMBH.  Stars moving in the disc have orbits characterised by a
pericentre $r_p=0.1-0.8\pc$ and apocentre $r_a=1-3\pc$. Nearly $30\%$
of the stellar BHs move within the disc.

If the GC hosts a population of stellar BH binaries, these will
distribute around the SMBH and possibly undergo the Lidov-Kozai
mechanism, which can boost their coalescence depending on their
orbital parameters \citep{antonini12,hoang17}.

In order to determine the probability for a stellar BHB to merge once
it is left orbiting around the SMBH, we performed simulations with the
\texttt{ARGdf} code \citep{ASCD17b}, an updated version of the
\texttt{ARCHAIN} code developed by \citet{mikkola08}.
These are ``few-body'' codes, suited to model the evolution of a few particles undergoing strong stellar encounters and, possibly, collision and GW emission. Our updated version allows including the effects of the galaxy gravitational field and to model in a semi-analytic way dynamical friction, a crucial feature to model the evolution of massive satellites moving in a galactic environment. These codes employ algorithmic regularization \citep{mikkola99} to allow for a high accuracy integration of the motion of particles undergoing strong interactions.

We considered stellar BHBs with initial semi-major axis in equally
spaced logarithmic bins between $20$ and $2\times 10^4$ AU and
eccentricity drawn from a thermal distribution $P(e)de \propto ede$
\citep{jeans19}.  The masses of the two binary components are randomly
selected in the range $10-50\msun$.  The initial distance of the BHB
centre of mass from the SMBH is drawn randomly in the range [1-4] pc,
while the eccentricity of the BHB orbit with respect to the SMBH is
set equal to the GC orbital eccentricity in the case of model Bb.
Finally, we varied the inclination $i$ of the BHB centre of mass with
respect to the BHB-to-SMBH direction. Note that values $i>90^{\circ}$
correspond to retrograde orbits, while values $i<90^{\circ}$
correspond to prograde orbits. We performed 1000 simulations of this
type with the {\tt ARGdf} code, with a simulation time of 2.5 Myr,
corresponding to $\sim 70$ orbits around the SMBH. 
\begin{figure}
\centering
\includegraphics[width=8cm]{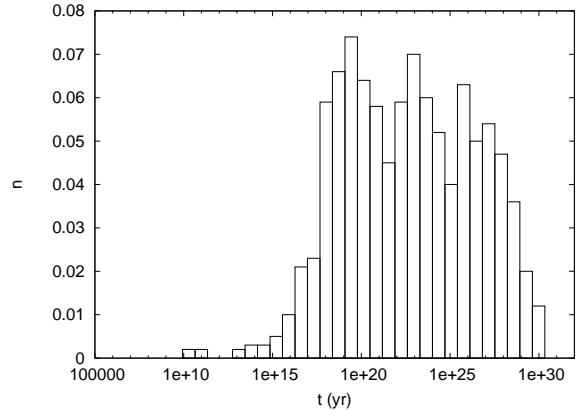}
\caption{GW time-scales for stellar BHBs orbiting the SMBH disc in model Bb.}
\label{gwhis}
\end{figure} 
We note that none of our initial configurations has a GW time-scale
smaller than $10^{10}$ yr (see Figure \ref{gwhis}), as calculated through equation \citep{peters64}

\begin{equation}
t_{\rm GW} \sim 6 {\rm Myr} \frac{(1+q)^2}{q} \left(\frac{a_{\rm
      BHB}}{0.01\pc}\right)^4 \left(\frac{M_{\rm BHB}}{10^8
    \msun}\right) (1-e^2)^{7/2},
\label{eq:tgw}
\end{equation}
where $a_{\rm BHB}$ is the binary semi-major axis, $e_{\rm BHB}$ its orbital eccentricity, while $M_{\rm BHB}$ and $q$ are the BHB total mass and mass ratio, respectively.

We found a $P_{\rm mer} = 5\%$ probability for the stellar BHB to
merge promptly, on time-scales shorter than the simulated $2.5$ Myr.

As shown recently in \cite{belckzynski17}, the average fraction of GCs
that are expected to contribute to the NSC assembly is $f_c = 0.11$,
provided that the total GCs initial mass is $0.01$ times the host
galaxy mass and the average initial GC mass is $M_{\rm GC,av} \sim
10^6\msun$ \citep{harris14,webb15}.

Simple stellar population models predict that the BH population
represents a fraction of $\sim 10^{-3}$ of the initial GC mass.
Assuming a binary fraction $\eta$ among the BHs and a sub-fraction
$\delta$ of binaries surviving long enough to be deposited in the
proximity of the SMBH, the number of BHBs expected to orbit the SMBH
in a galaxy with mass $M_g \sim 10^{10}\msun$,
similar to our model, is 
\begin{equation}
N_{\rm dec}  = \left(f_c \frac{0.01M_g}{M_{\rm GC,av}}\right) \left(10^{-3} \delta\eta N_{\rm GC,av}\right) \simeq 10^3 \eta\delta.
\label{ndec}
\end{equation}

If we assume that only $1\%$ of the initial BH population will still
be in the GC during the NSC formation, i.e. $\delta\eta=0.01$, we
obtain $\sim 10$ merging BHBs per galactic nucleus on a total
time-scale set by the GC dynamical friction time.

For GCs contributing to the NSC formation, i.e. with apocentre smaller
than $500\pc$, the typical dynamical friction time is $\tau_{\rm df}
\sim 0.1$ Gyr \citep{ASCD15He}.

 In order to calculate the rate at which BHB mergers are mediated by GC-SMBH interactions we
must estimate the number density of galaxies similar to the MW. 
The Illustris\footnote{\url{http://www.illustris-project.org/}} cosmological simulation 
represents one of the most reliable model of structure formation to date, 
allowing to properly model the galaxy distribution at low redshift. 
Using the Illustris public data release \citep{nelson15}, we calculated the  
number density $n_g$ of galaxies with stellar masses in the range $10^{10}-10^{11}\msun$, 
to which the MW belongs, finding $n_g = 0.008$ Mpc$^{-3}$.

We assume that all the galaxies in this mass range host a central SMBH, and that all of them 
witnessed at least one GC-SMBH interaction in the past. Note that this does not require necessarily
the presence of an NSC in the galaxy centre, as tidal forces can prevent its formation under certain
conditions \citep{antonini13,ASCD14b,ASCD17a}.

Hence, a rough estimate of the BHB merger rate for this channel can be obtained as
\begin{equation}
\Gamma_{\rm{BHB}} = \frac{P_{\rm mer}N_{\rm dec}n_g}{\tau_{\rm df}},
\label{rate}
\end{equation}
where $P_{\rm mer}$ is the BHB merger probability, $N_{\rm dec}$ is
the number of decaying GCs, $n_g$ is the number density of galaxies in
the local universe and $\tau_{\rm df}$ is the dynamical friction
time-scale.

GCs lose most of their BHBs on a $\sim 1$ Gyr time-scale. However, as
we have shown above, by this time GCs born closer to the galactic
centre will have already orbitally segregated, thus producing a larger
$\delta\eta$ parameter. In the most optimistic case in which $\sim
10\%$ of the whole BH population is in a binary and is brought to the
GC before ejection, $\delta\eta = 0.1$.

Substituting Eq.\,\ref{ndec} into Eq.\,\ref{rate} and considering both
an optimistic scenario for which $\delta\eta = 0.1$ and a pessimistic
one for which $\delta\eta = 0.01$, we find a merger rate
\begin{equation}
\Gamma_{\rm {BHB}} \sim 0.4-4\,{\rm yr}^{-1}\gpcc.
\label{rate}
\end{equation}

Similar estimates have been recently provided for globular clusters
($\Gamma \simeq 5$ yr$^{-1}\gpcc$,
\cite{rodriguez15,rodriguez16,askar17}), young massive clusters
($\Gamma \simeq 10$ yr$^{-1}\gpcc$,
\cite{banerjee16,banerjee18,mapelli16}), nuclear clusters ($\Gamma
\simeq 1.5$ yr$^{-1}\gpcc$, \cite{antonini16,hoang17} and around
SMBH in massive ellipticals ($\Gamma \simeq 1$ yr$^{-1}\gpcc$,
\cite{ASCD17b}).

All these different channels show that BHBs delivery by infalling GCs
can have a significant role in the production of GWs detectable by the
LIGO/VIRGO experiments.

\begin{figure}
\includegraphics[width=8cm]{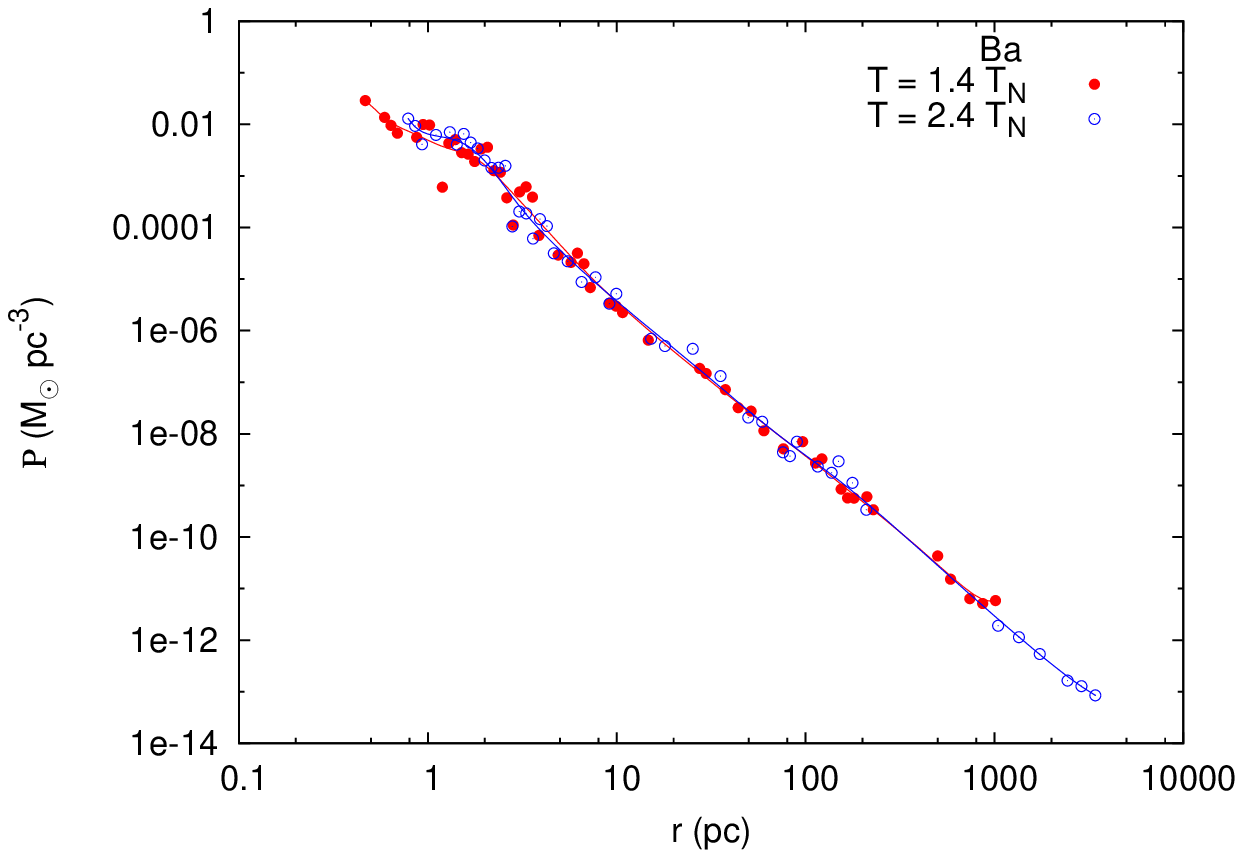}\\
\includegraphics[width=8cm]{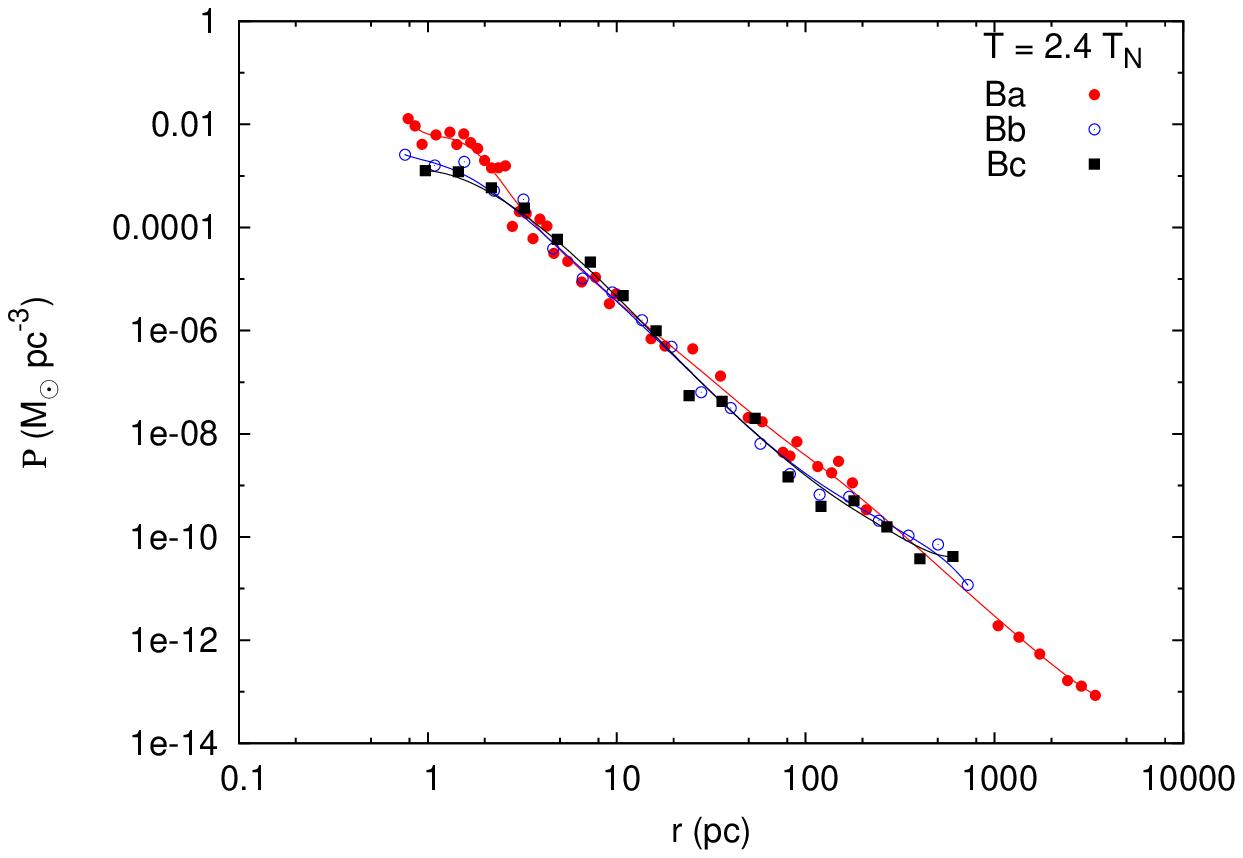}\\
\caption{Density profile of the stellar BHs at two different times in
  model Ba (top panel) and after 2.4 times the GC disruption time in
  models Ba, Bb and Bc. }
\label{densBH}
\end{figure} 
After cluster inspiral the BH population distributes
around the SMBH following a density profile that scales approximately
as $\rho_{\rm BH}(r) \propto r^{-3}$, independently of the GC
orbit, as shown in Figure \ref{densBH}. The distribution weakly varies
after the GC disruption.
In model Ba the BHs density profile is characterised by a small core,
extending out to $1.5-2\pc$, with an $r^{-3}$ scaling on larger scales.
In models Bb and Bc, instead, the central density is lower and
flattens beyond $\sim100\pc$. However, we caution that this feature
may be a result of the low number of objects in the innermost regions.

\begin{figure*}
\centering
\includegraphics[width=5.5cm]{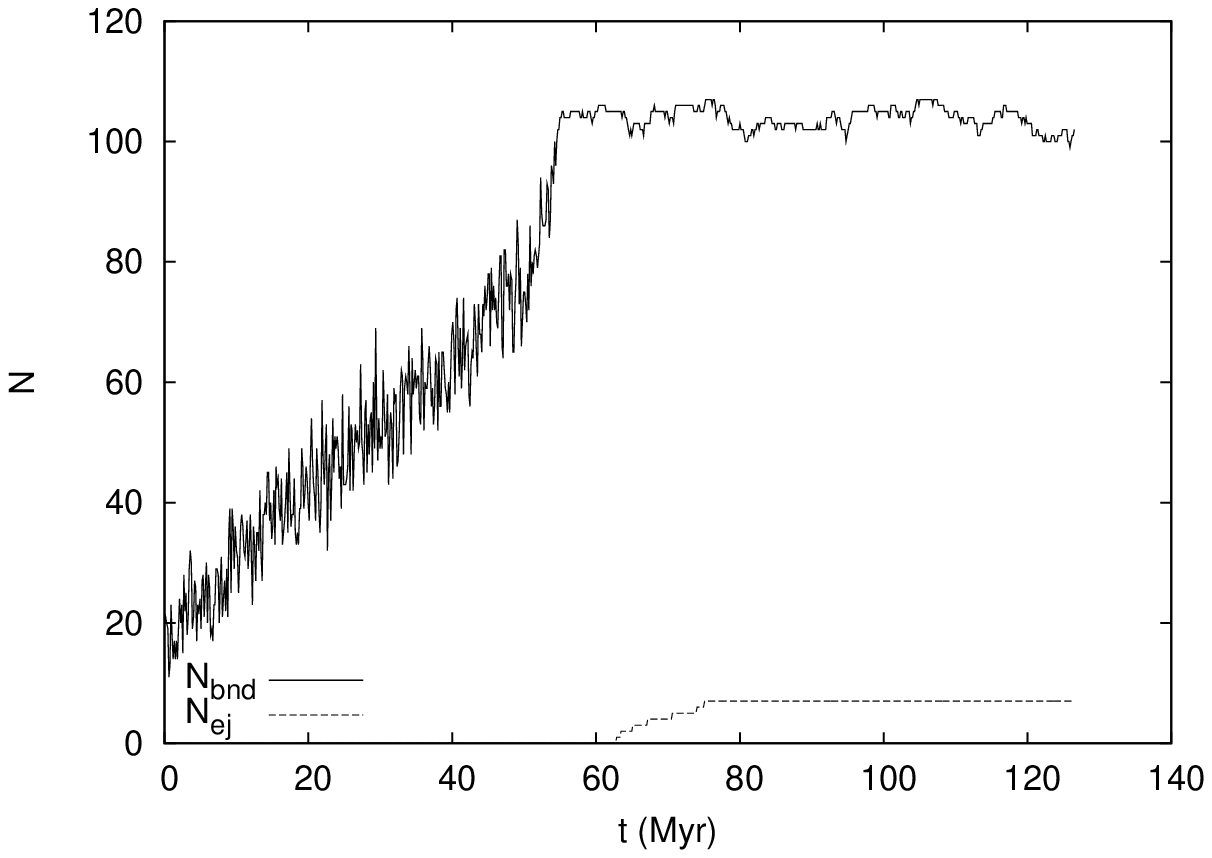}
\includegraphics[width=5.5cm]{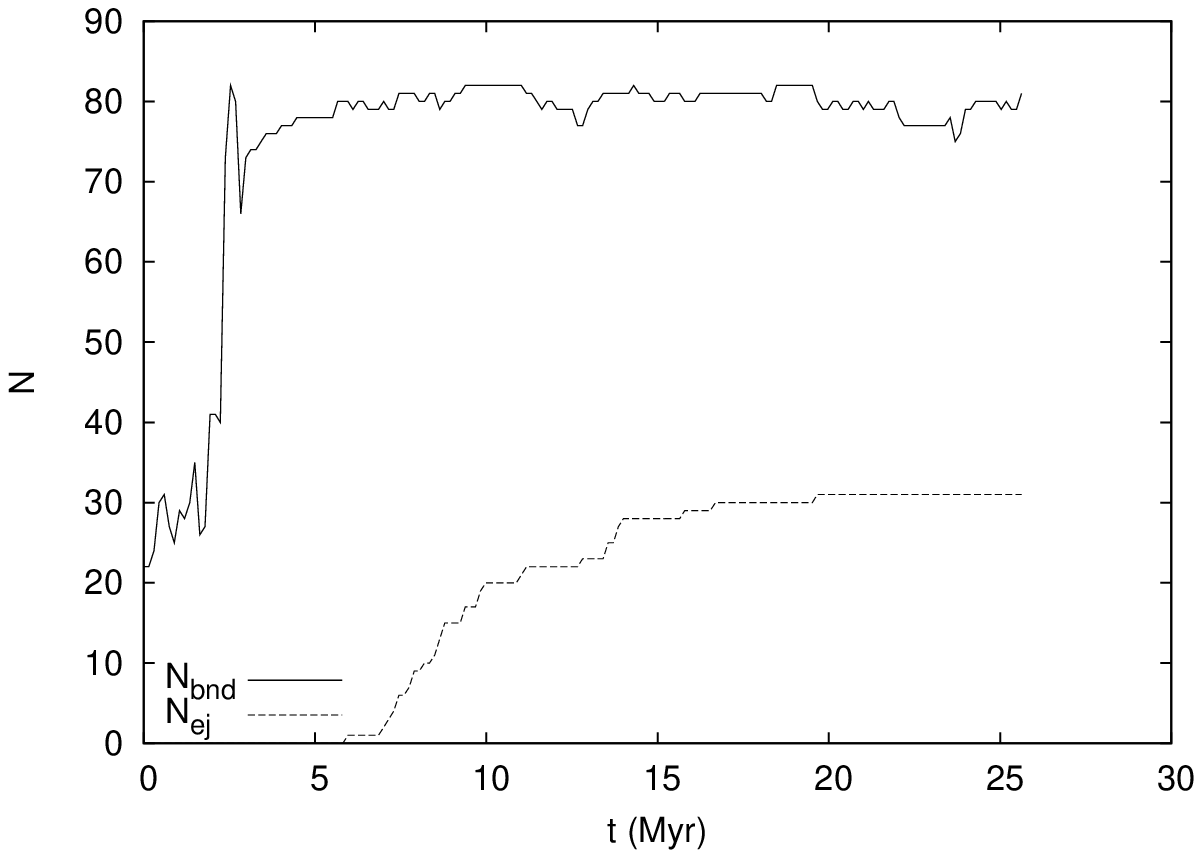}
\includegraphics[width=5.5cm]{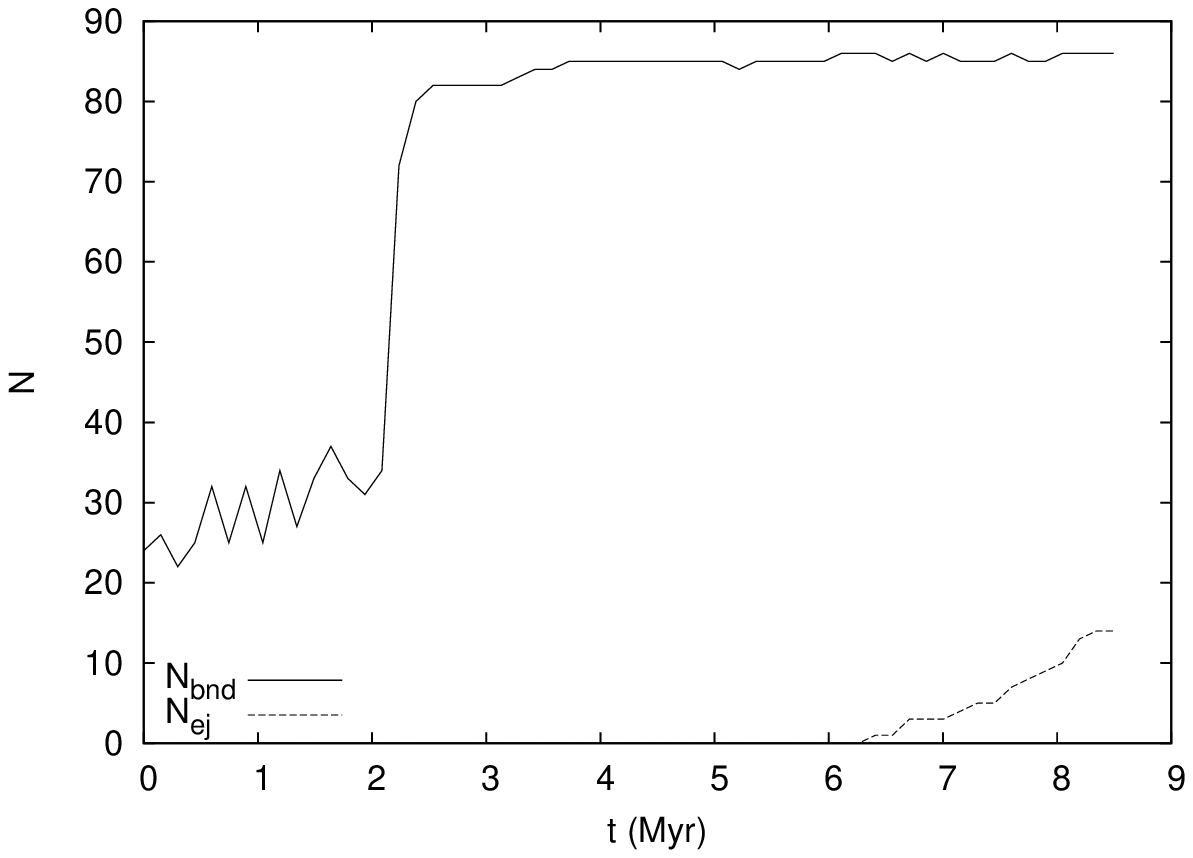}
\caption{Number of stellar BHs bound to the SMBH (straight line) and
  unbound and escaping (dotted line) as a function of time during the
  cluster inspiral, for models Ba (left), Bb (middle) and Bc (right).}
\label{BHbnd}
\end{figure*}

During the evolution, BHs can either be captured and become bound to
the SMBH or become unbound. Figure \ref{BHbnd} shows the number of BHs
bound and unbound to the SMBH as a function of time in models with
different initial orbital eccentricity. Unbound BHs are selected as
those BHs that are unbound to the SMBH and have reached a distance of
$500\pc$. We find that during the early, dynamical friction dominated,
phase of the inspiral, the number of bound BHs increases rapidly in
all models, though this effect is more visible in model Ba which is
characterised by a longer inspiral time. There is then a short but
efficient phase of further capture of BHs by the SMBH, which
terminates roughly at the time of cluster disruption. By this time,
the fraction of bound BHs is quite high, going from about $70-75\%$
for the radial and eccentric case to about $85\%$ for the circular
case.  After cluster disruption, the number of bound BHs remains
unchanged. It is at or after this time, however, that some BHs acquire
sufficient velocities to become unbound and escape.
Ejected BHs can reach distances of $100\pc$ or more in $10-20$ Myr,
depending on the model, though this is rare and limited to a small
number of objects. Potentially more interesting is the minimum
distance from the SMBH attained by the BHs over the inspiral.  This
ranges from about $0.5\pc$ for model Ba to about $0.1\pc$ for model
Bc. We will discuss the possibility of capture by the SMBH in the next section.

Our results are in good agreement with \cite{antonini14}, who model
the evolution of stellar BHs delivered by infalling star clusters
around an SMBH. Despite a similar numerical setup, there are important
differences in the two sets of simulations.  The galaxy models used by
\citet{antonini14} are much steeper and concentrated than our B
models, which are aimed to represent the MW centre before the
formation of the NSC.  While \cite{antonini14} consider only GCs
moving on circular orbits with small initial radius ($r_0=20\pc$), we
set a larger initial apocentre ($r_0=50\pc$) and allow for three
different values of the GC's orbital eccentricity ($e=0,0.7,1$).  The
number of BHs in the clusters is also smaller in our case.  While is
not trivial to predict the effect of such different initial
conditions, we expect that allowing for non-circular orbits can lead
to the ejection of some BHs by the slingshot mechanism
\citep{ASCDS15,CF2015}, thus reducing the number of BHs that bind to
the SMBH and leading to an expansion of the the BHs half-mass radius.
The fact that our GCs move on a wider orbit and contain an initially
unsegregated population of BHs can increase the effect of mass loss
and, consequently, increase the possibility that some BHs are tidally
lost along the inspiral.

\section{Gravitational wave sources}
\label{sec:gws}
We now investigate the formation of potential GW sources in the
simulations of GC inspiral with either an IMBH or a cluster of stellar
mass BHs.

\subsection{Simulations of GCs with a central IMBH}
%% IMBH CASE 
If the cluster contains an IMBH in the centre, the evolution of the
SMBH-IMBH system can be divided in three main phases. The first phase
is driven by dynamical friction exerted on the GC by the galactic
background, and is largely insensitive to the IMBH mass. Dynamical
friction becomes less and less efficient as the cluster is distorted
by tidal effects and nominally ends when the GC can be considered
disrupted. The second phase is characterised by dynamical friction
acting on the IMBH itself, and is therefore sensitive to the IMBH
mass. At the same time, close encounters with background stars start
contributing to the orbital decay of the SMBH-IMBH system, leading to
the formation of a binary. If the flux of interacting stars is
sufficient to harden the binary to the separation where emission of
GWs becomes dominant, a third phase ensues driven by GW losses in
which the binary quickly shrinks and circularises until the black
holes coalesce. 

 However, internal GC dynamics plays a crucial role in determining
  whether an IMBH can survive tidal forces and reach the inner
  galactic nucleus. IMBH formation is a largely debated process, which
  is thought to occur mainly through two channels: FAST and SLOW
  \citep{Giersz15}.

  The FAST scenario occurs in extremely dense GCs, having initial
  central densities of order $\sim10^8\msun\pcc$. In these
  extreme environments, BHs segregate rapidly to the core, where they
  form a compact subsystem. The large densities favour frequent
  single-binary and binary-binary BH interactions, driving the
  formation and growth of an IMBH seed with mass $\simeq 100\msun$. During
  these events, the IMBH seed undergoes multiple mergers with stellar
  mass BHs, possibly experiencing strong natal kicks due to
  anisotropic GWs emission. The recoil velocity is a function of the
  BH-to-IMBH mass ratio $q=m_\imbh/m_\bh$ through the parameter $\eta
  = q/(1+q)^2$ \citep{schnittman07}.  For typical parameters, $\eta
  \sim 0.05-0.22$.  The recoil velocity peaks in correspondence of
  $\eta \simeq 0.2$, with values of $100-500\kms$, thus sufficiently
  large to eject the IMBH seed from ordinary GCs.  We note, however,
  that the extreme densities required for the FAST channel correspond
  to escape velocities larger than $\sim 100\kms$ and that $\eta
  \gtrsim 0.2$ implies a mass ratio $q>0.4$, a condition that is only
  satisfied at early times, when all the most massive BHs are still in
  the cluster core. The FAST scenario is likely more relevant for NSCs
  than GCs, given their larger central densities.

  In the SLOW scenario IMBH formation can occur in sparser
  environments with central densities $\sim 10^5\msun\pcc$ if
  dynamical interactions and supernovae explosions are sufficiently
  efficient to eject all stellar BHs but one. In this case, the
  surviving BH undergoes a slow growth process, lasting $\sim 1-10$
  Gyr, reaching a mass of up to $10^4\msun$. Dynamical interactions
  are sufficiently energetic to eject the BHs, but generally do not
  lead to BH-BH mergers inside the cluster, thus avoiding merger
  recoils. IMBHs forming through the SLOW channel appear later in the
  GC lifetime, generally after the post-core collapse phase. As
  suggested by \cite{Giersz15}, this scenario is more probable for
  IMBH formation in GCs due to the lower central densities that it
  requires.

  Our models assume either that the cluster reaches the galactic
  centre before the NSC is fully assembled, i.e. on time-scales $\sim
  0.1-1$ Gyr, or after the NSC formation, $>1$ Gyr. Hence, our S
  models can, in principle, represent clusters where the IMBHs form
  through the FAST channel, and were deposited into the galactic
  centre promptly after their formation, while the M models better
  represent the SLOW scenario, with the IMBH forming at a later stage
  and the parent cluster reaching the Galactic Centre on time-scales
  longer than 1 Gyr.

\begin{figure*}
\centering
\includegraphics[width=5.5cm]{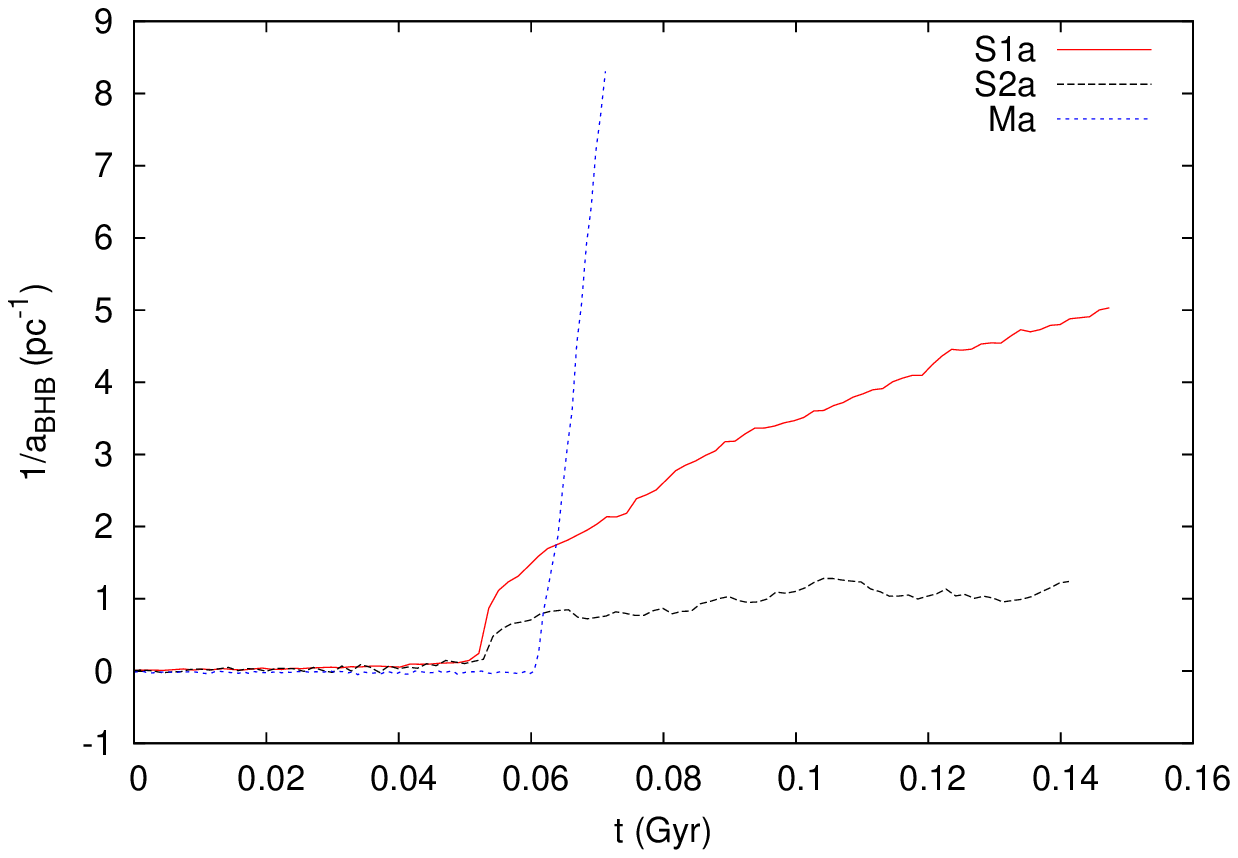}
\includegraphics[width=5.5cm]{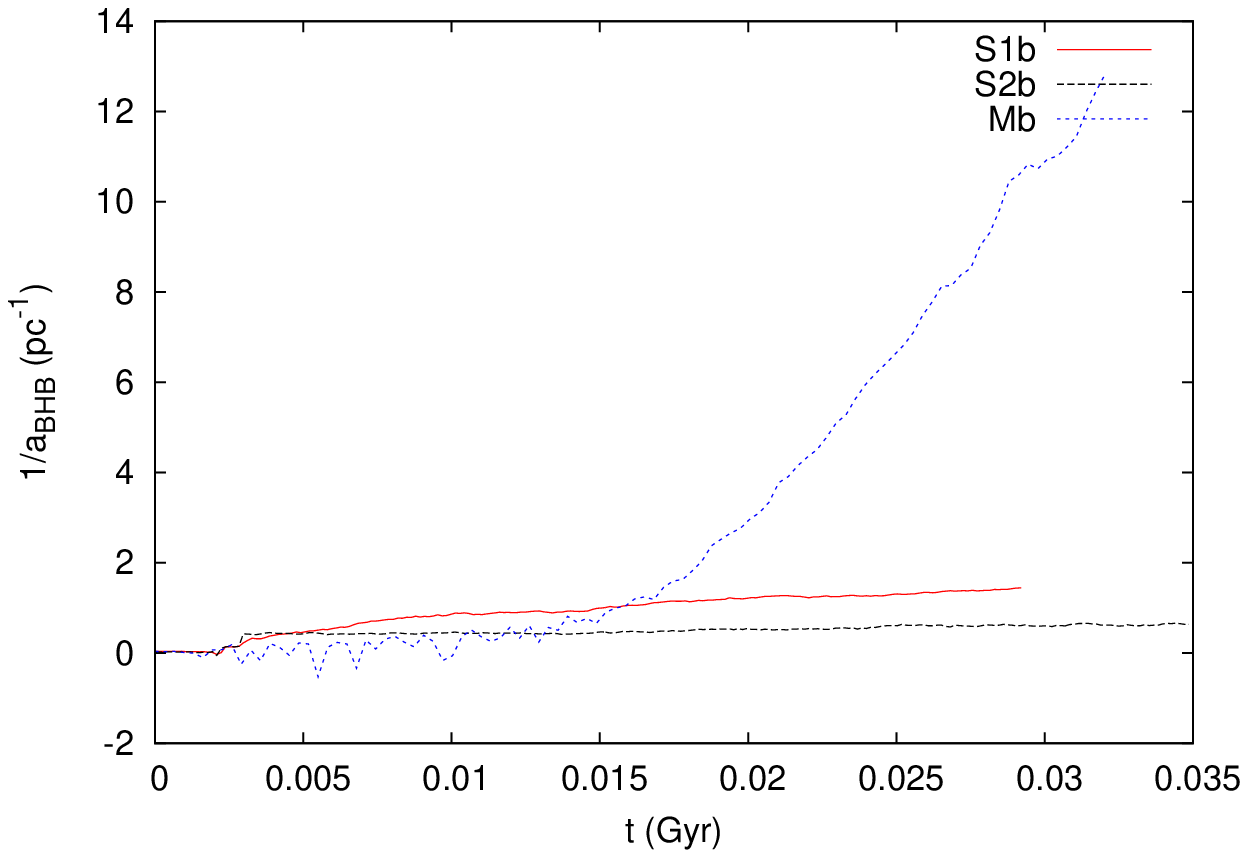}
\includegraphics[width=5.5cm]{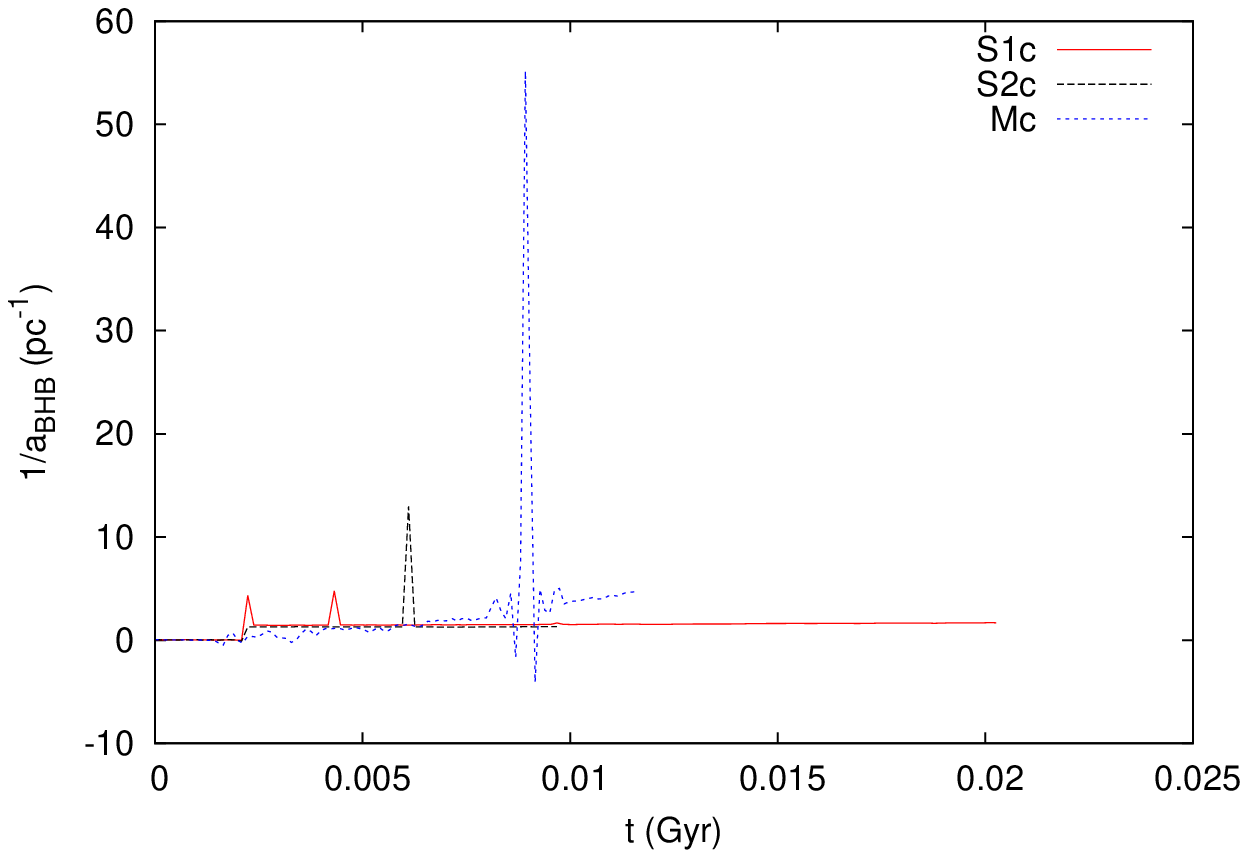}
\caption{The inverse semi-major axis of the BHB in all the
  models considered. Two distinct phases are evident, the first
  driven by dynamical friction on the cluster and insensitive to the
  IMBH mass and the second driven by dynamical friction on the IMBH
  and stellar encounters.}
\label{fig:semi}
\end{figure*}
The inverse semi-major axis of the SMBH-IMBH binary (hereafter
BHB) is shown in Fig.\,\ref{fig:semi} for all models hosting an IMBH.
The time of GC disruption at about 50 Myr (circular orbits) and about
3 Myr (radial and eccentric orbits) is clearly visible in all models,
and separates the first phase of binary evolution driven by dynamical
friction on the cluster from the second phase, driven by friction on
the IMBH and close encounters with intersecting stars. This second phase is
efficient and leads to a fast decay of the binary separation,
especially for the circular models for which this phase lasts the
longest.  

The orbital decay, as well as the hardening rate of the BHB, defined
as the derivative of the inverse semi-major axis, clearly depend on
the slope of the background galaxy density profile, the initial
cluster orbit and the IMBH mass. For clusters orbiting in a galaxy
with a shallow profile, the decay slows down significantly in the
third phase or even stalls, while it continues with high efficiency
until the simulations are terminated for the models orbiting in a
steeper density background. Eccentric cluster orbits result in a much
faster inspiral and therefore in a less efficient decay and binary
hardening, with hardly any hardening in the case of radial orbits. The
dependence on IMBH mass is clearly visible in the case of circular and
eccentric orbits, with larger IMBHs resulting in more efficient
hardening.

The evolution of the BHB semi-major axis leads naturally to the
definition of two time-scales: a ``disruption time'', $t_{\rm ds}$,
which marks the moment at which the GC is disrupted, and a ``stalling
time'', $t_{\rm st}$, at which the inspiral of the BHB stalls.  These
time-scales are given in Table \ref{tab:tbhb} for all models, and
compared with the disruption time $t_{\rm N}$ determined from the mass
bound to the IMBH.
\begin{table}
  \begin{center}
    \caption{Characteristic time-scales in the black hole binary evolution:
      simulation name, disruption time evaluated from the semi-major
      axis evolution, disruption time evaluated from the mass bound to
    the IMBH, stalling time of the binary separation and stalling
    semi-major axis.}
    \label{tab:tbhb}.
\begin{tabular}{ccccc}
\hline
Model &  $t_{\rm ds}$ & $t_{\rm N}$ & $t_{\rm st}$  & $a_{\rm st}$\\
 & (Myr) & (Myr) & (Myr) & (pc)\\ 
\hline
S1a & $ 54\pm 2$   & $ 53.4\pm 0.9$& $ -$   & $-$     \\
S2a & $ 54\pm 2$   & $ 54.3\pm 0.2$& $ 100$ & $0.8$   \\
Ma  & $ 61.0\pm0.3$& $ 61.6\pm 0.7$& $-$    & $<1/8$  \\
S1b & $ 2.1\pm 0.1$& $  3.2\pm 0.4$& $ 10$  & $0.7$   \\
S2b & $ 2.5\pm 0.2$& $  3.1\pm 0.6$& $ 25$  & $1.6$   \\
Mb  & $ 2.2\pm 0.1$& $  3.1\pm 0.9$& $ -$   & $<1/13$ \\
S1c & $ 2.1\pm 0.1$& $  2.2\pm 0.1$& $ 2.3$ & $0.6$   \\
S2c & $ 2.2\pm 0.1$& $  2.2\pm 0.1$& $ 2.3$ & $0.8$   \\
Mc  & $ 1.7\pm 0.1$& $  1.7\pm 0.1$& $ -$   & $ <1/8$ \\
\hline
\end{tabular}
  \end{center}
\end{table}

\begin{figure}
\centering
\includegraphics[width=8cm]{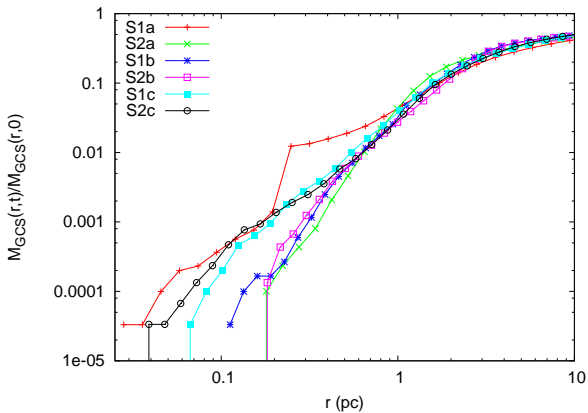}
\caption{Cumulative distribution of the GC mass at time $2.5 t_{\rm
    ds}$, normalised to the initial mass.}
\label{fig:mdiscmp}
\end{figure} 
The GC disruption time can be used to investigate how deeply the GC
relic penetrates within the galactic nucleus. Figure \ref{fig:mdiscmp}
shows the radial cumulative mass distribution of the GC after $\sim
2.5t_{\rm ds}$, in units of the initial GC mass. The mass distribution
is quite similar in the case of eccentric and radial orbits, showing
no significant dependence on IMBH mass. On the other hand, the
distributions for circular orbits are remarkably different, with stars
reaching distances of $0.03\pc$ or less in model S1, while reaching
$0.1\pc$ at most in model S2.  This is due to the dependence of the
number of stars bound to the IMBH on the mass of the IMBH itself. At
time $t=2.5t_{\rm ds}$ we find $10$ stars bound to the IMBH in model
S1a, but only $2$ in model S2a.  Scaling these results to a realistic
GC model, this would imply $\sim 300$ stars at distances $\lae
0.03\pc$. For our adopted galaxy mass model, the deposited stars would
represent a dominant component driving the BHB evolution at these distances.

\begin{figure*}
\includegraphics[width=5.5cm]{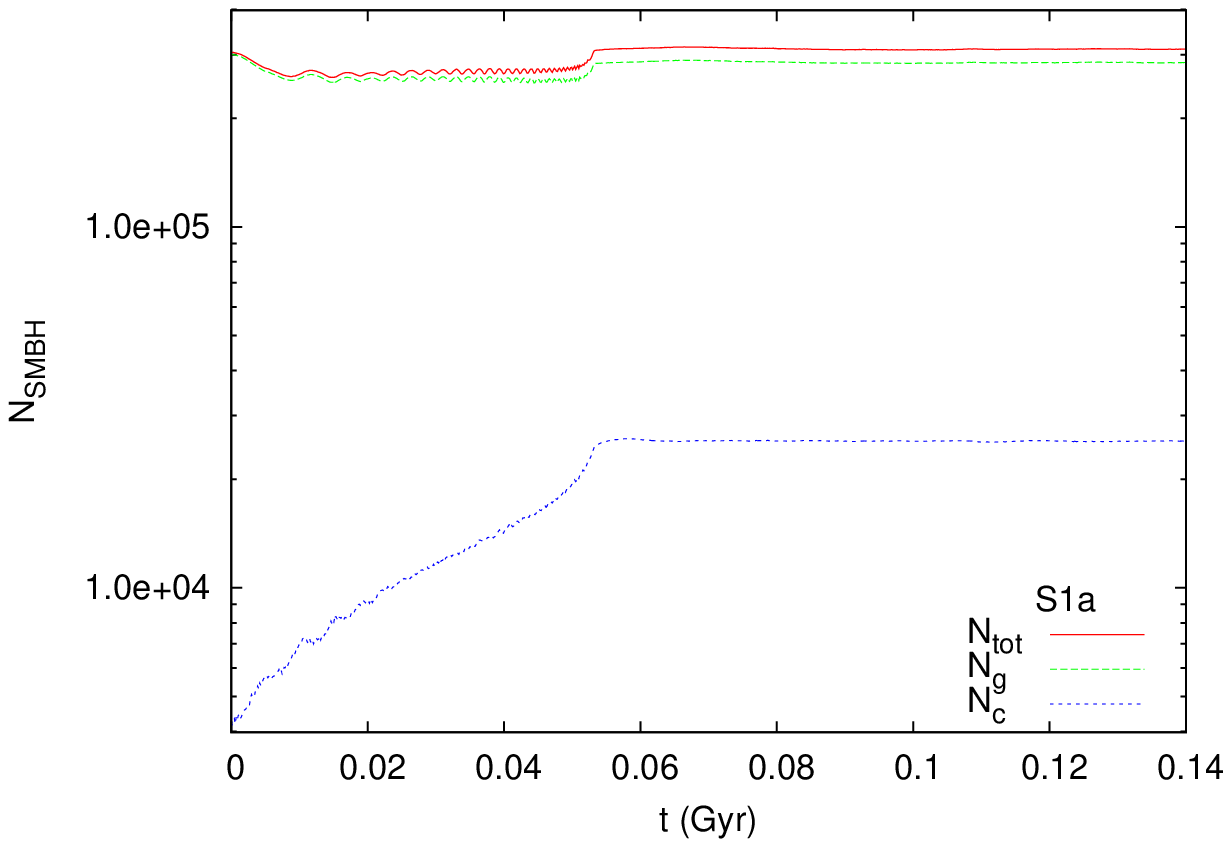}
\includegraphics[width=5.5cm]{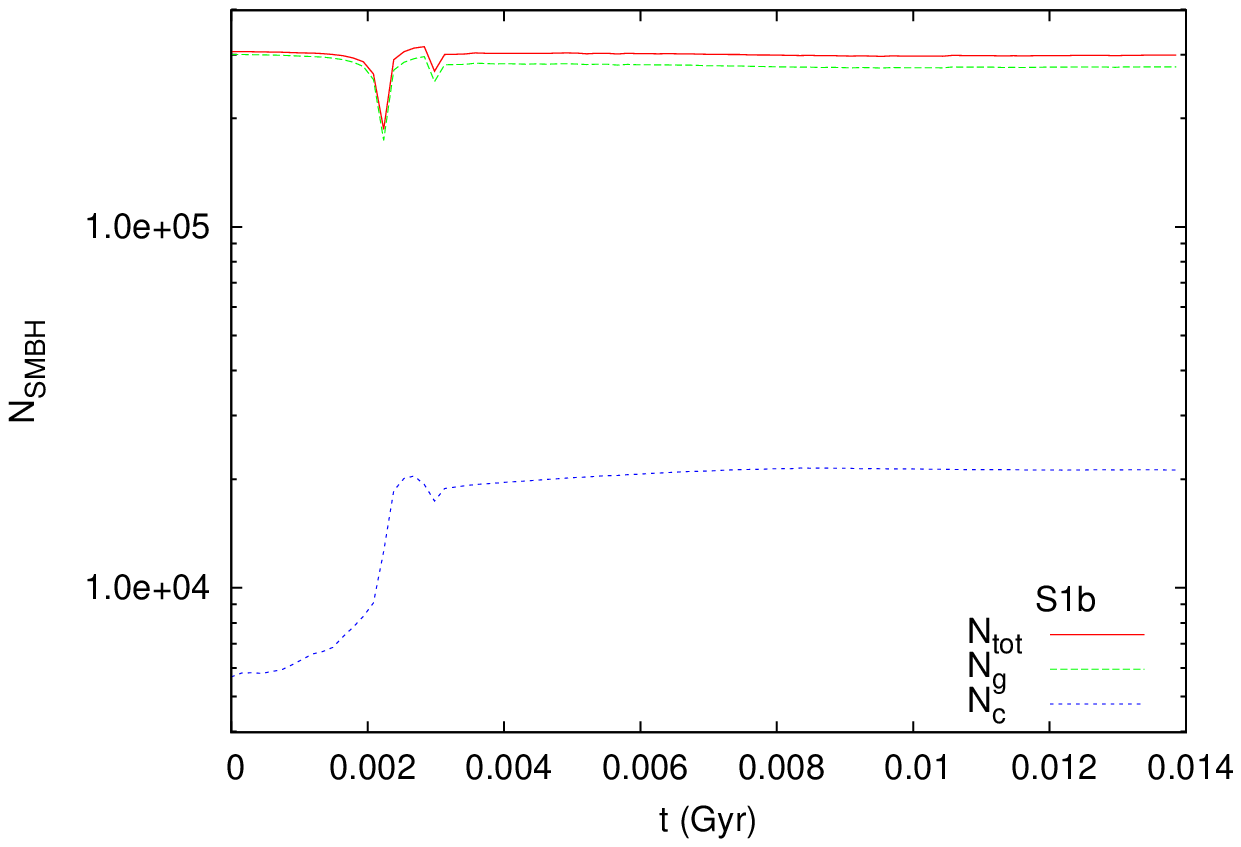}
\includegraphics[width=5.5cm]{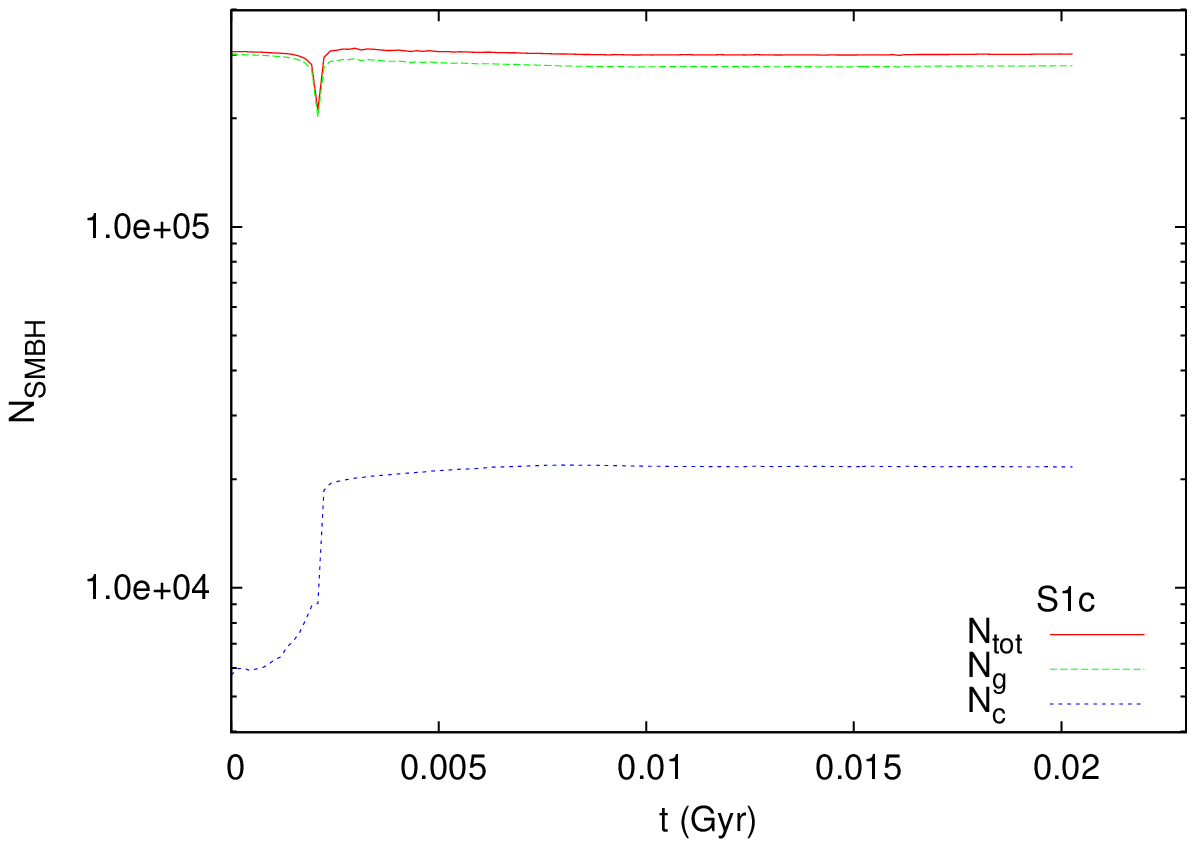}
\includegraphics[width=5.5cm]{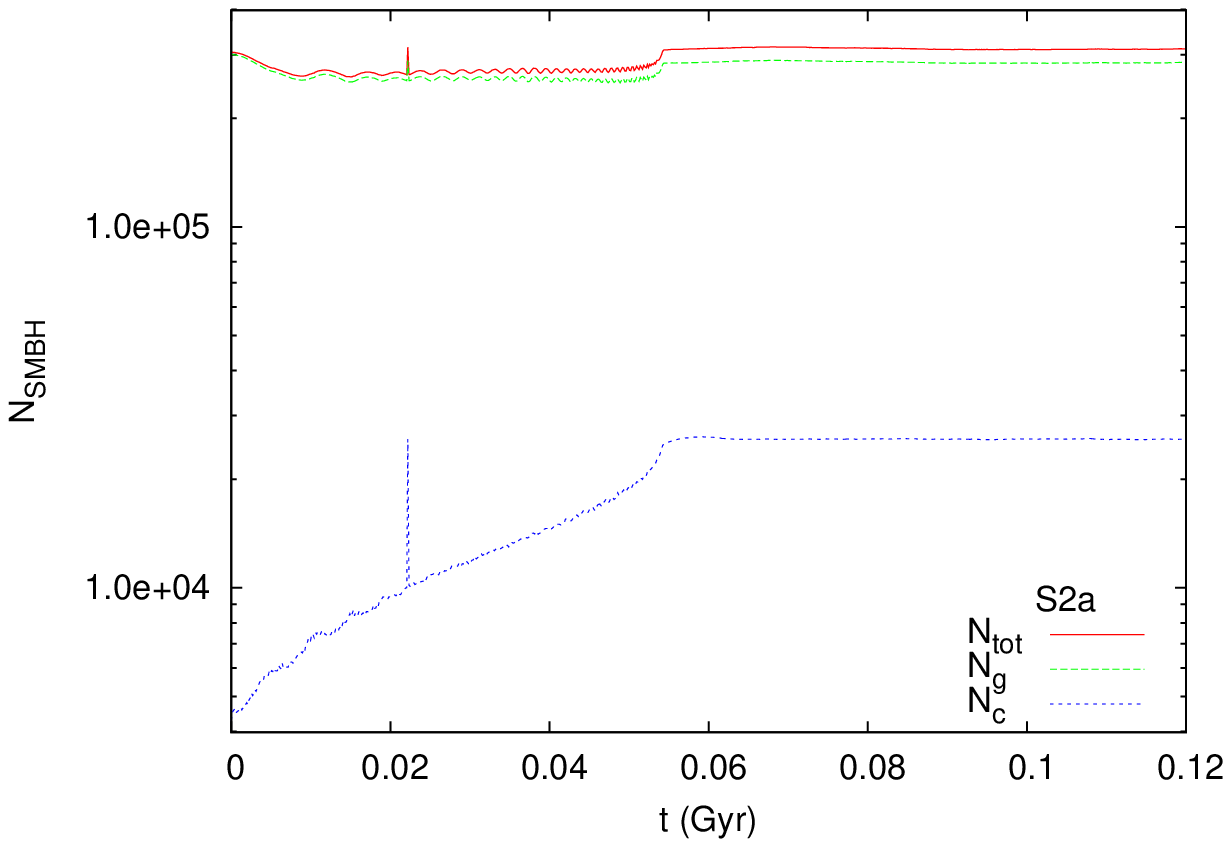}
\includegraphics[width=5.5cm]{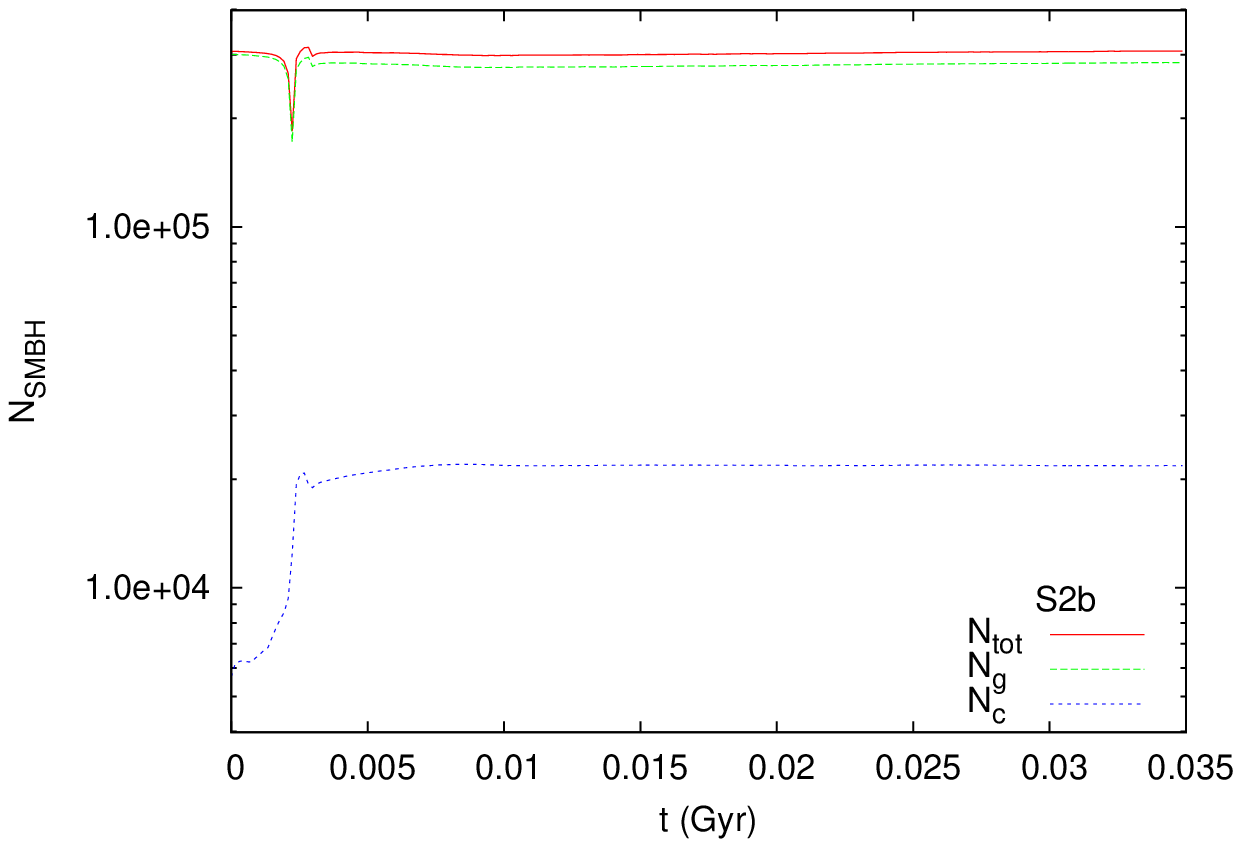}
\includegraphics[width=5.5cm]{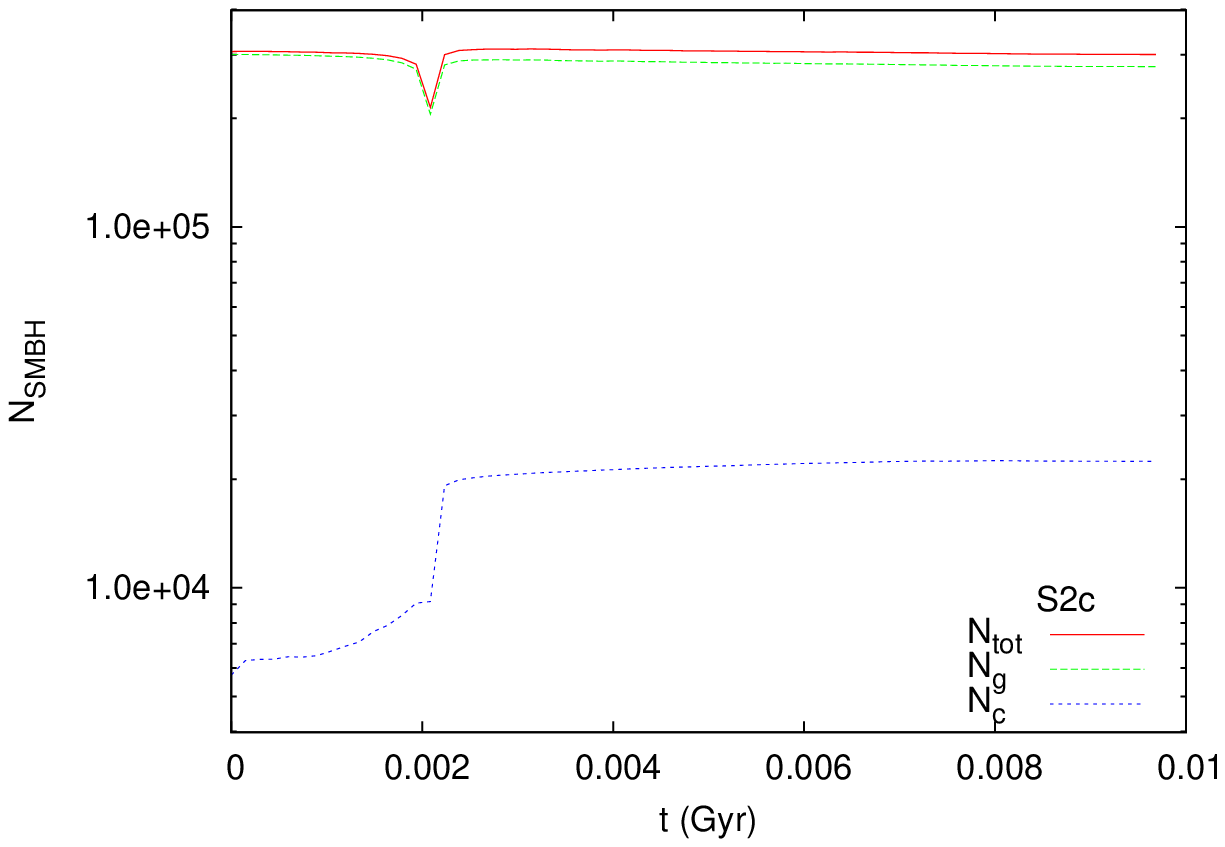}
\caption{Number of stars bound to the SMBH in
  model S1 (top panels) and S2 (bottom panels) as a function of time. From left to right
  models refer to the a, b and c cases, respectively, with different
  initial orbital eccentricity. In each panel, we show  the total
  number of stars bound to the SMBH (red/solid line), the number of
  bound stars initially belonging to the
  background galaxy (green/dashed line) and the number of bound stars
  initially  belonging to the GC (blue/dotted line).}
\label{fig:BHnum}
\end{figure*}

Figure \ref{fig:BHnum} shows the number of stars bound to the SMBH as
a function of time.
We find that, for circular orbits, the number of stars bound to the
SMBH oscillates during the phase of cluster infall and then increases
after the GC dissolution due to the contribution from GC stars. In the
cases of eccentric and radial orbits we observe a similar behaviour,
with the total number of bound stars reflecting the contribution from
cluster stars after disruption but with a dip at the time of
disruption.  This is likely due to the interaction between the
infalling cluster and the galactic nucleus when the GC reaches a
distance where the enclosed galactic mass becomes comparable to its
own mass. The acceleration imparted by the cluster results in stars
becoming, if only temporarily, unbound to the SMBH in the Keplerian
2-body sense. This effect is present in all simulations, but is more
significant in the case of eccentric and radial orbits.  A similar
effect can operate during the early evolution of dark matter dominated
galaxies and can have interesting consequences on the core/cusp
problem and the formation of SMBHs
\citep{sanchez06,goerdt2010,ASCD17a}.

\begin{figure*}
\includegraphics[width=5.5cm]{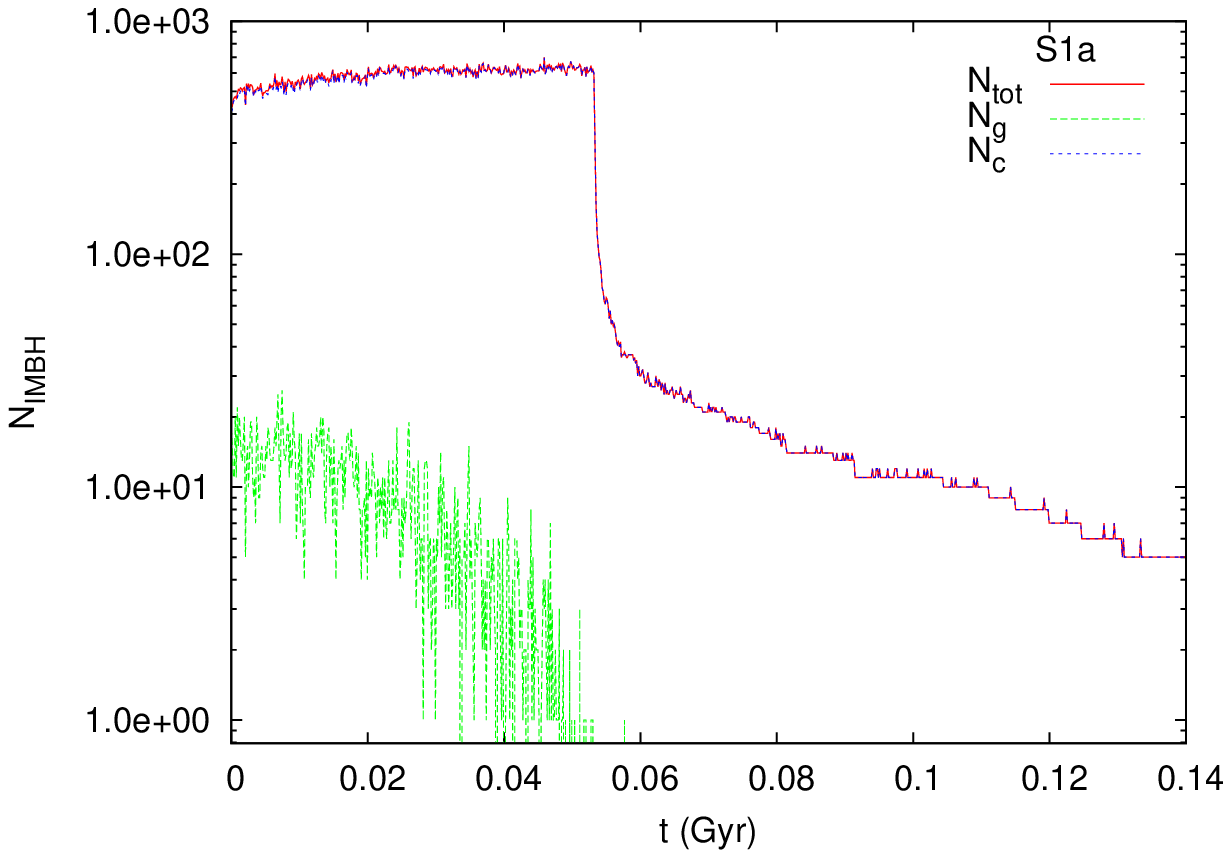}
\includegraphics[width=5.5cm]{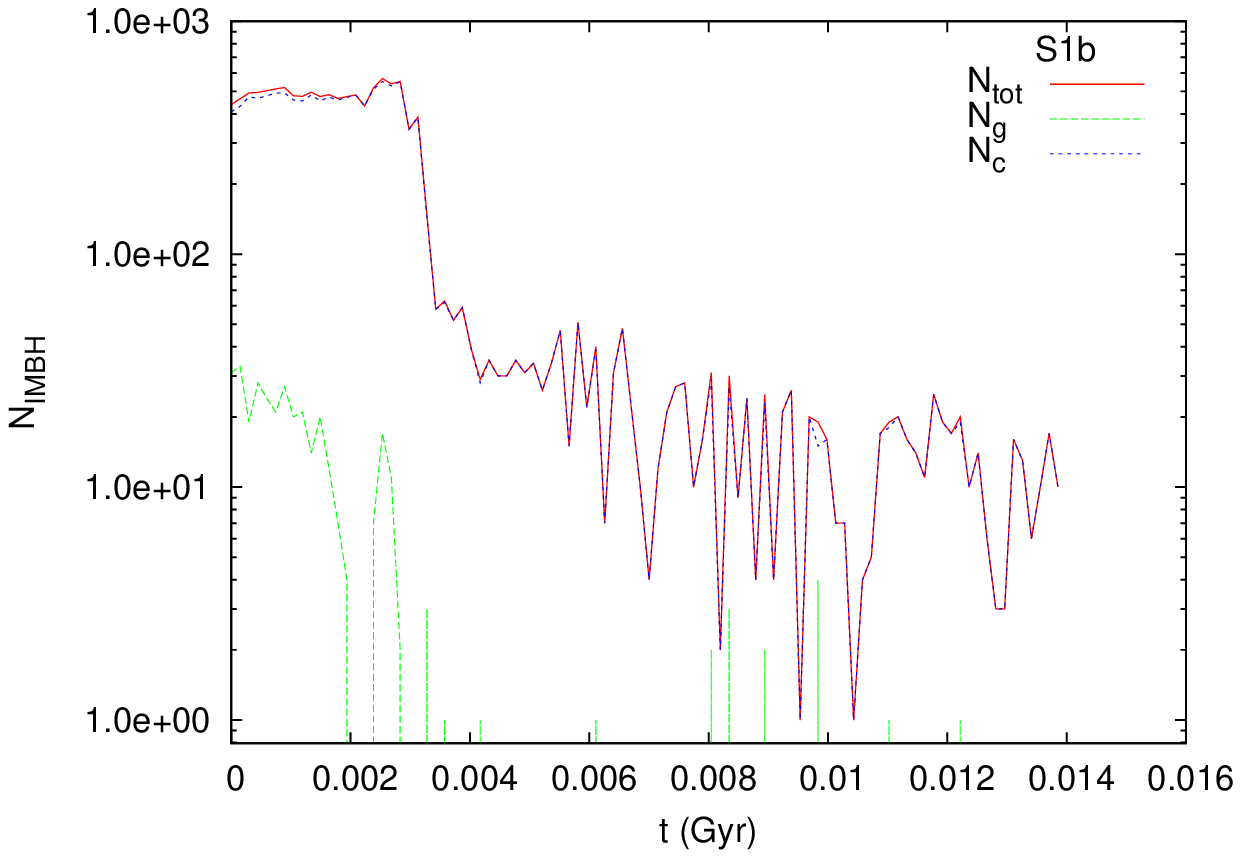}
\includegraphics[width=5.5cm]{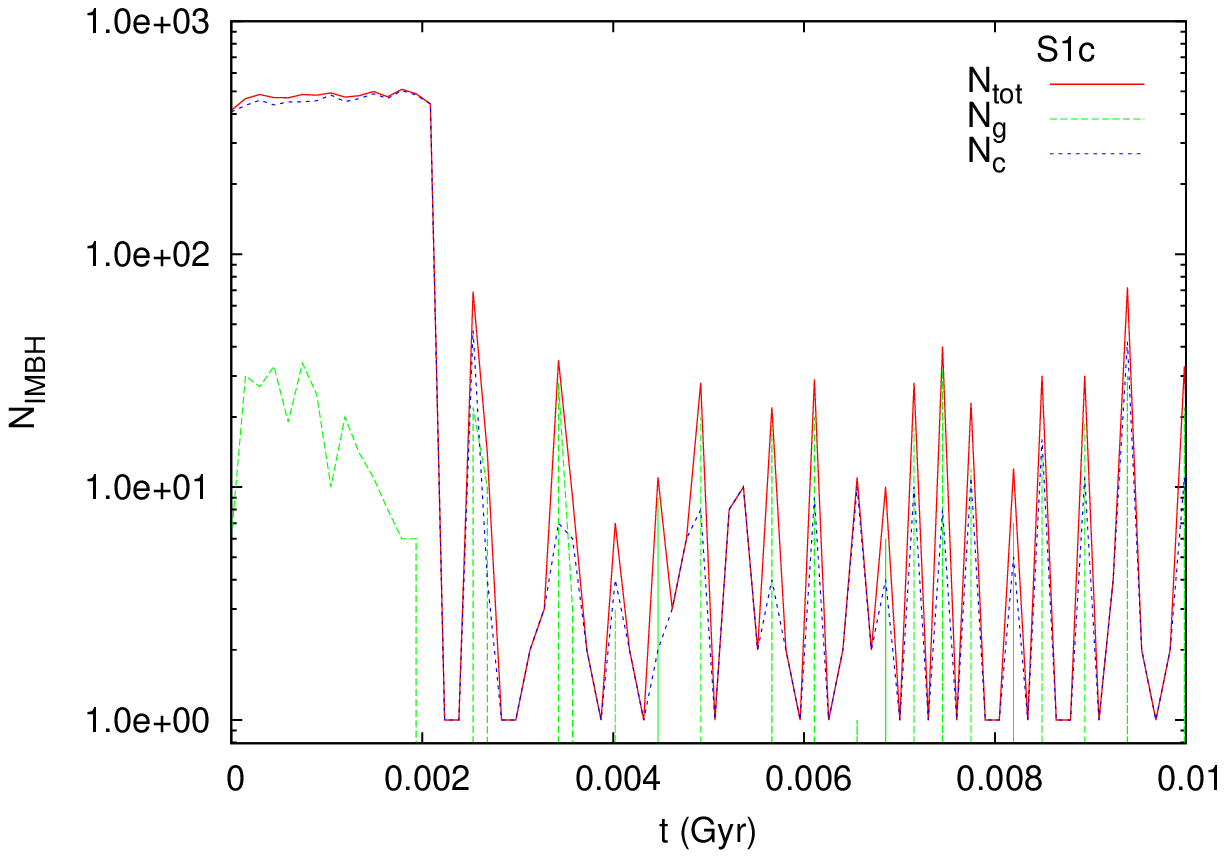}
\includegraphics[width=5.5cm]{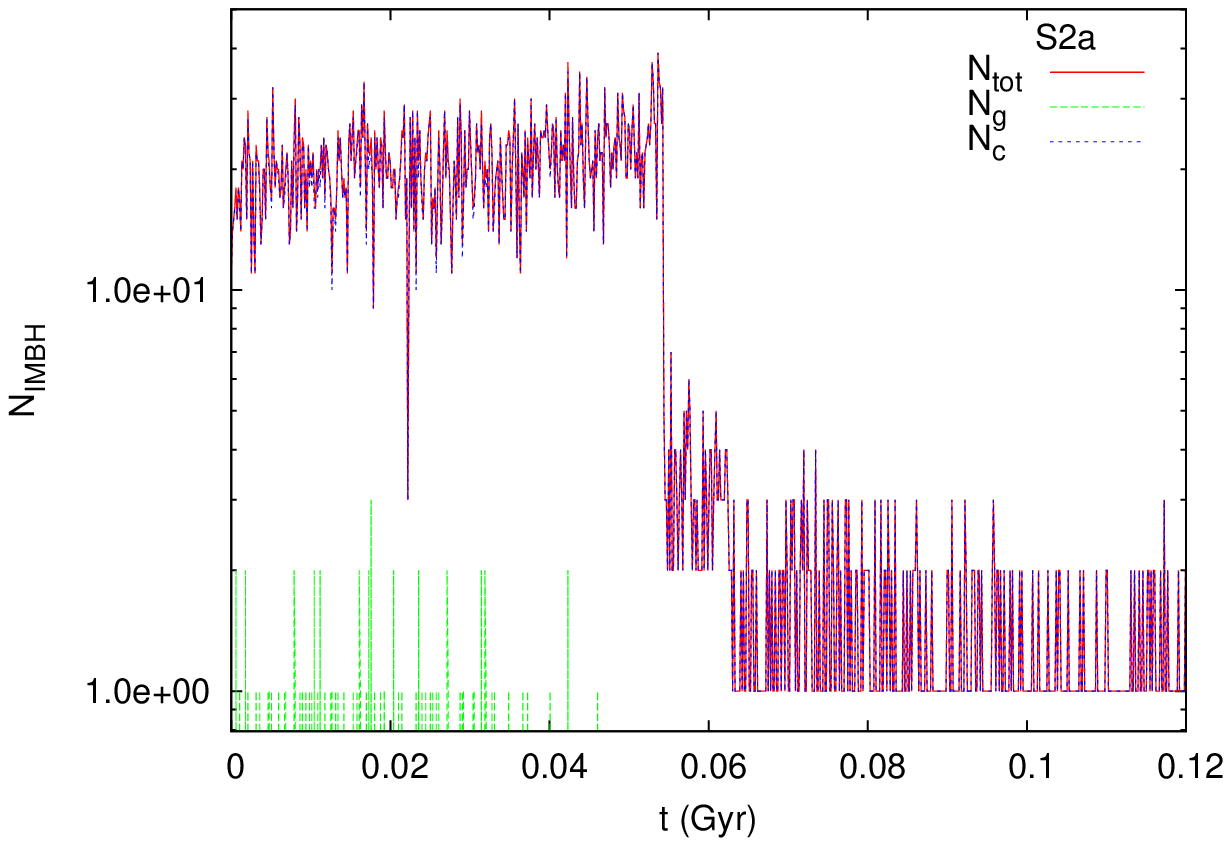}
\includegraphics[width=5.5cm]{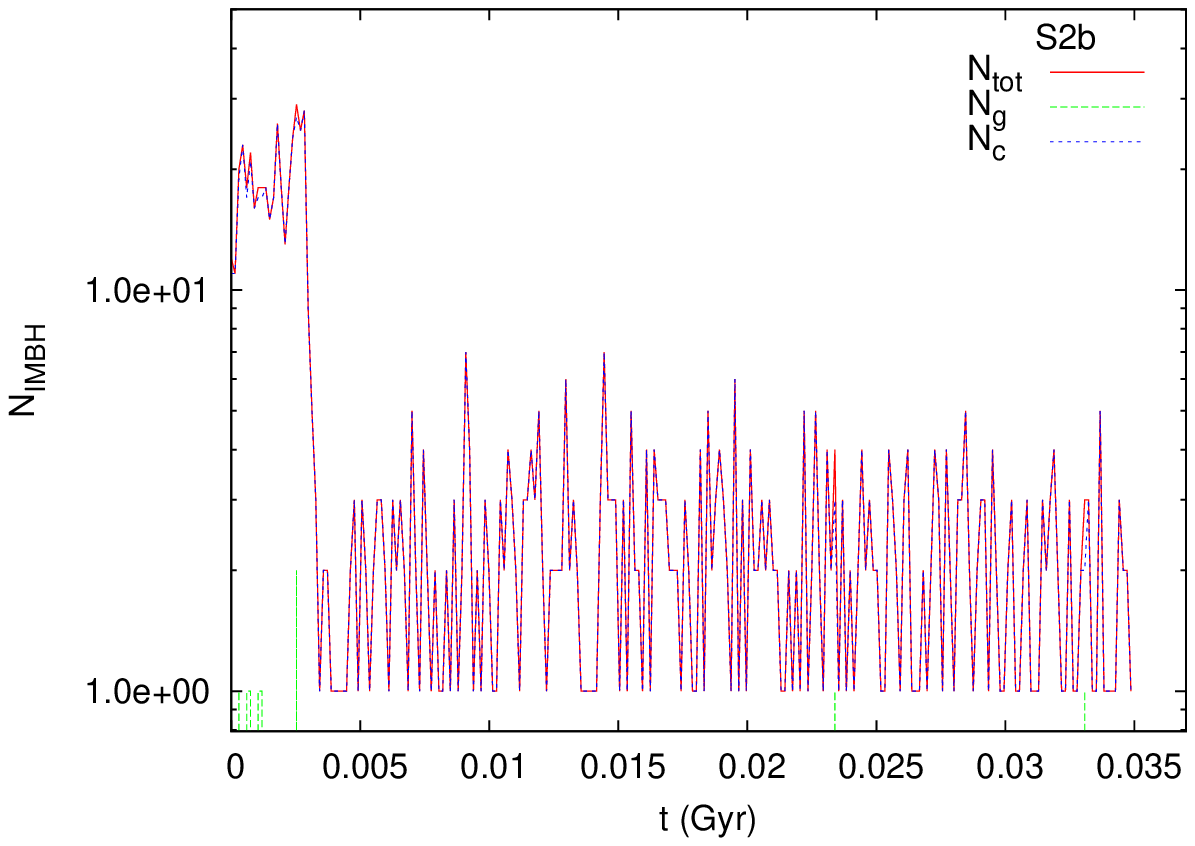}
\includegraphics[width=5.5cm]{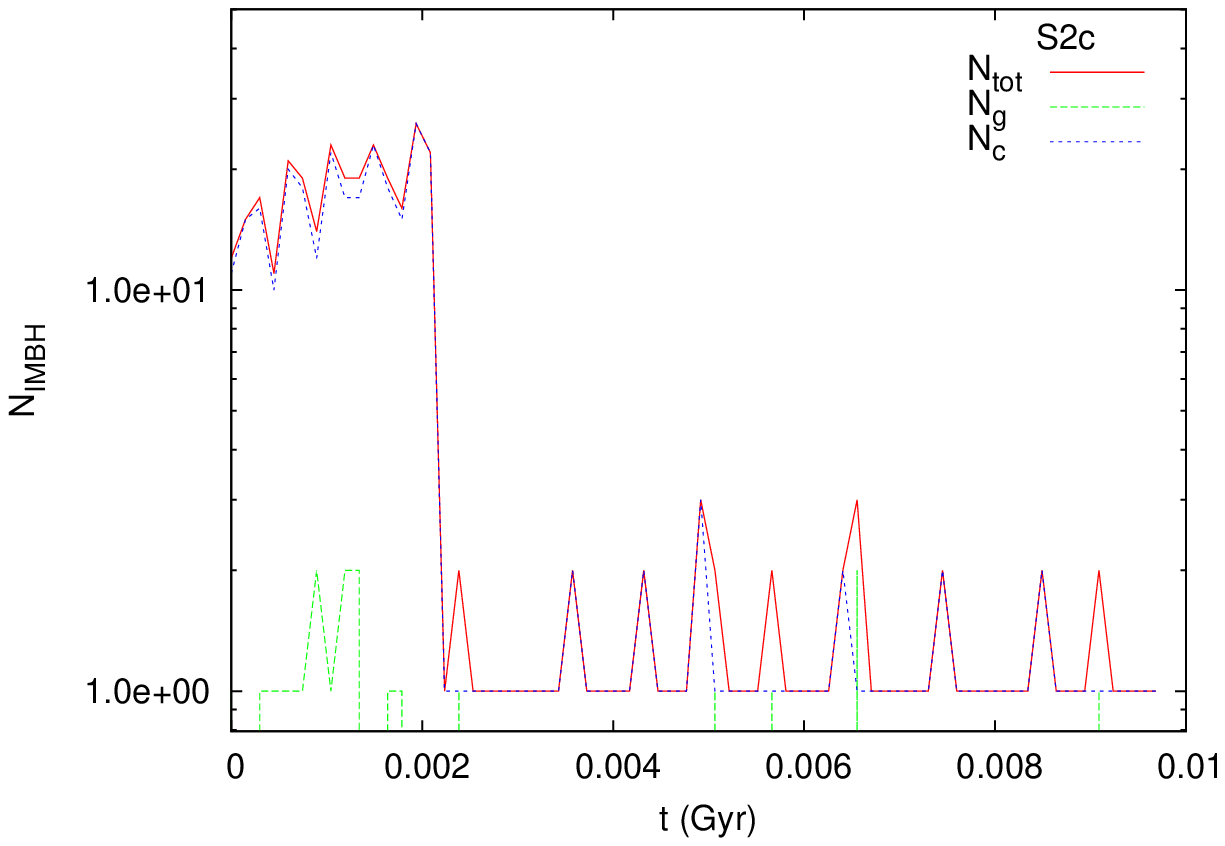}
\caption{Number of stars bound to the IMBH in model S1 (top panels) and S2 (bottom
  panels) as a function of time. Line-styles are as in Fig.\ref{fig:BHnum}.}
\label{fig:BHnum2}
\end{figure*}

Figure \ref{fig:BHnum2} shows instead the number of stars bound to the IMBH in all models.
We find a few hundred stars bound to the IMBH in model S1 during the
inspiral and a few tens in model S2, regardless of orbital
eccentricity. The vast majority of these stars originate from the GC.
After GC dissolution, the number of bound stars rapidly decreases to a
few tens (model S1) or zero (model S2), making the IMBH a freely
floating object, whose orbit is affected only by gravitational
encounters with stars.  A similar behaviour in the number of stars
bound to the SMBH and the IMBH is found in models M.

The time at which the number of stars bound to the IMBH drops, which
we name $t_N$, represents another way to estimate the disruption
time-scale cited above, $t_{\rm ds}$. Indeed, we expect that these two
time-scale should be quite similar since they are strictly connected
with the GC tidal disruption. It is worth noting that after the GC
disruption, more than $10^4$ GC stars are bound to the SMBH,
regardless of the GC initial orbit and the IMBH mass. On the other
hand, at most a few galactic stars are bound to the IMBH at any time.

\begin{figure}
\centering
\includegraphics[width=8cm]{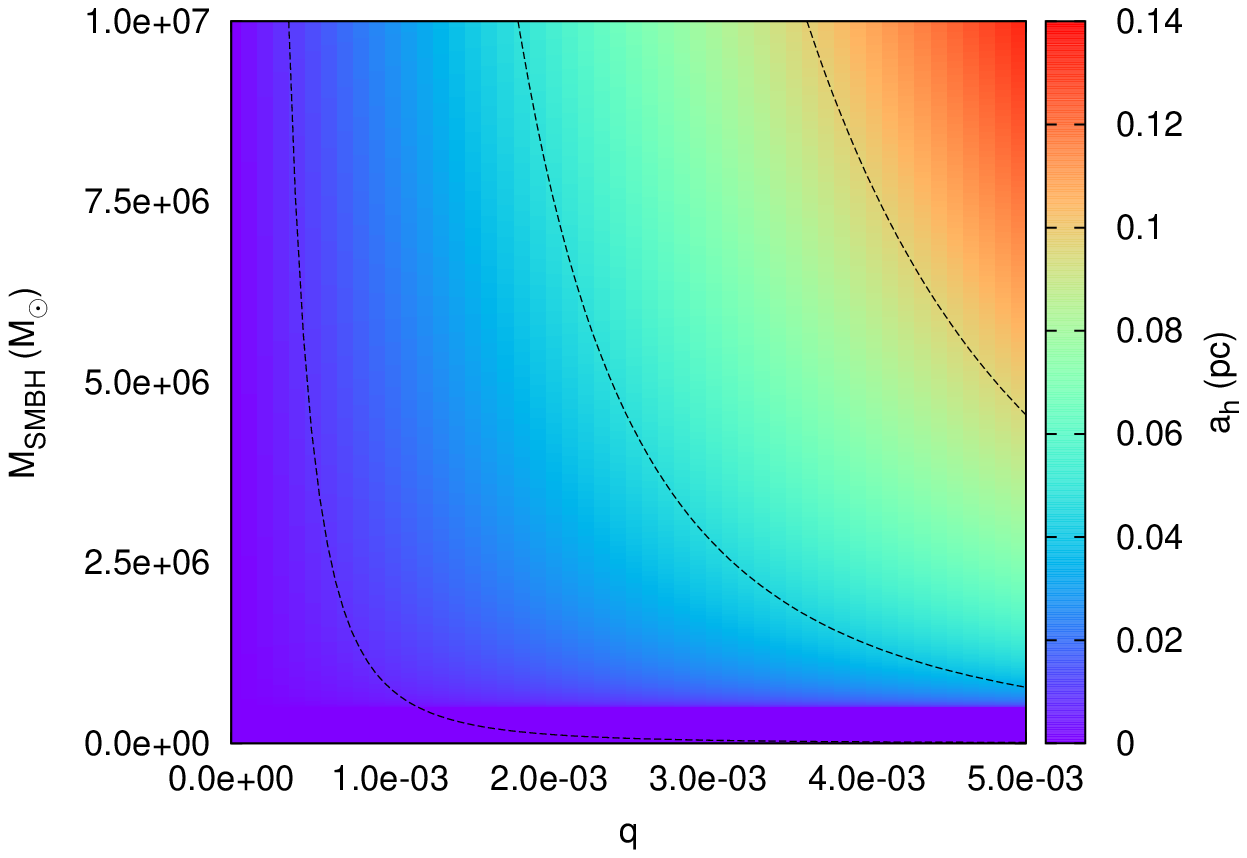}
\caption{Hard-binary separation as a function of the SMBH mass and the
  binary mass ratio.}
\label{rstall}
\end{figure}

The orbital evolution shown in Fig.\,\ref{fig:semi} indicates a
slowing down or even stalling of the inspiral in some of the models.
A binary is defined as ``hard'' when it reaches a separation at which
its binding energy per unit mass becomes comparable to $\sigma^2$,
where $\sigma$ is the stellar velocity dispersion. The corresponding
separation, called the ``hard-binary separation'', is given by
\citep{merritt06}
\begin{equation}
a_h \approx \frac{GM_2}{4\sigma^2}\,.
\label{ah}
\end{equation}
In an isothermal model, this is equivalent to
\begin{equation}
a_h=\frac{q}{\left(1+q\right)^2}\frac{1}{4}r_h,
\end{equation} 
where $q$ is the mass ratio and $r_h$ is the SMBH influence radius,
i.e. the radius containing twice the SMBH mass
\begin{equation}
M_g(r_h) = 2M_{\rm SMBH}\,.
\end{equation}
For our density profile \citep{Deh93} $r_h$ takes the simple form
\begin{equation}
r_h = r_g \left(\frac{\mathcal{M}}{1-\mathcal{M}} \right),
\label{rh}
\end{equation}
with 
\begin{equation}
\mathcal{M} = \left(\frac{2M_{\rm SMBH}}{M_g}\right)^{1/(3-\gamma)}\,.
\end{equation}
We obtain $r_h \simeq 83\pc$ for our models, which gives
$a_h=4.2\times 10^{-2}\pc$ for models S1 and $4.2\times 10^{-3}\pc$
for models S2.  Computing $r_h$ directly from the simulations
snapshots gives, at late times, $a_h=6.25\times 10^{-2}\pc$ for model
S1a and $a_h=6.04\times 10^{-3}\pc$ for model S2a, in good agreement
with the analytical estimate.  The dependence of the hard-binary
separation on the SMBH and IMBH masses is illustrated in
Fig.\,\ref{rstall}.

The hard-binary separation represents a sort of ``stalling radius'' 
for BHBs evolving in spherical galaxies \citep{merritt06} since the 
initial population of stars on low-angular orbits able to interact 
with the binary has been ejected in reaching $a_h$, and further 
orbital decay is due only to collisional repopulation of the binary's
losscone. 

In the M models, the BHB hardens much more efficiently and rapidly,
and no stalling is apparent in models Ma and Mb by the end of the
simulations.  This is due to the steeper inner density profile in
these models.  The stalling radius as evaluated directly from the
simulation is $a_h\simeq 1.6\times 10^{-2}\pc$ for model Ma, a few
times smaller than the S1a model.

\begin{figure}
\centering
\includegraphics[width=8cm]{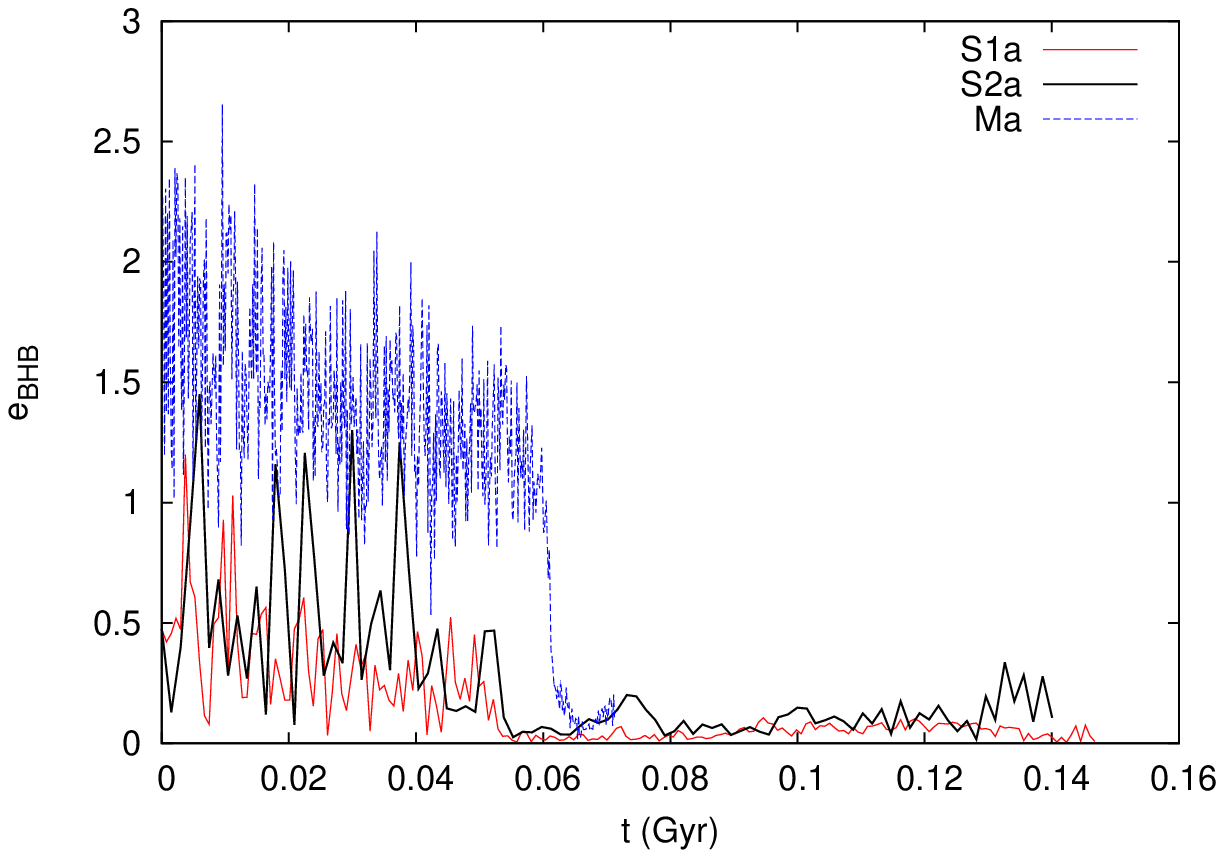}\\
\includegraphics[width=8cm]{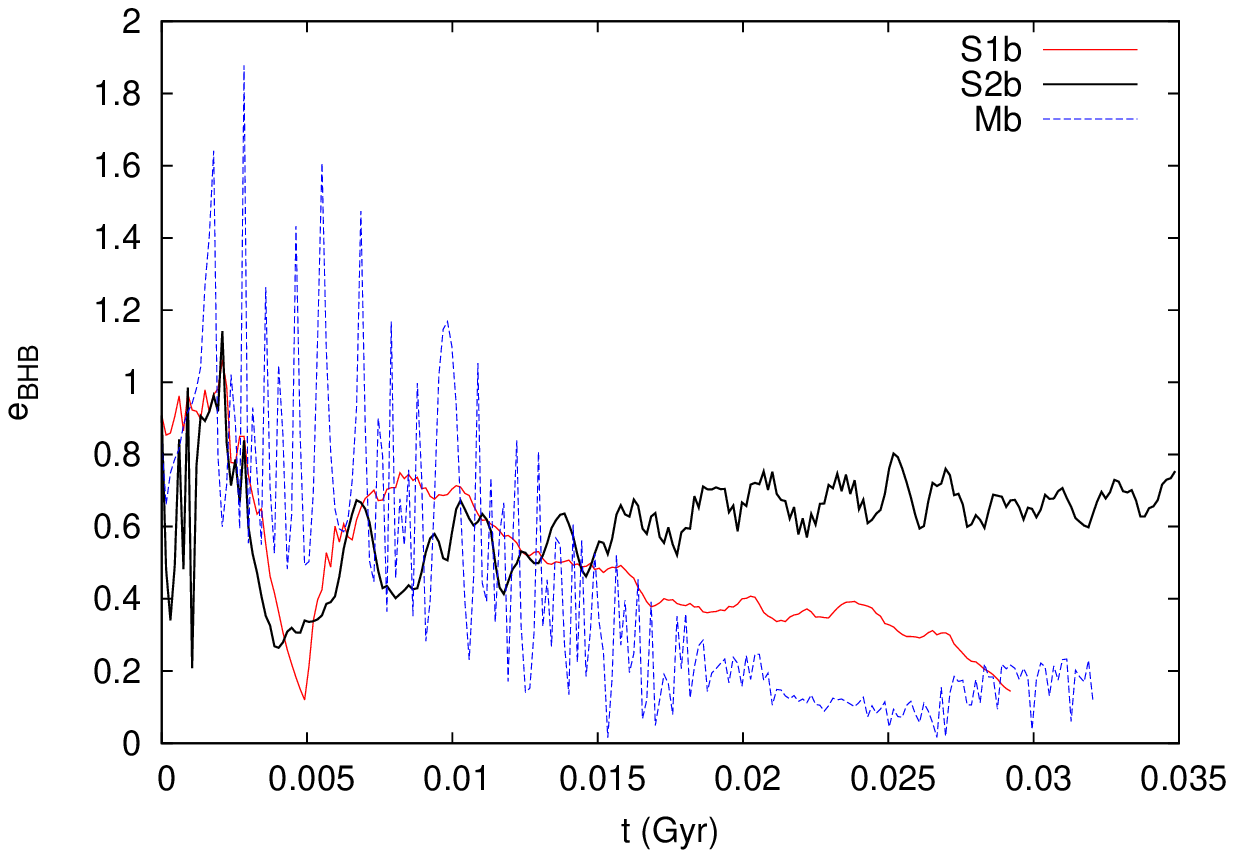}\\
\includegraphics[width=8cm]{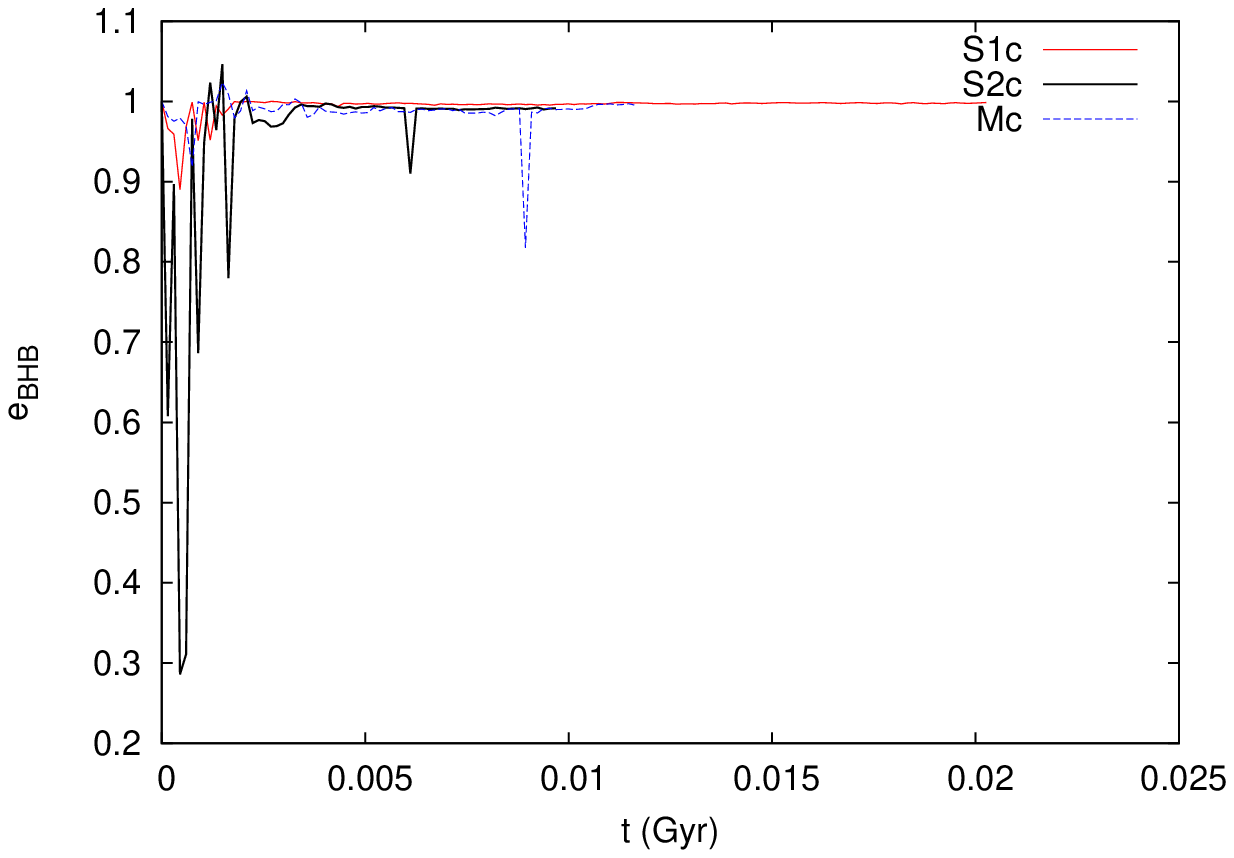}\\
\caption{Evolution of the BHB orbital eccentricity as a function of
  time. From top to bottom, panels refer to models ``a'' (circular
  orbit), ``b'' (eccentric orbit) and ``c'' (radial orbit).}
\label{fig:ecce}
\end{figure}

The evolution of the BHB eccentricity, $e_{\rm BHB}$, is shown in fig.\,\ref{fig:ecce} for all models. The effect of the initial GC orbit is clearly evident. When the GC moves on an initially circular orbit, the binary's eccentricity
evolves rapidly towards circularisation as the system hardens through
encounters. In the case of eccentric orbits, the binary tends to
circularise but is then subject to perturbations that lead to a
further growth in eccentricity. Finally, for GCs moving on radial
orbits the BHB retains a very large eccentricity in all cases.

Due to the high computational demands of the simulations presented
here we are not able to follow the evolution of the binary in the live
galaxy background up to 10 Gyr. We therefore estimate the merger
time-scale of the binaries by numerical integration of the equations
 governing the evolution of the semi-major axis and eccentricity, under
the assumption that this is due to both stellar encounters and, at
later times, emission of gravitational waves, i.e.
\begin{equation}
\frac{{\rm d}a}{{\rm d}t} = \frac{{\rm d}a}{{\rm d}t}\bigg|_* + \frac{{\rm d}a}{{\rm d}t}\bigg|_{\rm GW},
\label{top}
\end{equation} 
similarly to \citep{gualandris2012}.
Here the first term represents the evolution due to interactions with
stars and the second term that due to emission of GWs.

\begin{figure}
\centering
\includegraphics[width=8cm]{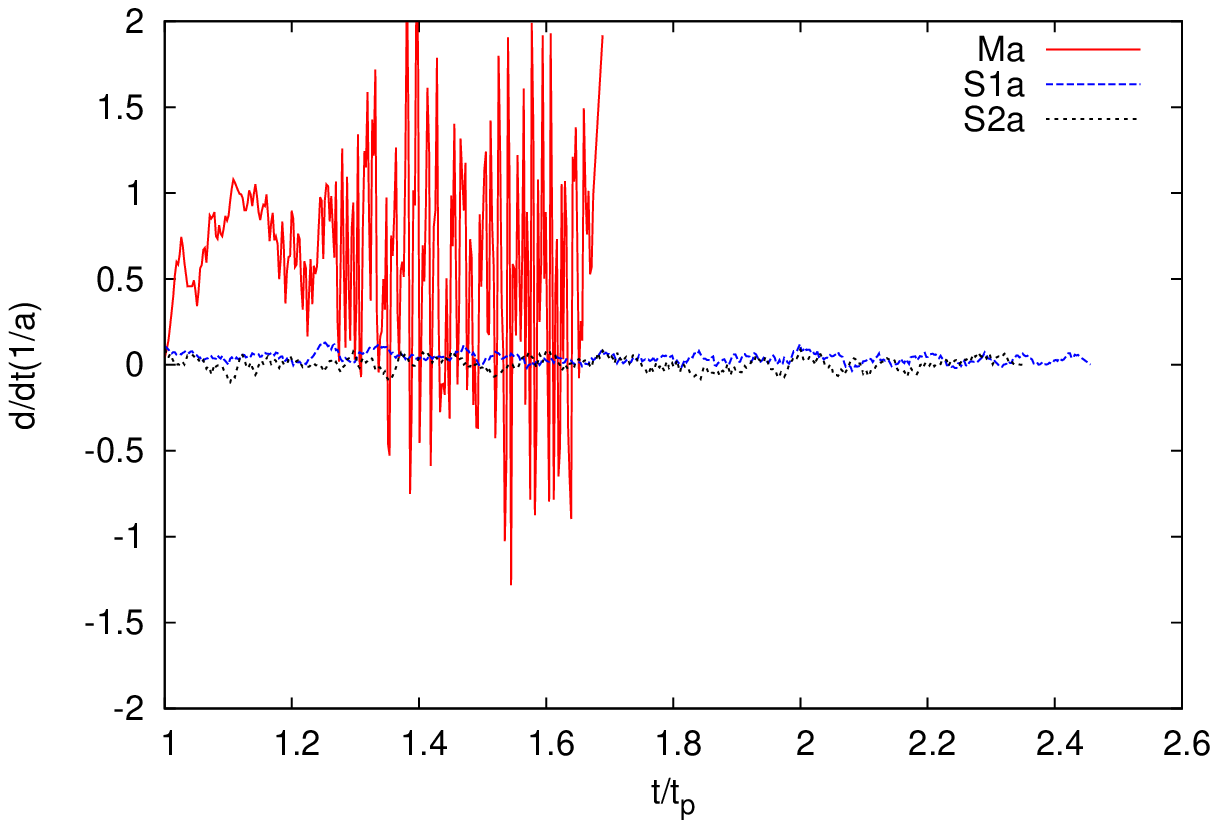}
\includegraphics[width=8cm]{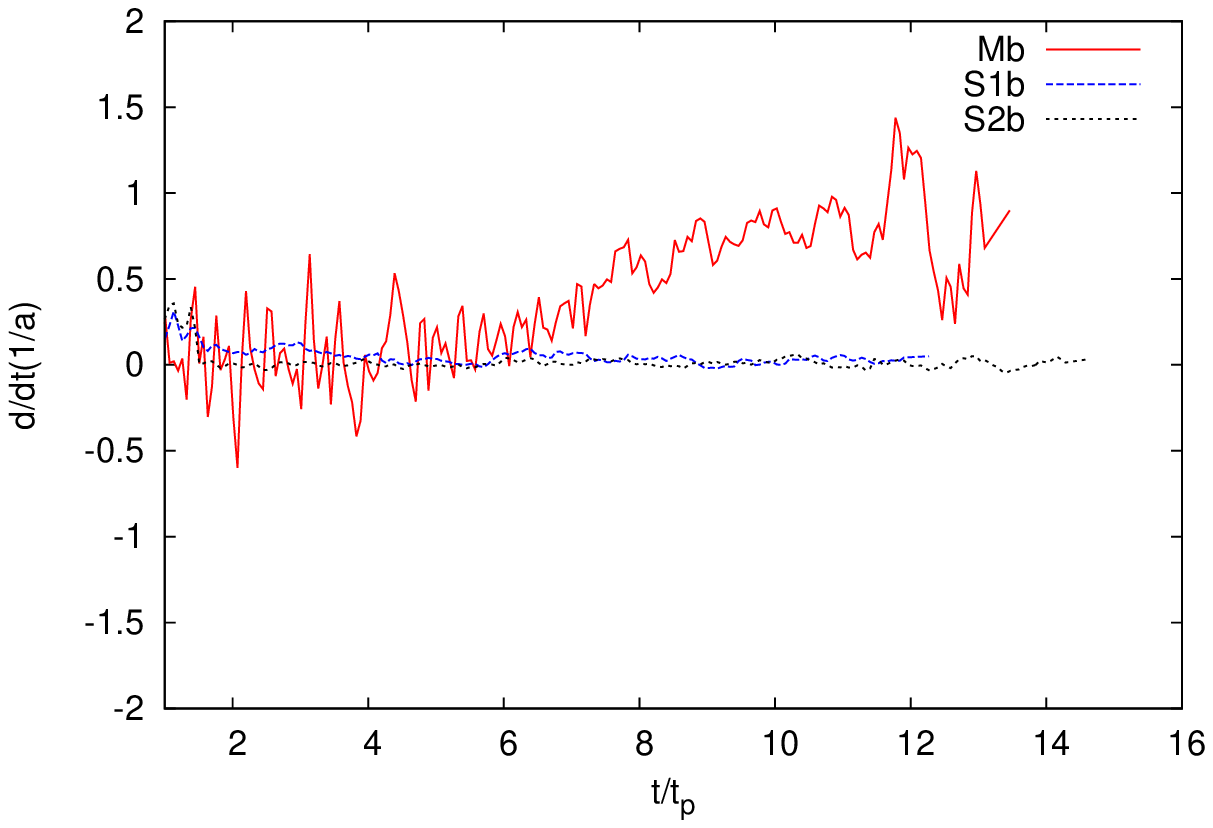}
\caption{Hardening rate as a function of time in models with a
  circular (top) and eccentric (bottom) orbit.}
\label{fig:hardrate}
\end{figure}

We compute the hardening rate of the binary
\begin{equation}
s = \frac{\rm d}{\rm dt}\left(\frac{1}{a}\right)
\label{eq:hardrate}
\end{equation}
by fitting $1/a$ over small time intervals with straight lines. The
time evolution of $s$ is shown in fig.\,\ref{fig:hardrate}. 
The hardening rate is roughly constant in time for models S1 and S2,
albeit rather small. On the other hand, it is much larger for models
Ma and Mb, though more noisy, due to the cuspier inner profile of the
galaxy model.

\begin{figure}
\centering
\includegraphics[width=8cm]{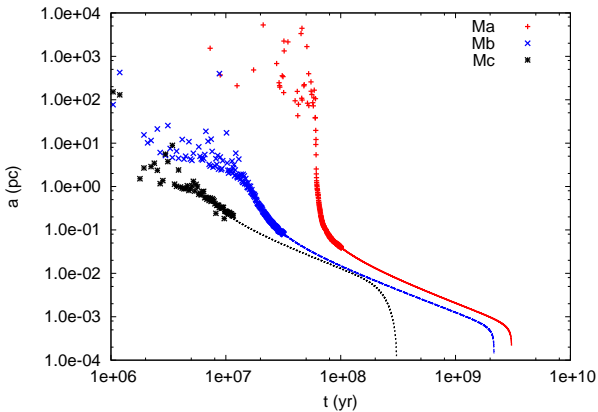}
\caption{Evolution of the BHB semi-major axis as obtained from the
  simulation (points) and the numerical integration (lines) in models
  M, in which the galaxy already harbors a central NSC during the GC
  infall.}
\label{fig:agw}
\end{figure}

Assuming that the BHB continues to harden at a constant hardening
rate, the term related to stellar interactions can be written as \citep{gualandris2012}
\begin{equation}
\frac{{\rm d}a}{{\rm d}t}\bigg|_* = -s^2 a(t),
\end{equation}
while the GW term can be obtained solving a system of two coupled differential equations \citep{peters64}
\begin{eqnarray}
\frac{{\rm d}a}{{\rm d}t}\bigg|_{\rm GW} &=& \frac{-64\beta}{5}\frac{F(e)}{a^3}\\
\frac{{\rm d}e}{{\rm d}t}\bigg|_{\rm GW} &=& \frac{-304\beta}{15}\frac{eG(e)}{a^4}
\end{eqnarray}
with
\begin{eqnarray}
F(e) &=& (1-e^2)^{-7/2}\left(1+\frac{73}{24}e^2+\frac{37}{26}e^4\right)\\
G(e) &=& (1-e^2)^{-5/2}\left(1+\frac{121}{304}e^2\right)\\
\beta &=& \frac{G^3}{c^5}M_1 M_2(M_1+M_2)\,.
\end{eqnarray}
The resulting evolution for the BHB semi-major axis is shown in
fig.\,\ref{fig:agw} for the M models, compared with the $N$-body data.

The merger time, i.e. the time from the beginning of GC inspiral to
binary coalescence, is given in table \ref{tab:tgw} for all models
containing an IMBH.
\begin{table}
\centering{}
\caption{Merger time for all the models investigated}
\label{tab:tgw}
\begin{tabular}{cc}
\hline
\hline
MODEL & $t_{\rm GW}$ (Gyr) \\
\hline
S1a & $26$\\
S1b & $2.2$\\
S1c & $1.3$\\
S2a & $679$\\
S2b & $77$\\
S2c & $6.6$\\
Ma  & $3.3$\\
Mb  & $2.0$\\
Mc  & $0.3$\\
\hline
\end{tabular}
\end{table}
If the galaxy has a shallow density profile, merger times are much
longer than a Hubble time, unless the cluster is initially on a highly
eccentric orbit. On the other hand, the presence of a central
over-density due to a NSC facilitates the hardening of the SMBH-IMBH
pair, leading the black holes to coalescence in a few billion years
from cluster disruption.

Our results can be used to infer the rate at which IMBH-SMBH
coalescence occurs in galaxies at low redshift.  In the context of the
dry-merger scenario, the number of GCs expected to segregate in a
Milky Way type like galaxy is $\sim 10$ \citep{AMB,ASCD15He,Tsatsi17}.
Due to the low IMBH formation probability, we assume that only one of
the clusters brought an IMBH to the galactic centre, $n_{\rm IMBH} =
1$.  In the assumption that the galaxy already hosts a NSC, we found a
$P_{\rm mer} = 100\%$ probability to have an IMBH-SMBH merger, thus
implying at least one event per nucleated galaxy. The fraction of
galaxies containing a NSC is $f_{\rm nc} = 0.7$, as suggested by many
observations \citep[][e.g.]{cote06,Turetal12}, although this
represents only an upper limit. As discussed above, the number density
of galaxies with stellar mass $10^{10}-10^{11}$ M$_\odot$ at redshift
0 is $n_g = 0.008$ Mpc$^{-3}$, while the time-scale of these events
can be obtained by table \ref{tab:tgw}, $t_{\rm mer} \simeq 2$ Gyr.
Hence, the rate of IMBH-SMBH mergers is given approximately by:
\begin{equation}
\Gamma_{\rm IMBH-SMBH} = \frac{f_{\rm nc}n_{\rm IMBH}P_{\rm mer}n_g}{t_{\rm mer}} = 
0.0028 ~{\rm yr}^{-1} ~{\rm Gpc}^{-1}.
\end{equation}
This relatively low number of events is due to the low
efficiency of IMBH formation, which is expected to be as low as $20\%$
\citep{Giersz15}, and to the small number of GCs that are expected to
contribute to the NSC formation in Milky Way type galaxies.

\subsection{Simulations of GCs with a central BH subsystem}
If instead the cluster contains stellar mass black holes in the
centre, BHs may be deposited close to the central SMBH as the cluster
inspirals.  
\begin{figure}
\centering
\includegraphics[width=8cm]{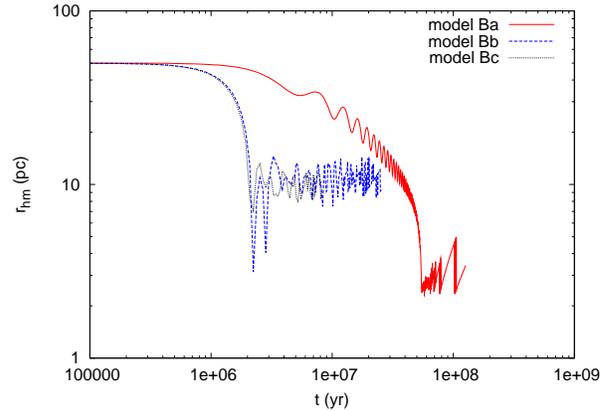}
\caption{Half-mass radius of the stellar mass black hole system in
  models Ba (solid line), Bb (dashed line) and Bc (dotted)  as a function of time during cluster inspiral.}
\label{fig:BHrhm}
\end{figure}
The half-mass radius of the BHs as a function of time is
shown in fig.\,\ref{fig:BHrhm} for all B models. BHs are deposited
closer to the SMBH in the case of a circular orbit.

\begin{figure}
\centering
\includegraphics[width=8cm]{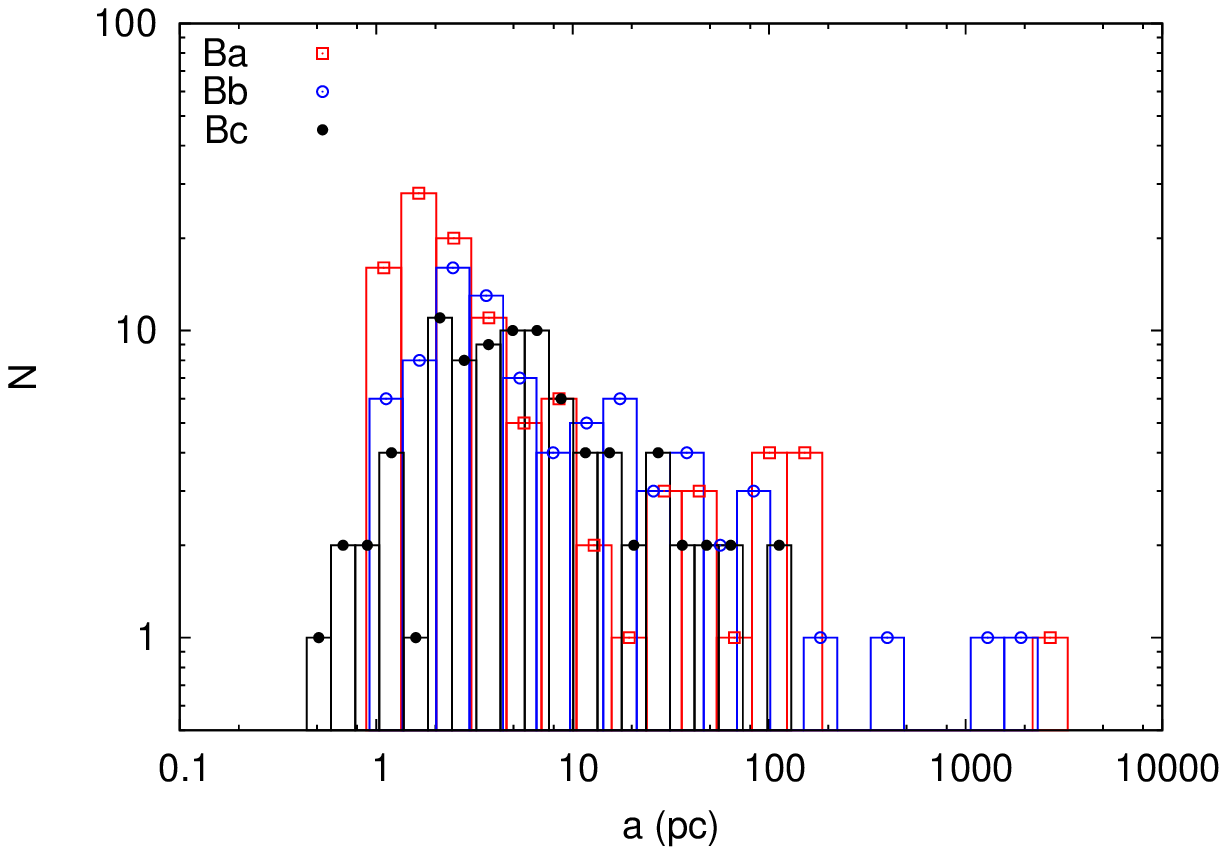}\\
\includegraphics[width=8cm]{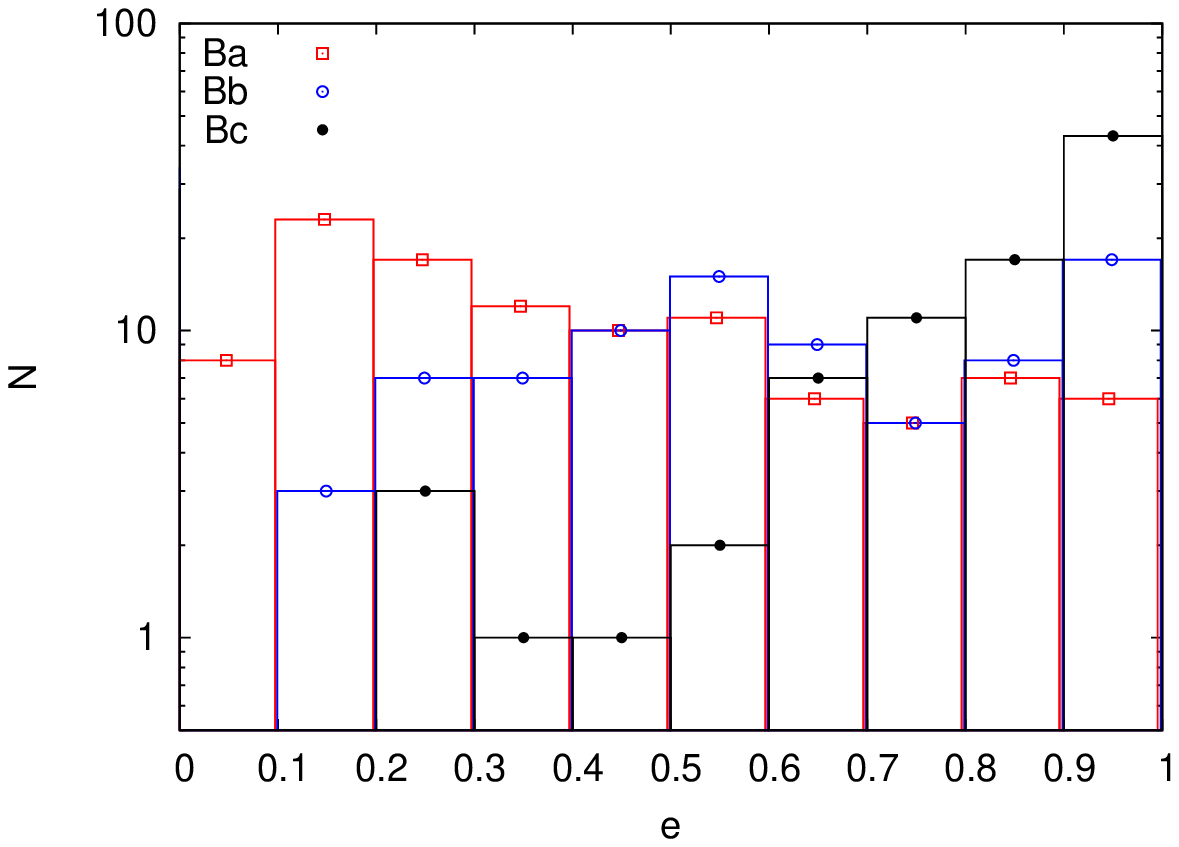}\\
\caption{Distribution of semi-major axis and eccentricity for all the
  bound BH-SMBH systems in the B models.}
\label{fig:MBHsemi}
\end{figure} 
The semi-major axis and eccentricity distribution for all SMBH-BH
bound pairs in the B models are shown in fig.\,\ref{fig:MBHsemi}.
Semi-major axes tend to be large, with a distribution that peaks at $a \sim 1-3\pc$ for the
circular orbit and around $5-7\pc$ for the eccentric and radial
orbits.  The eccentricity distribution, instead, partly reflects the
GC initial eccentricity: in model Ba the distribution is almost flat,
with a small peak at $e \simeq 0.15$; in model Bb the distribution
increases for values $e\lae 0.5$ and flattens beyond $e \sim 0.5$;
finally in model Bc the distribution is an increasing function of $e$.

We computed the time-scale $t_{\rm GW}$ to reach coalescence due to
emission of gravitational waves for all SMBH-BH binaries, according to Eq.\,\ref{eq:tgw}.
In our sample, only one binary in model Bb has $t_{\rm GW} <10$ Gyr.

We can estimate the probability to observe a stellar BH - SMBH
coalescence in a Hubble time through a simple relation:
\begin{equation}
P_{\rm mrg} = f_{\rm BH}\times f_{\rm orb},
\end{equation}
being $f_{\rm BH}=1/114$ the fraction of BHs with sufficiently small
$t_{\rm GW}$ and $f_{\rm orb}$ the fraction of GC with eccentricity in
the range $0.5-0.8$, compatible with our Bb model.  Assuming a thermal
distribution for the GC eccentricities, $P(e){\rm d}e = 2e{\rm d}e$
\citep{jeans19}, we find $f_{\rm orb}\simeq 0.39$. This implies
$P_{\rm mrg} = 0.34\%$ of observing a coalescence between a stellar BH
brought by an infalling cluster and an SMBH in the galactic nucleus.

Following \cite{seoane07}, we define a critical value for the semi-major axis below which this occurs
\begin{align}
\label{semitre}
a_{\rm EMRI} = & 5.3\times 10^{-2} \pc \, C_{\rm EMRI}^{2/3} \times   \nonumber  \\
 & \times\left(\frac{T_r}{1 {\rm Gyr}}\right)^{2/3} \left(\frac{m_\bh}{10\msun}\right)^{2/3}  \left(\frac{M_{\rm SMBH}}{10^6\msun}\right)^{-1/3}, 
\end{align}
being $C_{\rm EMRI}\lesssim 1$ and $T_r$ the relaxation time \citep{spitzer58,spitzer71}
\begin{equation}
T_r \simeq \frac{0.34\sigma_g^3}{G^2 {\rm ln}\Lambda \langle m\rangle \rho}.
\end{equation}
In our model, $\sigma \simeq 19.6\kms$ and $\rho \simeq
10^3\msun\pcc$ at $4\pc$ from the SMBH, the typical NSC length
scale. Assuming ${\rm ln}\Lambda = 6.5$ and $\langle m\rangle=0.62$
\Ms, typical of a Kroupa IMF \citep{kroupa01}, we find $T_r = 7.9$
Gyr.  For a $30\msun$ BH, this implies $a_{\rm EMRI} = 0.25\pc$, a
value comparable to the minimum value found in our B models.
Therefore, investigating a wider range of GC initial conditions may
allow to further investigate the possible formation of EMRIs through
this channel, which seems to be promising in galactic nuclei hosting
heavier SMBHs, $M_{\rm SMBH}\gtrsim 10^8\msun$ \citep{ASCD17c}.

 It must be noted that as long as the BH moves at distances
  $\sim 1\pc$ from the galactic centre, it will still subject to
  dynamical friction, which can cause a further decrease of the BH
  semi-major axis over a time-scale $\tau_{\rm df, BH}\propto
  m_\bh^{-0.67}$ \citep{ASCD14a}. This implies that our estimates
  above can slightly increase after $\tau_{\rm df, BH}$.  On the other
  hand, Eq.\,\ref{semitre} shows that $a_\emri$ scales with mass as
  $m_\bh^{2/3}$. For a population of lighter BHs with $M_\bh \sim
  10\msun$, typical of higher metallicities environments, the
  corresponding semi-major axis would decrease by $50\%$, $a_\emri
  \simeq 0.12\pc$, thus reducing the probability of EMRIs formation by
  infalling GCs.  Further investigations of this channel would
  require more realistic simulations, in which the full stellar mass
  spectrum is covered for both the infalling GC and the Galactic
  nucleus.

\begin{figure}
\centering
\includegraphics[width=8cm]{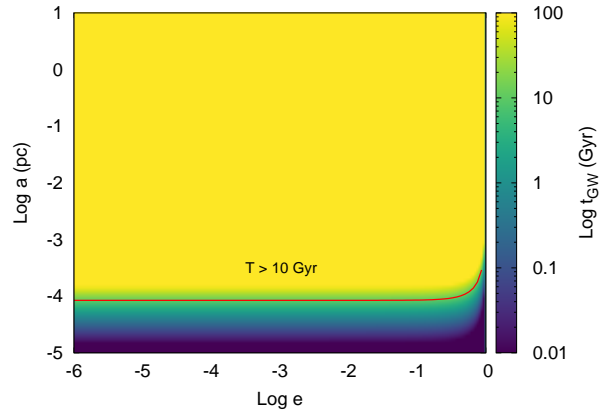}
\caption{Surface map showing how the GW time-scale for a stellar
  BH-MBH varies as a function of the orbital semi-major axis and
  eccentricity. Here we assumed $M_{\rm SMBH} = 5\times 10^6\msun$ and
  $M_{\rm *BH}=30\msun$.}
\label{mapgw}
\end{figure}

According to Eq.\,\ref{eq:tgw}, a stellar $30\msun$ BH falling toward
a SMBH would emit a burst of GW within a Hubble time only for orbital
eccentricities $e_{\rm *MBH}\gtrsim 0.7$ and semi-major axis $a_{\rm
  *MBH}<10^{-3}\pc$, as shown in Fig.\,\ref{mapgw}.

We estimate the number of BHs that can be captured as extreme mass-ratio inspirals (EMRIs) and merge within a Hubble time as
\begin{equation}
N_{\rm EMRI} = P_{\rm mrg} \left(0.001N_{\rm GC,av}\right)
\left(f_c\frac{0.01M_g}{M_{\rm GC,av}}\right)\simeq 3800,
\end{equation}
where the first term represents the probability for EMRIs coalescence
as found in our models, the second term is the number of BHs per
typical GC, and the third term is the number of GCs that reach the
galactic centre. This takes place on a relaxation time, which at the
edge of the SMBH influence radius can be calculated as $t_r = 10^9
{\rm yr} (M_{\rm SMBH}/10^5\msun)^{5/4} \simeq 10^2$ Gyr \citep{oleary12}.

As discussed above, the number density for galaxies with masses around
$10^{10}\msun$ in the local Universe is $n_g=0.008$ Mpc$^{-3}$, thus
implying an EMRIs rate from this channel:
\begin{equation}
\Gamma_{\rm EMRI}= \frac{N_{\rm EMRI}n_g}{t_r} = 0.25 {\rm yr}^{-1} \gpcc, 
\end{equation}
a value comparable to the EMRIs rate predicted for strongly segregated
BH populations around SMBHs similar to the Milky Way's SMBH \citep{seoane11}. 
Indeed, infalling GCs contribute to the enrichment of compact stellar
remnants in the galactic centre, acting on the GC dynamical friction
time-scale, which is much shorter than for stellar BHs.

\section{Conclusions}
\label{end}
Inspiralling globular clusters represent a leading scenario for the
formation of nuclear star clusters. Here, we model the formation and
evolution of gravitational wave sources during the inspiral of a
cluster containing either an intermediate mass black hole or a cluster
of stellar mass black holes. The cluster infall is followed by means
of 12 state-of-the-art direct summation $N$-body simulations, in which
the background galaxy is also integrated on a star-by-star basis. The
initial conditions are chosen to model different cluster orbits and
both a shallow and steep galaxy density profile.  Our main results can
be summarized as follows:
\begin{itemize}
\item In clusters hosting an IMBH, a IMBH-SMBH bound system forms
  after the GC is disrupted depositing stars in the galactic
  nucleus. The evolution and hardening of the binary depends
  sensitively on the slope of the galaxy density profile.  In shallow
  galaxy models, hardening proceeds slowly and eventually stalls at
  separations too large for emission of GWs to become dominant. Merger
  time-scales are longer than a Hubble time, unless the GC is on a
  highly eccentric orbit. In such cases, the black holes are expected
  to merge within $1-7$ Gyr.  In steep galaxy models, hardening
  proceeds quickly and leads the black holes to coalesce within $\sim
  3-4$ Gyr, depending on the GC's initial orbit.
\item The rate of formation of IMBH-SMBH binaries in nucleated galaxies
  is $\Gamma_{\rm IMBH-SMBH} = 2.8\times 10^{-3}$ yr$^{-1}\gpcc$.
\item In clusters hosting a population of stellar mass black holes,
  BHs are transported to the galaxy centre, where a fraction bind to
  the SMBH and form EMRIs. The merger rate for these systems is
  estimated as $\Gamma_{\rm BH-SMBH} = 0.25$ yr$^{-1}\gpcc$.
\item Stellar black holes also bind in binaries, about $2.5\%$ of which
  coalesce within $\sim 3$ Myr due to the tidal field from the
  SMBH. The corresponding merger rate for this channel is $\Gamma_{\rm
    BHB,SMBH} = 0.4-4$ yr$^{-1}\gpcc$, depending on the number
  of BHBs deposited by the infalling GC.
\end{itemize}

\section*{Acknowledgements}
MAS acknowledges the University of Rome Sapienza, which funded part of
this research through the grant 52/2015, in the framework of the
``MEGaN project: modelling the evolution of galactic nuclei''. MAS
also acknowledges the Sonderforschungsbereich SFB 881 ``The Milky Way
System'' (subproject Z2) of the German Research Foundation (DFG) for
the financial support provided.  
Part of the numerical simulations presented here were performed on the Milky Way supercomputer, which is funded by the Deutsche Forschungsgemeinschaft (DFG) through the Collaborative Research Center (SFB 881) "The Milky Way System" (subproject Z2) and hosted and co-funded by the J\"ulich Supercomputing Center (JSC).
We thank James Petts, Fabio Antonini
and Monica Colpi for interesting discussions.  We thank Martina
Donnari for having assisted us in obtaining the galaxy stellar mass
function from the Illustris data release.  We thank Roberto
Capuzzo-Dolcetta for giving us access to the high-performance
computing workstations ASTROC9, ASTROC15 and ASTROC16b, hosted in the
Physics Department of the University of Rome "Sapienza". 

%\clearpage
\footnotesize{
\bibliographystyle{mn2e}
\bibliography{bblgrphy}
}

\end{document}